\pdfoutput=1
\documentclass[a4paper,11pt]{article}
\usepackage{jheppub} 
\usepackage[T1]{fontenc} 
\usepackage{graphicx,slashed,multirow}

\newcommand{\feynrules}{{\sc FeynRules}}
\newcommand{\madgraph}{{\sc MadGraph}}
\newcommand{\pythia}{{\sc Pythia}}
\newcommand{\nc}{\newcommand}
\nc{\be}{\begin{equation}}   \nc{\ee}{\end{equation}}
\nc{\bea}{\begin{eqnarray}}   \nc{\eea}{\end{eqnarray}}
\nc\bpm{\begin{pmatrix}}      \nc\epm{\end{pmatrix}} 
\def\bsp#1\esp{\begin{split}#1\end{split}}
\def\d{{\rm d}}
\newcommand{\met} {\ensuremath{E\!\!\!\!/_T}}

\newcommand{\ie}{{\it i.e.}}

\renewcommand{\sl}{{\tilde\ell}}
\newcommand{\st}{{\tilde\tau}}
\nc{\thst}{{\theta_\st}}

\preprint{CERN-PH-TH/2013-240, MS-TP-13-35}

\title{Revisiting slepton pair production at the Large Hadron Collider}
\author[a,b]{Benjamin Fuks,}
\author[c]{Michael Klasen,}
\author[c]{David R.\ Lamprea}
\author[c]{and Marcel Rothering}

\affiliation[a]{Theory Division, Physics Department, CERN, CH-1211 Geneva 23,
  Switzerland}
\affiliation[b]{Institut Pluridisciplinaire Hubert Curien/D\'epartement
  Recherches Subatomiques, Universit\'e de Strasbourg/CNRS-IN2P3,
  23 rue du Loess, F-67037 Strasbourg, France}
\affiliation[c]{Institut f\"ur Theoretische Physik, Westf\"alische
  Wilhelms-Universit\"at M\"unster,\\ 
  Wilhelm-Klemm-Stra\ss{}e 9, D-48149 M\"unster, Germany}
\emailAdd{fuks@cern.ch}
\emailAdd{michael.klasen@uni-muenster.de}
\emailAdd{david.lamprea@uni-muenster.de}
\emailAdd{marcel.rothering@uni-muenster.de}

\abstract{
 Motivated by the shift in experimental attention towards electroweak supersymmetric particle
 production at the CERN LHC, we update in this paper our precision predictions at
 next-to-leading order of perturbative QCD matched to resummation at the next-to-leading
 logarithmic accuracy for direct slepton pair production in proton-proton collisions.
 Simplified models, now commonly adopted by the experimental collaborations for selectrons and
 smuons as well as mixing staus, are used as benchmarks for total cross sections at achieved
 and future center-of-mass energies. They are presented together with the corresponding scale
 and parton density uncertainties in graphical and tabular form for future reference.
 Using modern Monte Carlo techniques, we also reanalyze recent ATLAS and CMS slepton searches
 in light of our precision cross sections and for various assumptions on the decomposition of
 the sleptons and their neutralino decay products.
}


\begin{document}

\maketitle
\flushbottom


\section{Introduction}
After almost fifty years of experimental tests,
the Standard Model has proved to be
a very successful theory of the fundamental particles and interactions.
Nevertheless, it leaves many important
questions, like the hierarchy of the fundamental mass scales,
unanswered, and it is widely believed to represent a low-energy
limit of a more fundamental theory.
In particular, the recent discovery of a Higgs
boson~\cite{Aad:2012tfa,Chatrchyan:2012ufa}, still to be confirmed
as being \textit{the} Standard Model
Higgs boson, has rather reinforced our expectation to find physics
beyond the Standard Model
at the Large Hadron Collider (LHC) at CERN. This discovery is very likely the first
observation of a particle intrinsically unstable with respect to quantum corrections,
which requires either unnatural fine-tuning or stabilization
from new particles lying around the TeV scale.
This issue is addressed by large classes of new physics theories and in particular
by weak-scale supersymmetry~\cite{Nilles:1983ge,Haber:1984rc},
which also solves a considerable
number of other problems inherent in the Standard Model. For instance, it hints
at the possible unification of the gauge symmetries at high energies and provides
a candidate particle explaining the presence of dark matter in the Universe.

As a consequence, searches for supersymmetry constitute one of the key topics
of the present experimental program in high-energy particle physics. Up to now,
both the ATLAS and CMS collaborations have mainly focused on dedicated analyses of
signatures arising from the strong production of squarks and gluinos,
the partners of the strongly interacting quarks and gluons. However,
no evidence for such particles has been found, so that squark and
gluino masses are constrained to lie at higher and higher
scales~\cite{atlassusy,cmssusy}. The pair production of the electroweak superpartners,
\textit{i.e.}, the neutralino, chargino and slepton eigenstates, has
therefore recently received more and more
attention from both collaborations~\cite{Aad:2012pxa,Aad:2012hba,Aad:2012jja,%
ATLAS:2013yla,ATLAS-CONF-2013-049,%
Chatrchyan:2012pka,Chatrchyan:2012mea,Chatrchyan:2011ff,CMS:aro,CMS-PAS-SUS-13-006}.
In addition, it has also
been shown that the weak channels
can be interesting probes for distinguishing minimal from non-minimal
supersymmetry~\cite{Krauss:2012ku,%
Frank:2012ne,Bharucha:2013epa,Alloul:2013fra,Cerdeno:2013qta}
and could also provide an explanation for the recent anomalous multilepton
events observed by the CMS collaboration~\cite{D'Hondt:2013ula}.

All these phenomenological and experimental studies so far rely on theoretical
predictions valid either at the leading order \cite{Dawson:1983fw,Baer:1993ew,%
Bozzi:2004qq,Bozzi:2007me,Debove:2008nr} or at the next-to-leading order
\cite{Berger:1999mc,Beenakker:1999xh,Berger:2000iu,Spira:2002rd} of
perturbative QCD, which may lead to large theoretical uncertainties.
For more accurate predictions, and subsequently for more precise
limits on or determination of the electroweak
superpartner masses, the next-to-leading order results must be supplemented by QCD
resummation, since soft-gluon radiation can give rise to large logarithmic terms
that have to be resummed to all orders in the strong coupling
constant~\cite{Bozzi:2006fw,Bozzi:2007qr,Bozzi:2007tea,%
Debove:2009ia,Debove:2010kf,Debove:2011xj,Fuks:2012qx}. In addition,
a matching with fixed order computations is mandatory for consistent
predictions in all kinematical regions.

Motivated by these observations, we have studied in a previous work
resummation effects on chargino and neutralino pair production at the LHC for
the already achieved
center-of-mass energies of 7~TeV and 8~TeV \cite{Fuks:2012qx}.
Additionally, we have also very recently
released a user-friendly computer code, dubbed {\sc Resummino},
allowing to perform such precision
calculations for arbitrary supersymmetric scenarios and
collision energies~\cite{Fuks:2013vua}.
In this paper, we turn to slepton pair production and present accurate
predictions for total cross sections at the LHC, focusing on benchmark
scenarios based on simplified models as currently employed by both the
ATLAS and CMS collaborations. We provide several cross section tables for
LHC collisions at center-of-mass energies of 7 TeV, 8 TeV, 13 TeV and
14~TeV at the leading-order and next-to-leading order of perturbative QCD,
and
after matching the results with soft-gluon resummation. We also take the
opportunity of performing a detailed study of the theoretical
uncertainties, their control being largely improved by soft-gluon
resummation. Including this information in the tables, we collect in this
way the most precise cross section reference values so that they are
available for all future ATLAS and CMS analyses of slepton pair production
and decay.

In a second step, we reinterpret two of the most recent slepton
LHC analyses. On the one hand, we include resummed predictions for total
cross sections. On the other hand, we investigate the dependence of the
predictions on the left-handed or right-handed nature of the slepton
eigenstates, on its flavor, as well as on the bino, wino or higgsino nature
of the lightest supersymmetric particle, assumed to be the lightest
neutralino, which sleptons decay into. Experimental results indeed usually
include either a sum over the slepton flavors or a sum over both slepton
chiralities and focus on some specific neutralino decomposition. We
demonstrate, in this work, that while the exact definition of the
neutralino nature can be safely ignored, the expected limit strongly
depends on the chosen patterns for the slepton nature. Therefore, making
available results where only one specific slepton type is considered could
be useful in the future, in particular when one aims for a recasting
of the experimental analyses in the context of models where only one given
slepton is light.

This work is organized as follows. In Section \ref{sec:bench}, we first describe a set of
benchmark scenarios appropriate for slepton pair production studies. Next, we
investigate in Section~\ref{sec:xsec} slepton production cross sections
and the corresponding scale and parton density function uncertainties. We then reinterpret,
by means of state-of-the-art Monte Carlo simulations, two of the more recent slepton LHC analyses
in Section~\ref{sec:MC} and focus on the variations of the results with respect to both
precision predictions and the exact nature of the supersymmetric particles. Our results are summarized
in Section~\ref{sec:conclusions}, and in Appendix~\ref{app:highcms} we list
total cross section expectations at next-to-leading logarithmic accuracy for
future center-of-mass energies of 13~TeV and 14~TeV.

\section{Benchmark scenarios for slepton pair hadroproduction and decays}
\label{sec:bench}

In order to probe the usually large parameter spaces of several
beyond the Standard Model theories simultaneously,
simplified production and decay schemes have been developed. In this approach,
model-independent search strategies are designed on the basis of
models consisting of extensions of the Standard Model with minimal
additions in terms of particles and interactions. These simplified models
are constructed upon a very specific final state topology so that one can test
generic features possibly common to several new physics models. Moreover, this framework aims
for an easy comparison of theoretical predictions with data and reinterpretation of
the experimental results in the case of complete theories.

\begin{figure}
  \centering
  \includegraphics[width=.30\columnwidth]{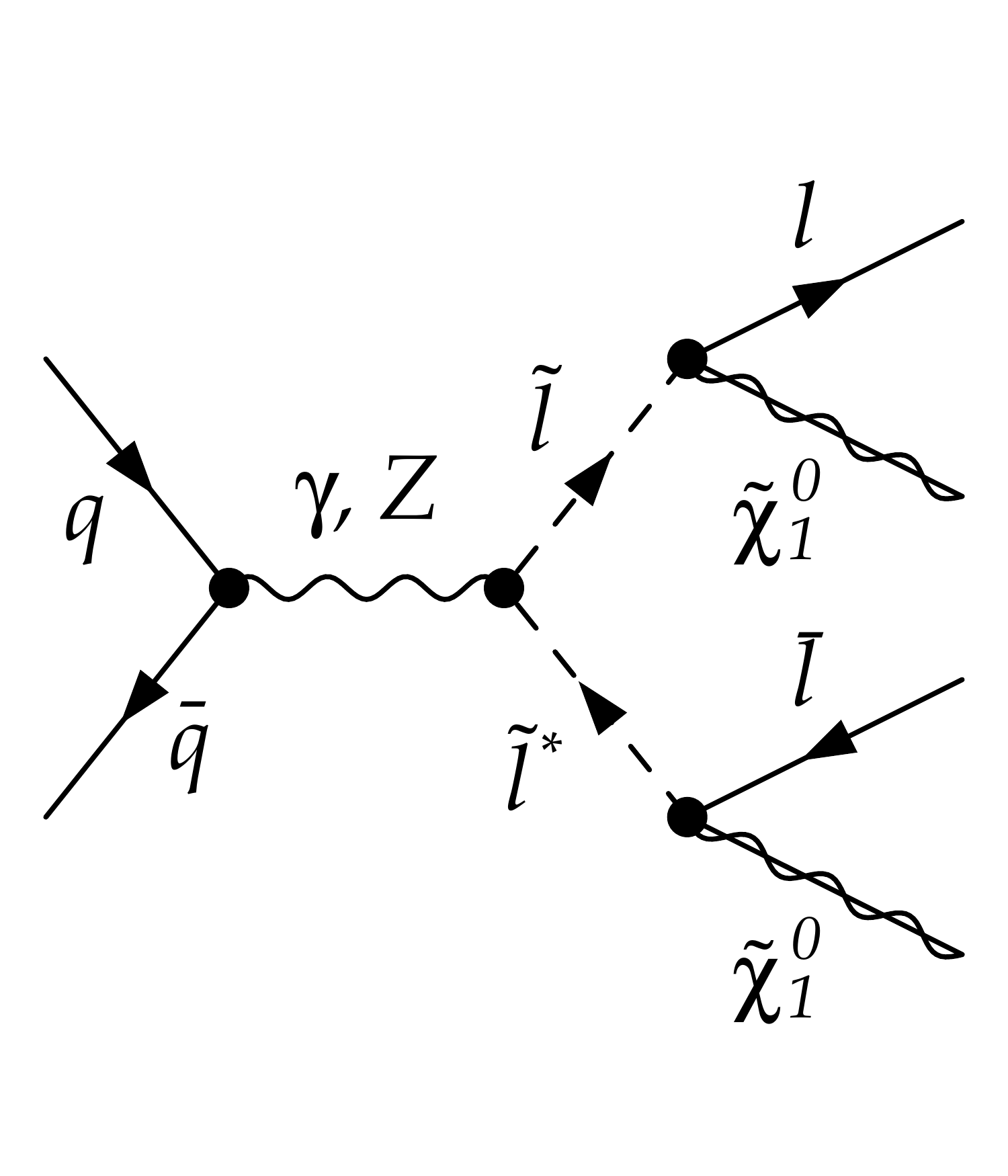}
  \vspace{-.2cm}
\caption{\label{fig:diag}Leading-order Feynman diagram depicting slepton pair production
  and decay in the simplified models under consideration.
  The symbol $\sl$ stands for any generation of (s)leptons.}
\end{figure}

In this section, we present two series of simplified models relevant for slepton
pair production and decay at the LHC. Inspired by the constrained version of the Minimal
Supersymmetric Standard Model (MSSM), sleptons $\tilde\ell$ are assumed to be
light and always decay in
a flavor-conserving way into the lightest supersymmetric particle, taken as the lightest
neutralino $\tilde\chi_1^0$,
and the corresponding Standard Model lepton $\ell$. This pattern,
illustrated in Figure~\ref{fig:diag}, has motivated
several experimental analyses where sleptons are searched for in
final states comprised of a same-flavor lepton pair
produced in association with missing energy \cite{ATLAS:2013yla,ATLAS-CONF-2013-049,CMS:aro}.

In the considered scenarios, the Standard Model is supplemented by two supersymmetric particles,
a charged slepton $\tilde\ell$ and the lightest neutralino $\tilde\chi^0_1$,
whereas the other supersymmetric particles are decoupled. This holds in particular
for the strong superpartners
associated with negative search results~\cite{atlassusy,cmssusy}.
The flavor and left-handed or right-handed nature of the considered
slepton, together with the gaugino or higgsino nature of the lightest neutralino,
allow us to define several series of benchmark scenarios which we describe in the rest of this section.
First and second generation sleptons are addressed in Section~\ref{sec:bench12}, while
third generation sleptons are considered in Section~\ref{sec:bench3}.

\subsection{First and second generation sleptons}
\label{sec:bench12}

We start with the Standard Model field content and its $SU(3)_C \times
SU(2)_L \times U(1)_Y$ gauge group. We then add a four-component
Majorana spinor $\Psi_N$ representing the lightest neutralino $\tilde\chi_1^0$
as well as one charged selectron or smuon field
that we generically denote by $\sl_L$ or $\sl_R$. The $L/R$ subscript refers to the chirality
of the slepton Standard Model partner that is described by the Dirac field $\Psi_\ell$.
The associated Lagrangians are extracted from the MSSM
Lagrangian and read,
in the mass eigenbasis,
\be\bsp
  {\cal L}^{(L)} =&\ {\cal L}_{\rm SM} +\partial_\mu\sl^\dag_L \partial^\mu\sl_L
    + \frac{i}{2} \bar\Psi_N\gamma^\mu\partial_\mu\Psi_N
    + i e \Big[\partial_\mu \sl_L^\dag \sl_L  - \sl_L^\dag \partial_\mu \sl_L\Big] A_\mu
\\ &\
    - \frac{i e}{s_W c_W} {\cal C}^{(L)}_Z
        \Big[\partial_\mu \sl_L^\dag \sl_L  - \sl_L^\dag \partial_\mu \sl_L\Big] Z_\mu
    +\bigg[
      \frac{1}{2 c_W s_W} \bar\Psi_N {\cal C}^{(L)}_N P_L \Psi_\ell \sl_L^\dag
      + {\rm h.c.} \bigg]\ , \\
  {\cal L}^{(R)} =&\ {\cal L}_{\rm SM} +\partial_\mu\sl^\dag_R \partial^\mu\sl_R
    + \frac{i}{2} \bar\Psi_N\gamma^\mu\partial_\mu\Psi_N
    + i e \Big[\partial_\mu \sl_R^\dag \sl_R  - \sl_R^\dag \partial_\mu \sl_R\Big] A_\mu
\\ &\
    - \frac{i e}{s_W c_W} {\cal C}^{(R)}_Z
        \Big[\partial_\mu \sl_R^\dag \sl_R  - \sl_R^\dag \partial_\mu \sl_R\Big] Z_\mu
    +\bigg[
      \frac{1}{2 c_W s_W} \bar\Psi_N {\cal C}^{(R)}_N P_R \Psi_\ell \sl_R^\dag 
      + {\rm h.c.} \bigg] \ ,
\esp\ee
for left-handed and right-handed sleptons, respectively.
Additional interactions, such as
four-scalar or neutralino-gauge interactions,
are allowed by gauge invariance and supersymmetry. Moreover, contributions
to the neutralino-lepton-slepton vertices proportional to the
muon or electron Yukawa couplings are also present in the MSSM.
However, in the context of the current work, these interactions are
either phenomenologically irrelevant or negligibly small, so that
they have been omitted for simplicity.

We have introduced, in the previous Lagrangians,
the photon and $Z$-boson fields $A_\mu$ and $Z_\mu$, the sine and cosine of the electroweak
mixing angle $s_W$ and $c_W$ and the electromagnetic coupling constant $e$. Furthermore,
in our conventions, the different
coupling strengths are given by
\be\label{eq:lagparam12}\bsp
  {\cal C}^{(L)}_Z = &\ -\frac12 + s_W^2 \ , \\
  {\cal C}^{(L)}_N = &\ \sqrt{2} e \Big[s_W N_1^\ast + c_W N_2^\ast \Big]\ , \\
  {\cal C}^{(R)}_Z = &\ s_W^2 \ , \\
  {\cal C}^{(R)}_N = &\ -2 \sqrt{2} e s_W N_1 \ ,
\esp\ee
where the complex quantities $N_1$ and $N_2$ stand for the bino and wino component
of the lightest neutralino mass-eigenstate.
For further reference, we denote the two higgsino components
of the lightest neutralino by $N_3$ and $N_4$,
and the four mixing parameters are thus constrained by
the unitarity relation
\be
  |N_1|^2 + |N_2|^2 + |N_3|^2 + |N_4|^2 =1 \ .
\label{eq:unitarity}\ee

This allows us to define a
particular series of benchmark models as follows. First,
the two parameters $M_\sl$ and $M_{\tilde\chi^0_1}$ representing
the slepton and neutralino masses must be fixed, recalling that the neutralino has to be
lighter than the slepton.
Next, the flavor and chirality\footnote{Strictly
speaking, a scalar field has no chirality and
we always refer to the chirality of the Standard
Model partner.} of the slepton state have to be defined. Finally,
it is also necessary to specify the gaugino/higgsino nature of the neutralino
by fixing three of the four $N_i$ quantities, the fourth one
being derived by Eq.~\eqref{eq:unitarity}.

\subsection{Simplified models for staus}
\label{sec:bench3}
In order to account for third generation sleptons, the simplified model described above
has to be slightly  modified. In contrast to the first two
generations, the superpartners of the left-handed and right-handed taus are expected to
mix as
\be
  \bpm \st_1 \\ \st_2\epm = 
  \bpm \phantom{-}\cos\thst & \sin\thst\\ -\sin\thst & \cos\thst\epm
  \bpm \st_L \\ \st_R\epm \ ,
\ee
where the mass eigenstates $\st_1$ and $\st_2$ are mass-ordered from the lighter to the heavier.
Consequently, we focus on the production of the lightest of the two states, for which the
associated cross section is larger.

To describe the interactions of the $\st_1$ state,
we construct a simplified model similarly to the one introduced in Section~\ref{sec:bench12}
and supplement the Standard Model with a state $\Psi_N$
representing the lightest neutralino and a
scalar field $\st_1$ denoting the lighter supersymmetric
partner of the tau lepton, described by the
Dirac spinor $\Psi_\tau$. The associated Lagrangian reads
\be\bsp
  {\cal L}^{(\tau)} =&\ {\cal L}_{\rm SM} +\partial_\mu\st_1^\dag \partial^\mu\st_1
    + \frac{i}{2} \bar\Psi_N\gamma^\mu\partial_\mu\Psi_N
\\ &\
    + i e \Big[\partial_\mu \st_L^\dag \st_1  - \st_1^\dag \partial_\mu \st_1\Big] A_\mu
    - \frac{i e}{s_W c_W} {\cal C}^{(\tau)}_Z
        \Big[\partial_\mu \st_1^\dag \st_1  - \st_1^\dag \partial_\mu \st_1\Big] Z_\mu
\\ &\
    +\frac{1}{2 c_W s_W} \bigg[
      \bar\Psi_N {\cal C}^{(\tau,L)}_N P_L \Psi_\tau \st_1^\dag
      + \bar\Psi_N {\cal C}^{(\tau,R)}_N P_R \Psi_\tau \st_1^\dag 
      + {\rm h.c.} \bigg]\ ,
\esp \ee
where we have again omitted all the interactions irrelevant for stau pair production and decay at the LHC.
The Lagrangian parameters can be deduced from Eq.~\eqref{eq:lagparam12} after accounting for
appropriate gauge-eigenstate mixings,
\be\bsp
  {\cal C}^{(\tau)}_Z = &\ \Big[-\frac12 + s_W^2\Big] \cos^2\thst + \Big[  s_W^2  \Big] \sin^2\thst\ , \\
  {\cal C}^{(\tau,L)}_N = &\ \sqrt{2} e \Big[s_W N_1^\ast + c_W N_2^\ast \Big] \cos\thst
    - \Big[2 c_W s_W N_3^\ast y_\tau\Big] \sin\thst\ , \\
  {\cal C}^{(\tau,R)}_N = &\ \Big[-2 \sqrt{2} e s_W N_1\Big] \sin\thst 
    - \Big[2 c_W s_W N_3 y_\tau\Big] \cos\thst\ ,
\esp\ee
where $y_\tau$ denotes the tau lepton Yukawa coupling, which in this
case cannot be neglected.

Simplified models describing
tau sleptons are thus defined by two mass parameters,
namely the lightest stau and lightest neutralino masses $M_{\st_1}$ and $M_{\tilde\chi_1^0}$,
three out of the four neutralino mixing parameters (the $N_i$ quantities)
as well as the stau mixing angle $\thst$. The fourth neutralino component is again
extracted using Eq.~\eqref{eq:unitarity}.

\section{Precision predictions for slepton pair production at the LHC}
\label{sec:xsec}

In this section, we present total cross sections relevant for direct slepton searches
as performed by the ATLAS and CMS experiments. We focus on both past LHC runs at
center-of-mass energies of 7~TeV and 8~TeV, while results for possible future runs at
center-of-mass energies of 13~TeV and 14~TeV are presented in Appendix~\ref{app:highcms}.
Further predictions
can be made available by employing the {\sc Resummino} package~\cite{Fuks:2013vua},
that can be downloaded from the webpage

\verb+http://www.resummino.org+

\noindent or be obtained from the authors upon request.

\subsection{General features}\label{sec:xsecgeneral}
In the pioneering and subsequent more recent works, slepton pair production has been
studied at the leading-order~\cite{Dawson:1983fw,Baer:1993ew,Bozzi:2004qq}
and next-to-leading order~\cite{Beenakker:1999xh} of perturbative QCD.
All of these cross section calculations rely on the QCD factorization theorem.
Rendering the dependence on the slepton-pair invariant mass $M$
explicit, the results are obtained by
convoluting the partonic cross section $\d\hat\sigma_{ab}$ with the
universal distribution functions $f_a$ and $f_b$ of partons
$a$ and $b$ carrying the momentum fractions $x_a$ and $x_b$ of the colliding hadrons,
\be\label{eq:fixed}\bsp
  M^2 \frac{\d \sigma}{\d M^2}(\tau) =&\ \sum_{ab}
   \int_0^1 \! \d x_a \d x_b \d z\  \delta(\tau \!-\! x_ax_bz) \Big[x_a f_a(x_a,\mu_F^2) \Big]
   \Big[x_b f_b(x_b, \mu_F^2)\Big]\\&\quad \times
   \Big[z \d\hat\sigma_{ab}(z,M^2,\mu_F^2,\mu_R^2)\Big] \ .
\esp\ee
In this equation, $\mu_R$ and $\mu_F$ stand for the factorization and
renormalization scales and $\tau=M^2/S$, $\sqrt{S}$ being the hadronic center-of-mass energy.
Although the partonic cross section $\d\hat\sigma_{ab}$ exhibits logarithmic terms that are
large close to the production threshold, $z\lesssim 1$, the convergence
of the perturbative series is recovered at the hadronic level
thanks to the vanishing parton densities
when the momentum fraction is close to one. This however yields significant scale uncertainties
making the theoretical predictions less precise.

These logarithmic terms are not surprising and originate from soft and collinear
radiation of partons by the initial
state particles. It has been proved long ago that in this case, once transformed into
the appropriate conjugate space, the cross section can be refactorized. Consequently,
the large logarithms can be resummed to all orders
in the strong coupling, which implies an important reduction
of the theoretical uncertainties by including in the predictions
higher order effects in a consistent way.
More specifically, this threshold resummation formalism is based on the fact that
we can write the cross section, in the conjugate Mellin $N$-space,
as
\be
  M^2 \frac{\d\sigma}{\d M^2}(N-1)=\sum_{ab} f_a(N,\mu_F^2)
    f_b(N,\mu_F^2) \hat\sigma_{ab}(N,M^2,\mu_F^2, \mu_R^2)\ ,
\ee
where the Mellin $N$-moments of the quantities $F = \sigma$, $\hat\sigma_{ab}$ and $f_{a,b}$
are given, in our conventions, by
\be
  F(N) = \int_0^1 \d y\ y^{N-1}\ F(y)  \ ,
\ee
with $y = \tau$, $z$ and $x_{a,b}$, respectively. Reorganizing the different
contributions to the partonic
cross section, the latter can be recast in a closed exponential form~\cite{Sterman:1986aj,%
Catani:1989ne,Catani:1990rp,Berger:1996ad,Kidonakis:1997gm,Berger:1997gz,
Kidonakis:1998bk,Vogt:2000ci,
Kramer:1996iq, Catani:2001ic, Kulesza:2002rh, Almeida:2009jt},
\be
  \hat\sigma_{ab}(N,M^2,\mu_F^2, \mu_R^2) =
  {\cal H}_{ab}(M^2, \mu_F^2,\mu_R^2) \exp\Big[{\cal G}_{ab}(N,M^2,\mu_F^2,\mu_R^2)\Big] \ .
\label{eq:resumpart}\ee
This expression contains two functions, the hard function ${\cal H}$ and the Sudakov
form factor ${\cal G}$. The hard function
can be derived from the perturbative results by first isolating the terms
independent of the Mellin variable
$N$ and then collecting the dominant $1/N$-terms
stemming from universal collinear radiation of the initial state partons.
In contrast, the Sudakov form factor embeds the dominant dependence on $N$,
\ie, it contains soft and collinear parton emission and
absorbs in this way all large logarithmic contributions.
For more details on the explicit form
of the ${\cal H}$ and ${\cal G}$ functions, we refer to Ref.~\cite{Fuks:2013vua}.

Far from threshold, the perturbative computation, only partially accounted for
in Eq.~\eqref{eq:resumpart}, is expected to be reliable. Therefore,
predictions valid in all kinematical regions require a consistent matching of the
fixed order with the resummed results. This is achieved by
subtracting from the sum of both predictions their overlap
calculated by expanding analytically the resummed cross section at
the same order in the strong coupling as the one employed for the fixed order calculation.

Finally, contrary to the fixed order results that are computed in
physical space using Eq.~\eqref{eq:fixed}, the resummed
component and its expansion at a given order are evaluated in Mellin space. Therefore,
it is necessary to transform them back to the physical space.
Care must be taken with the singularities
possibly arising at the level of the $N$-space cross section. To this aim, we choose an
integration contour inspired by the principal value procedure and the minimal
prescription~\cite{Contopanagos:1993yq,Catani:1996yz}, which satisfies two properties.
First, the poles associated with the Regge singularity of the
Mellin moments of the parton densities lie to the left of the integration
contour. Second, the Landau pole related to the running of the strong
coupling lies to its right.

Threshold resummation has been initially performed in the context of
slepton pair production several years ago. This computation combines
soft gluon resummation at the next-to-leading logarithmic accuracy and
a fixed order calculation at the next-to-leading order~\cite{Bozzi:2007qr}.
It should be noted that soft gluon resummation can also be performed
when the transverse momentum of the produced slepton pair $p_T$
is small \cite{Bozzi:2006fw}, as this gives also rise to possibly large logarithmic terms,
or simultaneously in both regimes, \ie, when $p_T$ is small and
$M$ close to threshold~\cite{Bozzi:2007tea}.

\subsection{Total cross section computations}

\begin{figure}
  \centering
  \includegraphics[width=.656\columnwidth]{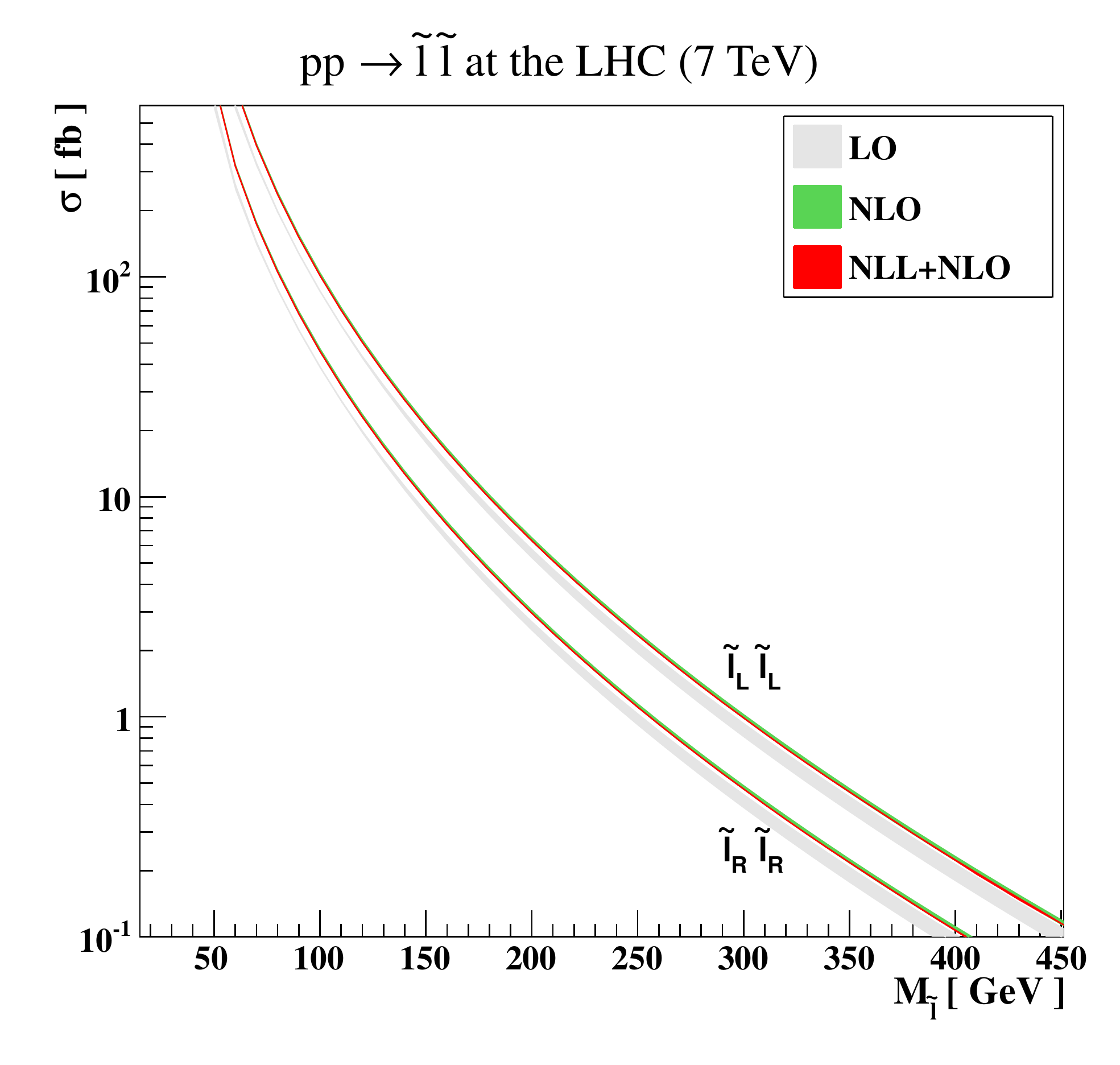}
  \includegraphics[width=.656\columnwidth]{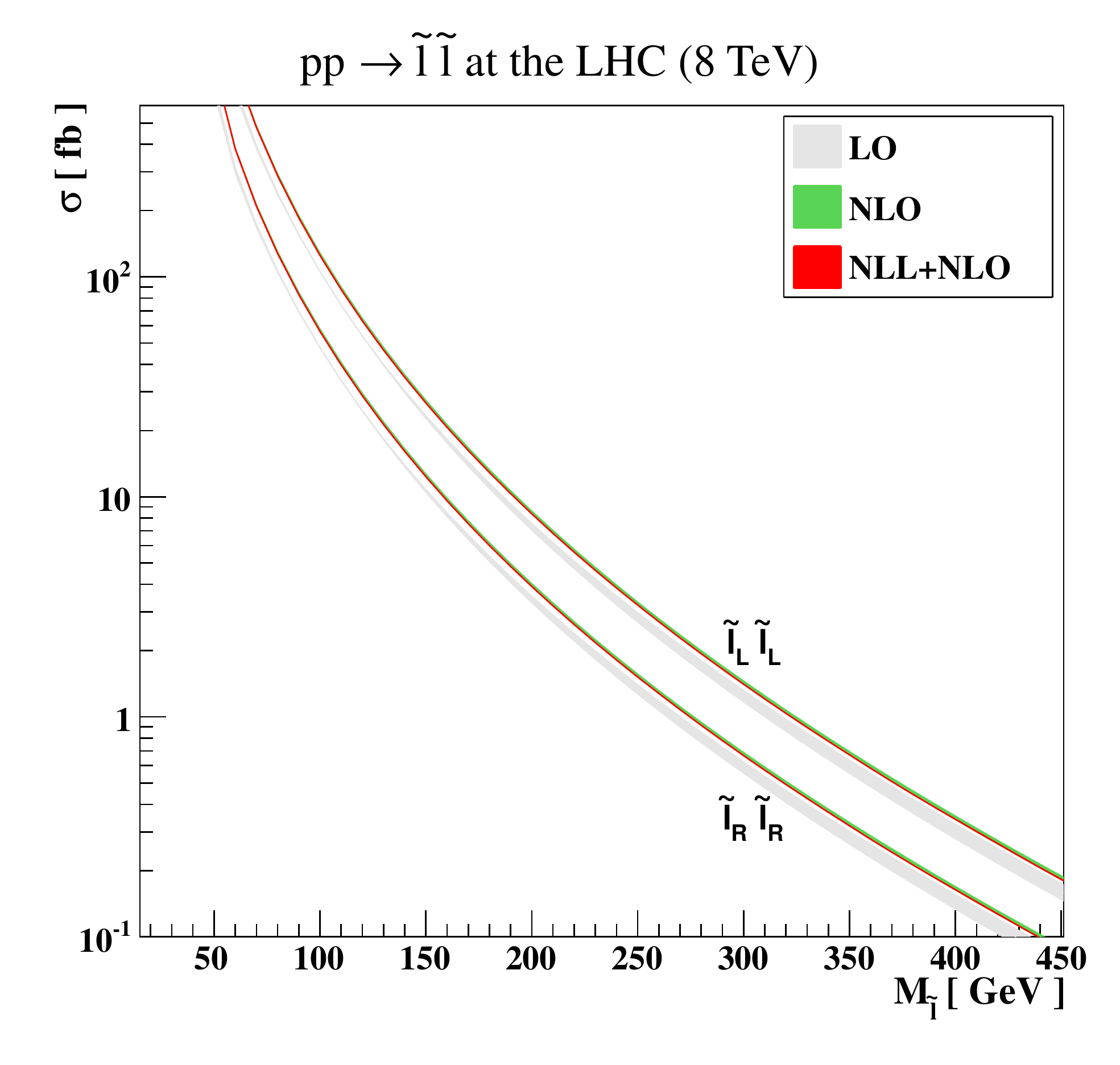}
\caption{\label{fig:massplot}Total cross sections for
  slepton pair production at the LHC, running at
  center-of-mass energies of 7~TeV (upper panel) and 8~TeV (lower panel). We present
  predictions as functions of the slepton mass $M_\sl$
  at LO (gray) and NLO
  (green) of perturbative QCD and after matching the NLO results with threshold
  resummation at the NLL accuracy (red). The
uncertainty bands correspond to variations induced by modifications of the
unphysical scales in the $[1/2 M_{\sl}, 2 M_\sl]$ range.}
\end{figure}

We dedicate this section to an overview of the behavior
of the total rate for slepton pair production at the LHC as a function
of the slepton mass $M_\sl$. We fix Fermi's coupling constant and
the masses and widths of the electroweak gauge bosons
to their latest measured values~\cite{Beringer:1900zz}
and compute the squared sine of the electroweak mixing
angle and the electromagnetic fine structure constant in the improved
Born approximation. We then numerically evaluate
total cross section predictions
at the leading order (LO) and next-to-leading order (NLO) of perturbative QCD,
and after matching the NLO results with a computation including the
resummation of the threshold
logarithms at the next-to-leading logarithmic accuracy (NLL+NLO). Focusing
on both past runs of the LHC, we present results for the
series of benchmark models introduced in Section~\ref{sec:bench12} in
Figure~\ref{fig:massplot} for
collider center-of-mass energies of 7~TeV (upper panel of the figure)
and 8~TeV (lower panel of the figure). While
the LO predictions are obtained after convoluting the partonic
results with the leading order fit of the CTEQ6 parton densities~\cite{Pumplin:2002vw},
we employ the more recent NLO CT10 fit of the CTEQ collaboration~\cite{Lai:2010vv} for our
NLO and NLL+NLO calculations. In all cases, both factorization
and renormalization scales are fixed to
the mass of the produced slepton $\mu_R = \mu_F = M_{\sl}$.

We present, in Figure~\ref{fig:massplot}, results for a
restricted slepton mass range of $M_\sl \in [0,450]$~GeV, which corresponds to
parameter space regions where
cross sections of at least about 0.1~fb are expected for both
the pair production of left-handed and right-handed slepton pairs.
Equivalently, in the case a supersymmetric model with light sleptons
is realized in Nature, those regions feature the presence of at least
a few slepton events in the 2011-2012 LHC data, although these events are
hidden in an overwhelming Standard Model
background. Furthermore, cross sections greater than about 10 fb
are found for left-handed
(right-handed) sleptons whose mass satisfies
$M_{\sl} \lesssim 200$~GeV (150~GeV), so that the corresponding
benchmark scenarios are appealing models for the possible observation of
hints for sleptons at the LHC.
This is addressed more into details in Section~\ref{sec:MC}.

In addition, Figure~\ref{fig:massplot} also shows that
large $K$-factors are obtained when including NLO corrections
to the predictions, reaching up to 1.25 in the low-mass region
($M_\sl~\gtrsim~50$~GeV)
and 1.15 for heavier sleptons ($M_\sl \gtrsim 200$~GeV). Matching
the NLO results to a computation including the threshold resummation
of the soft gluon radiation at the NLL accuracy
does not drastically modify the NLO $K$-factor,
although theoretical uncertainties estimated by varying the unphysical
scales by a factor of two around the central scale, chosen to be
the mass of the produced sleptons, are found to be largely reduced.
This property is addressed in greater detail in Section~\ref{sec:theouncertainties}.

\begin{figure}
  \centering
  \includegraphics[width=.66\columnwidth]{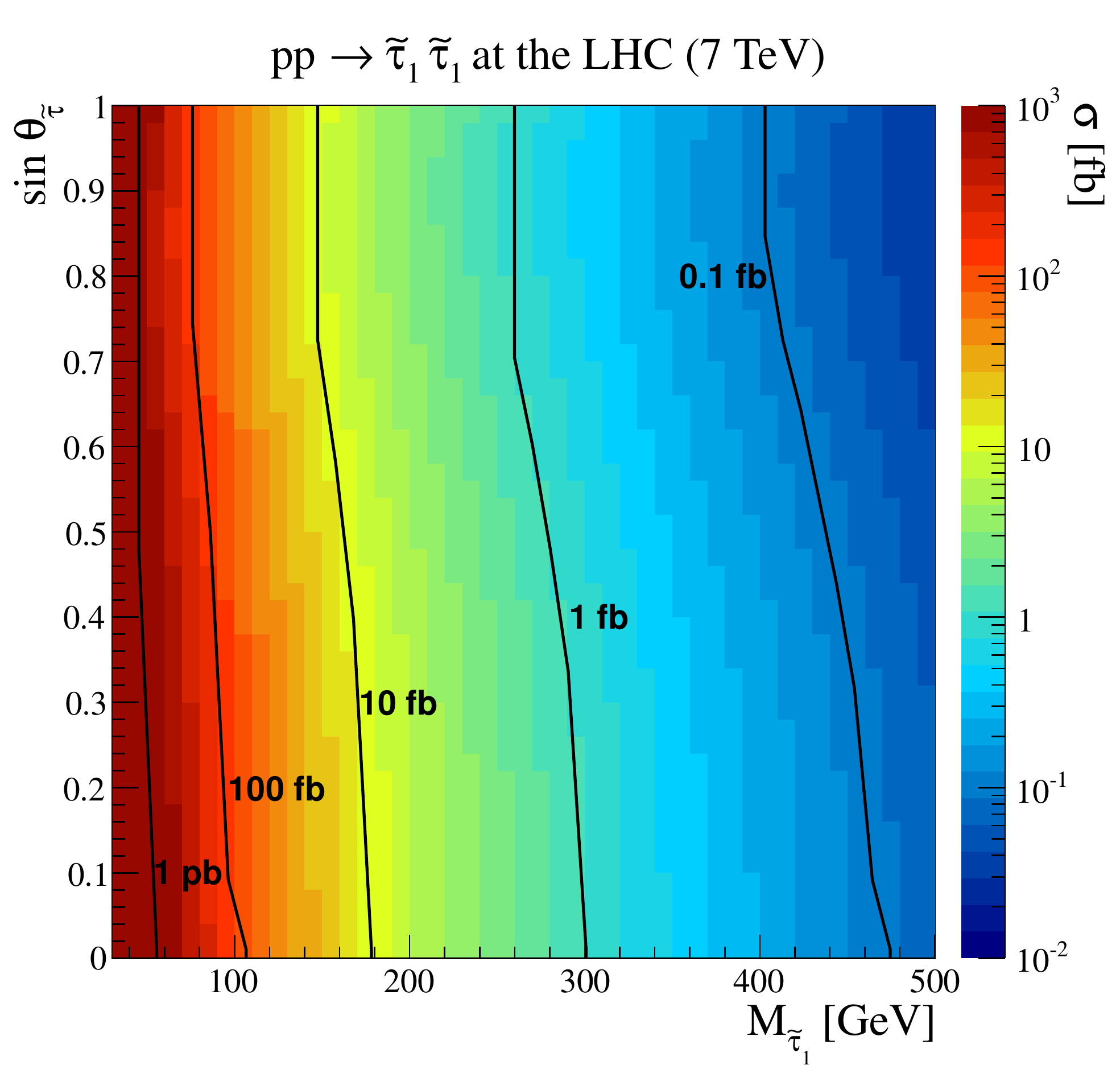}
  \includegraphics[width=.66\columnwidth]{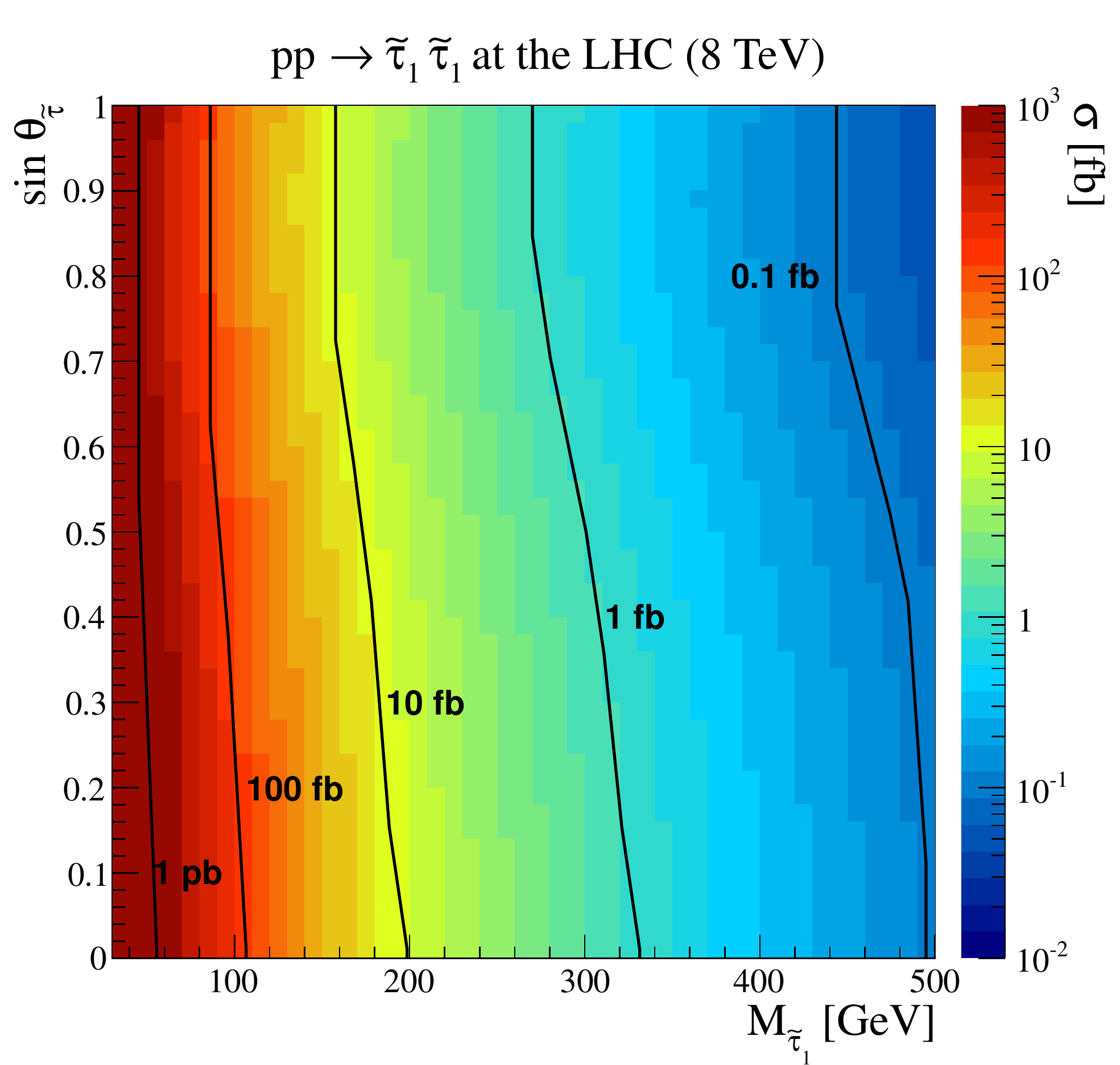}
\caption{\label{fig:massplotstau}Total cross sections for stau
  pair production at the LHC, running at
  center-of-mass energies of 7~TeV (upper panel) and 8~TeV (lower panel). We present
  predictions as functions of the stau mass and the stau mixing
  angle after matching the NLO results with threshold resummation at the NLL
  accuracy.}
\end{figure}

In Figure~\ref{fig:massplotstau}, we investigate
stau pair production in the framework of the simplified models
introduced in Section~\ref{sec:bench3}. We show the dependence of the
total cross section, computed after matching the resummed results
with the NLO predictions for both LHC center-of-mass
energies of 7~TeV (upper panel of the figure)
and 8~TeV (lower panel of the figure),
on both the stau mass $M_{\st_1}$ and the sine of the mixing
angle $\sin\theta_\st$. As before, we convolute the partonic
cross sections with the CT10 NLO parton
distribution functions (PDFs) and fix the unphysical scales to the mass
of the produced stau eigenstate $\mu_F = \mu_R = M_{\st_1}$.
Similarly to the first and second generation slepton cases,
cross sections larger than about 10~fb are obtained
for stau masses ranging up to about $150-200$~GeV for both center-of-mass
energies. However, from the perspective of the possible
observation of stau events at the LHC, the situation is expected to be
more complicated due to tau reconstruction efficiencies. We refer to
Section~\ref{sec:MC} for more details.

As already found for smuon and selectron pair production,
total cross sections associated with the production of a pair of
left-handed sfermions are larger than those related to the production
of a pair of right-handed sfermions.
Stau eigenstates mainly constituted of a left-handed component
are therefore more easily produced than when the
right-handed component is dominant. In the rest of this paper,
we adopt a scenario where the stau mixing is maximal, \ie,
$\theta_\st = \pi/4$.
One observes in Figure~\ref{fig:massplotstau} that this
implies a cross section value almost minimal for stau
lighter than 200~GeV and slightly larger than the minimal value
for heavier staus. A large mixing can however allow for sizable
loop-induced contributions to stau pair production from bottom-antibottom
or gluon-pair initial states~\cite{Lindert:2011td}.
This has not been considered in the present work.

\subsection{Theoretical uncertainties}\label{sec:theouncertainties}

\begin{figure}
  \centering
  \includegraphics[width=.65\columnwidth]{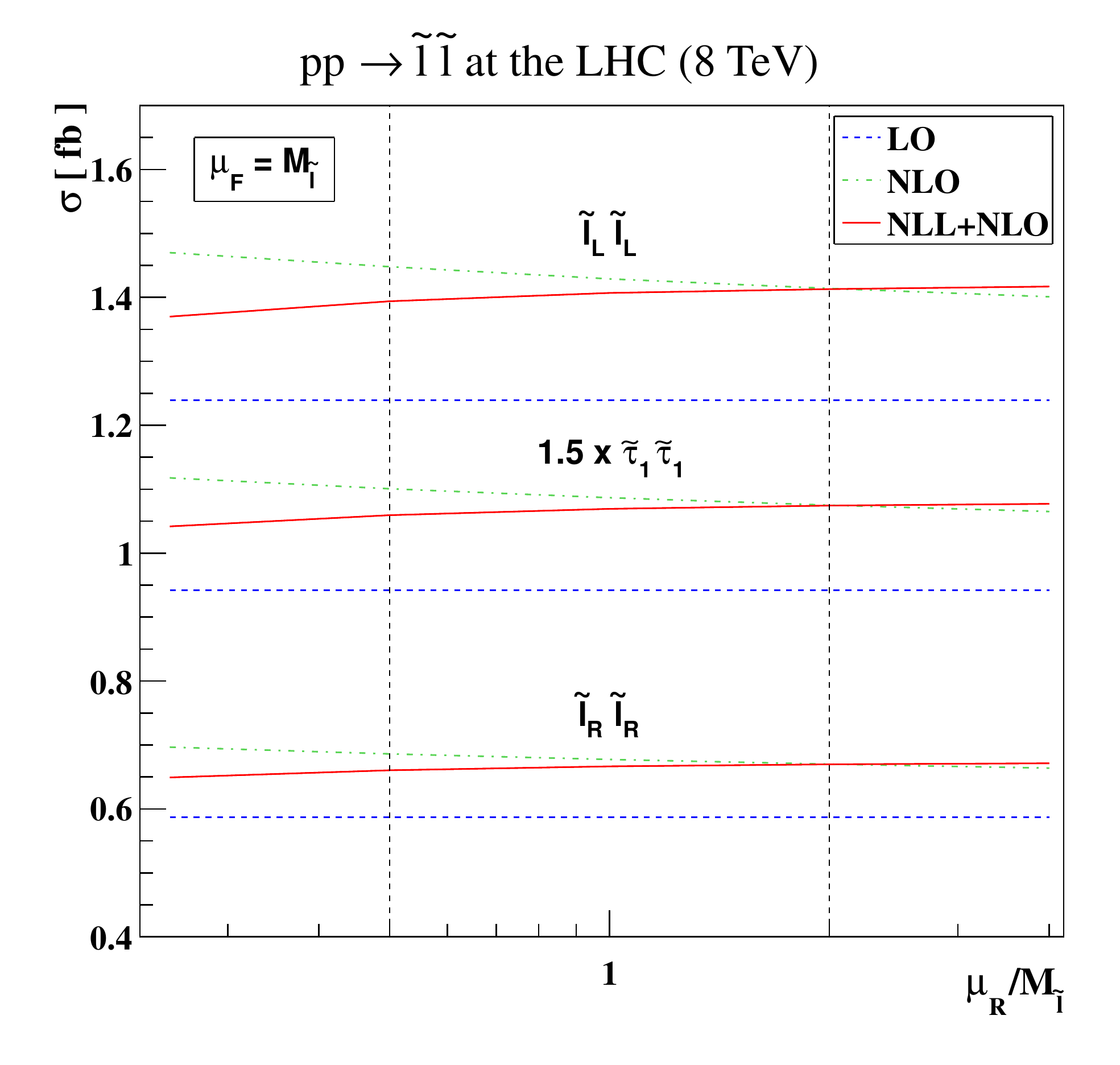}
  \includegraphics[width=.65\columnwidth]{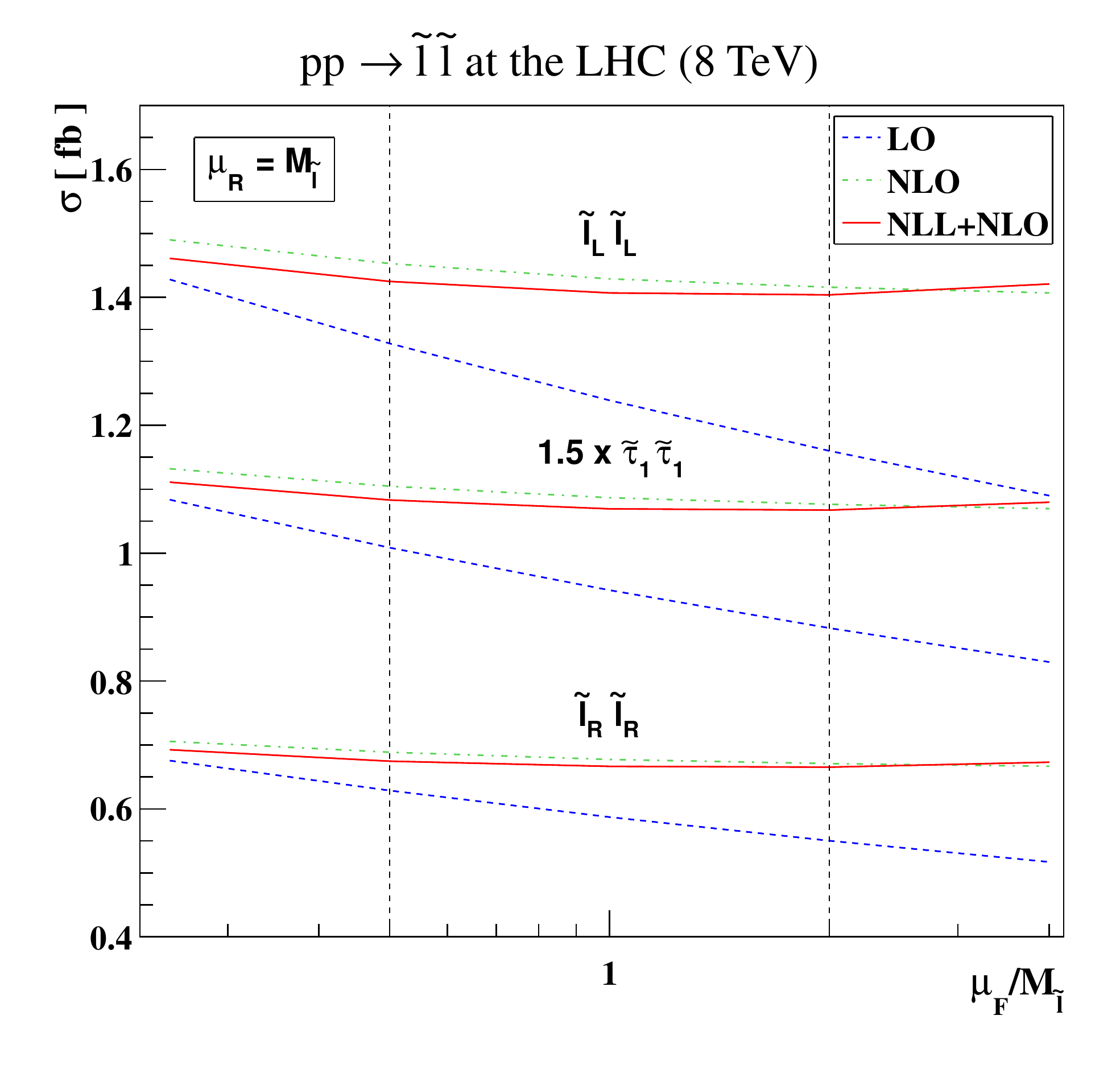}
\caption{\label{fig:scale}Total cross sections for the production of a pair of
  left-handed first or second generation sleptons, right-handed first or second
  generation sleptons and maximally mixing staus at the
  LHC, running at a center-of-mass energy of 8~TeV. We depict
  the dependence of the results on the renormalization (upper panel) and factorization
  (lower panel) scales at the LO (dashed blue), NLO (dashed-dotted green) and NLL+NLO
  (plain red) accuracy.}
\end{figure}

In the previous subsection, we have briefly mentioned that
slepton pair production total cross section predictions
are stabilized with respect to scale variations when resummation
techniques are employed.
The NLL+NLO prediction have indeed been found much less dependent on the common
unphysical scale
$\mu_R=\mu_F=M_\sl$ than the LO and NLO expectations, where the
central scale value is taken as the mass of the produced sleptons.
In this section, we investigate
this property more into detail and show, in Figure~\ref{fig:scale},
the scale dependence of the total cross sections for different subprocesses,
 a slepton mass fixed to 300~GeV
and a center-of-mass energy of 8~TeV.
We separately consider variations of the renormalization
(upper panel of the figure) and factorization (lower panel of the figure) scale
for the production of a left-handed first or second
generation slepton pair, a right-handed first or second generation slepton pair
and for a stau pair where the stau is defined as a maximal admixture
of the left-handed and right-handed stau gauge-eigenstates.

As mentioned above, we fix, on the upper panel of the figure,
the factorization scale to its central value, $\mu_F = M_\sl = 300$~GeV,
and vary the renormalization
scale $\mu_R$ by a factor of four around this value. We study the dependence
of the total rate, as a function of the ratio $\mu_R/M_\sl$,
at LO and NLO of perturbative QCD and after matching the NLO results
with resummation predictions at the NLL accuracy. As the tree-level
Feynman diagrams
do not exhibit any QCD vertex, the LO predictions are
independent of the renormalization scale $\mu_R$.
In contrast, logarithms of the
renormalization scale appear at higher orders
through the strong coupling constant $\alpha_s(\mu_R)$
entering in both virtual loop-diagrams and real emission diagrams, which
makes the NLO and NLL+NLO results
depending on $\mu_R$. The NLO cross sections
hence decrease with increasing values of the renormalization scale, whilst
at the NLL accuracy,
the resummation of soft gluon emissions attenuates this dependence, increasing
the size of the plateau region.

On the lower panel of Figure~\ref{fig:scale}, we conversely fix
the renormalization scale to its central value $\mu_R = M_\sl = 300$~GeV
and vary the factorization scale $\mu_F$ by a factor of four around this
value, the results being presented as
functions of the ratio $\mu_F/M_\sl$.
The LO estimates of the cross sections feature a very strong dependence
on the factorization scale, in particular due to the employed
leading order set of parton densities~\cite{Pumplin:2002vw}.
Using instead parton densities including NLO corrections in
their evolution and
extracted by making use of more precise Standard Model predictions~\cite{Lai:2010vv}
allows to largely tame this dependence\footnote{A part of the effects
is induced by the partonic cross sections
which depend on both $\mu_R$ and $\mu_F$ at NLO and NLL+NLO.}, which is
even further
reduced with soft gluon resummation although the same PDF sets are employed.

\begin{figure}
  \centering
  \includegraphics[width=.70\columnwidth]{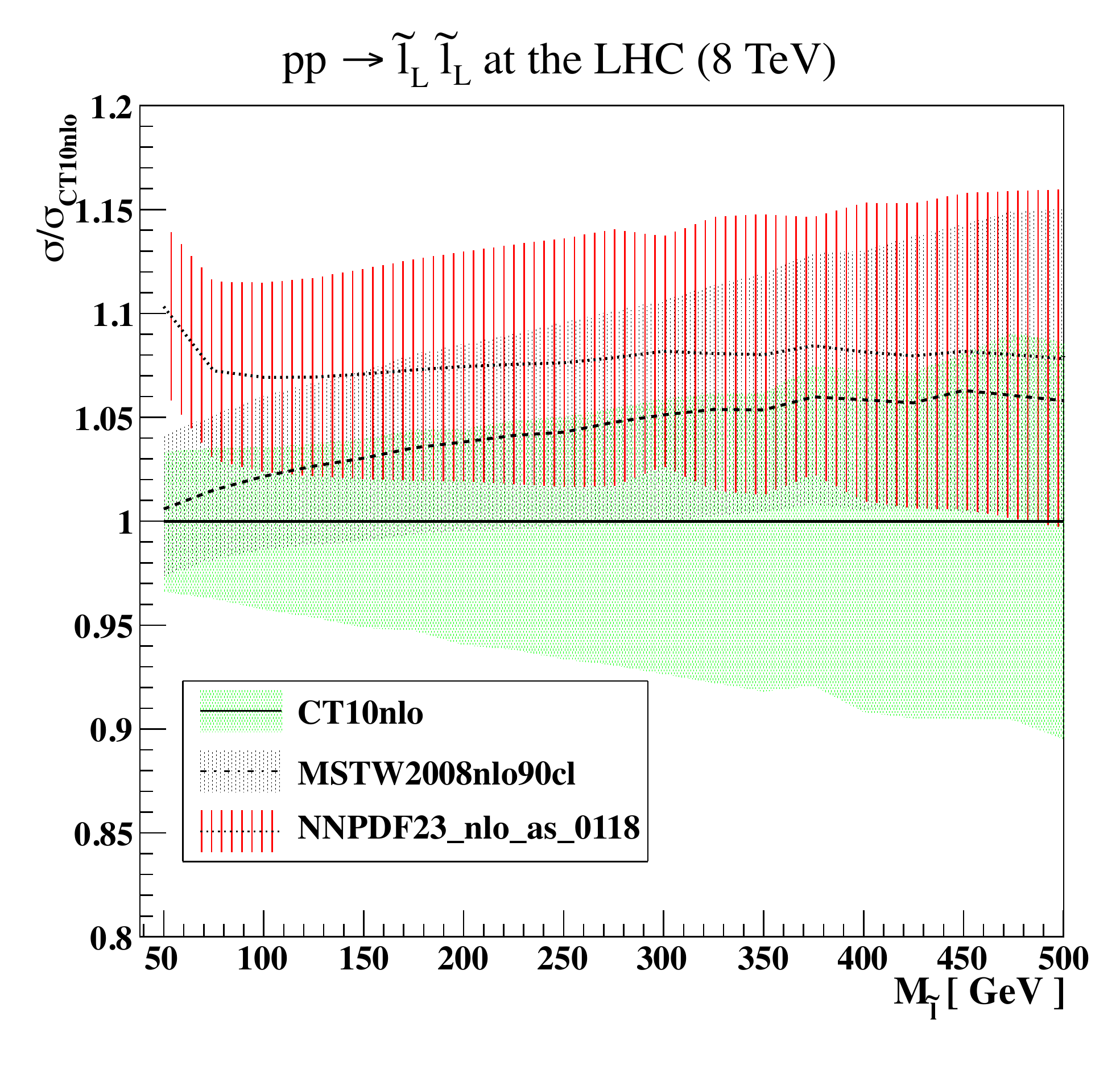}
\caption{\label{fig:pdf}Total cross sections for the production of a pair of
  left-handed sleptons at the LHC, running at a center-of-mass energy of 8~TeV.
  We study the dependence of the results on the parton density fits by
  showing predictions obtained with the central PDF fits as provided
  by the CTEQ (plain), MSTW (dashed) and NNPDF (dotted) collaborations, normalized to
  results obtained when using the best CT10 NLO fit.
  We indicate additionally the associated theoretical uncertainties
  for all PDF choices.}
\end{figure}

We now study another source of theoretical uncertainties
for cross section predictions and investigate the dependence of the results
on the choice of the parton density fits.
Up to now, we have computed NLL+NLO cross sections
using the best NLO fit provided
by the CTEQ collaboration, dubbed CT10~\cite{Lai:2010vv}.
In order to assess the uncertainties arising from this choice, we re-evaluate
the total rate for slepton pair production at the LHC
using each of the 52 parton density fits accompanying the CT10 best fit, which
allows to estimate the effects of $\pm 1\sigma$
variations along the $n=26$ eigenvectors of the parton density
fit covariance matrix. The positive
and negative variations $\Delta\sigma_{\rm up}$ and
$\Delta\sigma_{\rm down}$ from the central cross section value $\sigma_0$
are then computed by combining the 52 results,
\be\bsp
 \Delta\sigma_{\rm up} =&\ \sqrt{\sum_{i=1}^n\bigg[\max\Big(\sigma_{+i}-\sigma_0,
 \sigma_{-i}-\sigma_0,0\Big)\bigg]^2}\ , \\
 \Delta\sigma_{\rm down} =&\ \sqrt{\sum_{i=1}^n\bigg[\max\Big(\sigma_0-\sigma_{+i},
 \sigma_0-\sigma_{-i},0\Big)\bigg]^2} \ ,
\esp\ee
where $\sigma_{+i}$ and $\sigma_{-i}$ denote predictions obtained when
employing a PDF set including a positive and negative variation
along the $i^{\rm th}$ eigenvector of the covariance matrix, respectively.
We finally compare the obtained cross section envelope to the ones obtained
when making use of
the 90\% confidence level MSTW 2008 parton density fit~\cite{Martin:2009iq} and of
the 2.3 version of the NNPDF fit~\cite{Ball:2008by,Ball:2013hta}. In the first case,
the uncertainties $\Delta\sigma_{\rm up}$ and $\Delta\sigma_{\rm down}$
are computed as above while in the latter case, we extract the $2\sigma$ range
spanned by the 100 replica fits provided with the NNPDF set. As NNPDF densities cannot be properly
transformed in Mellin space, the method described in Section~\ref{sec:xsecgeneral} is not appropriate.
We instead rely, for a fixed slepton mass, on the ratio $\kappa$
of the NLL+NLO to NLO results which is expected to be
largely independent of the parton density choice if the two
calculations employ PDF sets evaluated at the
same perturbative order. This feature has been
verified with the CT10 and MSTW 2008 densities. Therefore, the NNPDF results
at the NLL+NLO accuracy are based on a pure NLO computation,
rescaled by the average $\kappa$ value
obtained when considering all CT10 or MSTW 2008 PDF sets.

The results are presented in Figure~\ref{fig:pdf}, where we focus
on the production of a pair of left-handed sleptons at the LHC running
at a center-of-mass energy of 8~TeV and study the dependence of
the predictions on different choices for the
slepton mass. We normalize the NLL+NLO predictions to the rate
obtained when using the best NLO PDF fit provided by the CTEQ
collaboration
and include the uncertainties as defined above for the CT10,
MSTW 2008 and NNPDF 2.3 fits. We observe that up to these uncertainties, all
predictions agree over the covered slepton mass range at the percent level. However, while
the theoretical errors associated with the production of a pair of light sleptons
are of a few percents, they increase with the slepton mass, reaching
about 10\% for 500~GeV sleptons. This is due to the relevant regions of
the $(x,\mu_F^2)$ parameter space that lead to PDFs less constrained by data
when the slepton mass increases.

\subsection{Summary tables}\label{sec:summarytables}

We summarize all the results computed in this section in Table~\ref{tab:7tev} and
Table~\ref{tab:8tev} where we provide slepton pair production cross sections 
at the LHC, running at a center-of-mass energy of 7~TeV and 8~TeV, respectively,
for different choices of the produced superparticle
masses and nature. We separately consider, in the framework
of the series of benchmark scenarios described in Section~\ref{sec:bench12},
the production of left-handed and right-handed first or second generation sleptons,
together with the production of a pair of maximally mixing staus in the context
of benchmark scenarios as introduced in Section~\ref{sec:bench3}.
We additionally indicate theoretical
uncertainties induced by variations of the unphysical scales $\mu_R = \mu_F$
by a factor of two around the central scale value taken as the mass of the produced
sleptons, as well as those related to variations along the 26 eigenvectors of the
covariance matrix associated with the CT10 parton density fit which we compute
as presented in Section~\ref{sec:theouncertainties}.

\begin{table}[!t]
\renewcommand{\arraystretch}{1.195}
\begin{center}
\begin{tabular}{| c | c || l | l | l |}
\hline
  $M_{\tilde \ell}$ [GeV] & Final state&  LO [fb] & NLO [fb] & NLL+NLO [fb] \\
\hline
  \multirow{3}{*}{$50$}&
    $\tilde\ell_L^+\tilde\ell^-_L$ & $ 1381.00^{+6.8\%}_{-27.4\%} $&$ 1740.00^{+16.4\%+3.2\%}_{-1.4\%-4.5\%} $&$ 1723.00^{-0.6\%+3.2\%}_{-0.1\%-4.5\%}$ \\
  & $\tilde\ell_R^+\tilde\ell_R^-$&$ 619.40^{+6.9\%}_{-27.4\%} $&$ 780.80^{+16.4\%+3.4\%}_{-1.5\%-4.5\%} $&$ 773.30^{-0.6\%+3.4\%}_{-0.1\%-4.5\%} $\\
  & $\tilde\tau_1^+\tilde\tau_1^-$&$ 407.20^{+5.9\%}_{-28.1\%} $&$ 512.90^{+17.0\%+3.9\%}_{-1.3\%-4.9\%} $&$ 506.90^{-0.6\%+3.9\%}_{-0.1\%-4.9\%} $\\
\hline

\multirow{3}{*}{$100$}&
   $\tilde\ell_L^+\tilde\ell^-_L$ &$ 85.97^{+0.8\%}_{-22.1\%} $&$ 102.60^{+16.0\%+3.5\%}_{-1.9\%-4.7\%} $&$ 101.20^{+0.3\%+3.5\%}_{-0.2\%-4.7\%}$ \\
 & $\tilde\ell_R^+\tilde\ell_R^-$ &$ 38.88^{+0.8\%}_{-22.2\%} $&$ 46.42^{+16.0\%+4.0\%}_{-1.8\%-5.1\%} $&$ 45.80^{+0.3\%+4.0\%}_{-0.3\%-5.1\%} $\\
 & $\tilde\tau_1^+\tilde\tau_1^-$&$ 39.82^{+0.8\%}_{-22.2\%} $&$ 47.53^{+16.0\%+4.1\%}_{-1.9\%-5.3\%} $&$ 46.88^{+0.3\%+4.1\%}_{-0.3\%-5.3\%} $\\
\hline

  \multirow{3}{*}{$150$}&
    $\tilde\ell_L^+\tilde\ell^-_L$&$18.06^{+3.4\%}_{-19.8\%} $&$ 21.13^{+11.6\%+4.0\%}_{-2.0\%-5.6\%} $&$ 20.80^{+0.5\%+4.0\%}_{-0.2\%-5.6\%}$ \\
  & $\tilde\ell_R^+\tilde\ell_R^-$ &$ 8.37^{+3.3\%}_{-19.9\%} $&$ 9.79^{+11.7\%+4.4\%}_{-2.0\%-6.3\%} $&$ 9.64^{+0.5\%+4.4\%}_{-0.2\%-6.3\%} $\\
  & $\tilde\tau_1^+\tilde\tau_1^-$&$ 8.81^{+3.3\%}_{-19.8\%} $&$ 10.31^{+11.7\%+4.7\%}_{-1.9\%-6.2\%} $&$ 10.16^{+0.6\%+4.7\%}_{-0.2\%-6.2\%} $\\
\hline

  \multirow{3}{*}{$200$}&
  $\tilde\ell_L^+\tilde\ell^-_L$ &$ 5.52^{+5.0\%}_{-18.9\%} $&$ 6.40^{+9.1\%+4.5\%}_{-1.9\%-6.5\%} $&$ 6.30^{+0.7\%+4.5\%}_{-0.0\%-6.5\%} $\\
 & $\tilde\ell_R^+\tilde\ell_R^-$&$ 2.59^{+4.9\%}_{-18.9\%} $&$ 3.00^{+9.2\%+5.2\%}_{-1.9\%-7.1\%} $&$ 2.95^{+0.7\%+5.2\%}_{-0.0\%-7.1\%} $\\
 & $\tilde\tau_1^+\tilde\tau_1^-$&$ 2.75^{+4.9\%}_{-18.9\%} $&$ 3.19^{+9.2\%+5.3\%}_{-1.9\%-7.2\%} $&$ 3.14^{+0.7\%+5.3\%}_{-0.0\%-7.2\%} $\\
\hline

  \multirow{3}{*}{$250$}&
    $\tilde\ell_L^+\tilde\ell^-_L$&$2.06^{+6.2\%}_{-18.5\%} $&$ 2.38^{+7.5\%+5.1\%}_{-1.8\%-7.3\%} $&$ 2.34^{+0.7\%+5.1\%}_{-0.0\%-7.3\%}$ \\
  & $\tilde\ell_R^+\tilde\ell_R^-$&$0.97^{+6.1\%}_{-18.5\%} $&$ 1.12^{+7.6\%+6.0\%}_{-1.8\%-7.8\%} $&$ 1.10^{+0.8\%+6.0\%}_{-0.0\%-7.8\%} $\\
  & $\tilde\tau_1^+\tilde\tau_1^-$&$ 1.04^{+6.1\%}_{-18.5\%} $&$ 1.20^{+7.7\%+5.9\%}_{-1.8\%-8.0\%} $&$ 1.18^{+0.8\%+5.9\%}_{-0.0\%-8.0\%} $\\
\hline

  \multirow{3}{*}{$300$}&
    $\tilde\ell_L^+\tilde\ell^-_L$&$0.87^{+7.1\%}_{-18.6\%} $&$ 1.00^{+6.5\%+5.8\%}_{-2.1\%-8.0\%} $&$ 0.99^{+0.9\%+5.8\%}_{-0.1\%-8.0\%}$ \\
  & $\tilde\ell_R^+\tilde\ell_R^-$&$0.41^{+7.1\%}_{-18.5\%} $&$ 0.48^{+6.6\%+6.7\%}_{-2.0\%-8.8\%} $&$ 0.47^{+0.9\%+6.7\%}_{-0.0\%-8.8\%} $\\
  & $\tilde\tau_1^+\tilde\tau_1^-$&$ 0.44^{+7.0\%}_{-18.6\%} $&$ 0.51^{+6.6\%+6.7\%}_{-2.0\%-8.9\%} $&$ 0.50^{+0.8\%+6.7\%}_{-0.1\%-8.9\%} $\\
\hline

  \multirow{3}{*}{$350$}&
    $\tilde\ell_L^+\tilde\ell^-_L$&$0.40^{+7.9\%}_{-18.9\%} $&$ 0.46^{+5.8\%+6.5\%}_{-2.2\%-8.8\%} $&$ 0.45^{+1.0\%+6.5\%}_{-0.2\%-8.8\%}$ \\
  & $\tilde\ell_R^+\tilde\ell_R^-$&$0.19^{+7.8\%}_{-18.8\%} $&$ 0.22^{+5.9\%+7.5\%}_{-2.1\%-9.8\%} $&$ 0.22^{+1.0\%+7.5\%}_{-0.2\%-9.8\%} $\\
  & $\tilde\tau_1^+\tilde\tau_1^-$&$ 0.20^{+7.8\%}_{-18.8\%} $&$ 0.24^{+5.9\%+7.7\%}_{-2.1\%-9.8\%} $&$ 0.23^{+1.0\%+7.7\%}_{-0.3\%-9.8\%} $\\
\hline

  \multirow{3}{*}{$400$}&
    $\tilde\ell_L^+\tilde\ell^-_L$ &$0.20^{+8.6\%}_{-19.3\%} $&$ 0.23^{+5.4\%+7.3\%}_{-2.6\%-9.4\%} $&$ 0.22^{+1.3\%+7.3\%}_{-0.3\%-9.4\%}$ \\
  & $\tilde\ell_R^+\tilde\ell_R^-$&$0.09^{+8.5\%}_{-19.2\%} $&$ 0.11^{+5.4\%+8.5\%}_{-2.7\%-10.8\%} $&$ 0.11^{+1.2\%+8.5\%}_{-0.3\%-10.8\%}$\\
  & $\tilde\tau_1^+\tilde\tau_1^-$&$ 0.10^{+8.5\%}_{-19.2\%} $&$ 0.12^{+5.4\%+8.9\%}_{-2.7\%-10.6\%} $&$ 0.11^{+1.2\%+8.9\%}_{-0.3\%-10.6\%} $\\
\hline

  \multirow{3}{*}{$450$}&
    $\tilde\ell_L^+\tilde\ell^-_L$&$0.10^{+9.2\%}_{-19.9\%} $&$ 0.12^{+5.1\%+8.1\%}_{-3.1\%-10.2\%} $&$ 0.11^{+1.3\%+8.1\%}_{-0.6\%-10.2\%}$ \\
  & $\tilde\ell_R^+\tilde\ell_R^-$&$0.05^{+9.1\%}_{-19.7\%} $&$ 0.06^{+5.1\%+9.9\%}_{-3.1\%-11.6\%} $&$ 0.06^{+1.3\%+9.9\%}_{-0.6\%-11.6\%} $\\
  & $\tilde\tau_1^+\tilde\tau_1^-$&$ 0.05^{+9.0\%}_{-19.7\%} $&$ 0.06^{+5.1\%+10.0\%}_{-3.1\%-11.7\%} $&$ 0.06^{+1.3\%+10.0\%}_{-0.6\%-11.7\%} $\\
\hline

  \multirow{3}{*}{$500$}&
    $\tilde\ell_L^+\tilde\ell^-_L$&$0.05^{+9.7\%}_{-20.5\%} $&$ 0.06^{+4.9\%+9.2\%}_{-3.3\%-10.7\%} $&$ 0.06^{+1.4\%+9.2\%}_{-0.9\%-10.7\%}$\\
  & $\tilde\ell_R^+\tilde\ell_R^-$&$0.03^{+9.6\%}_{-20.4\%} $&$ 0.03^{+4.9\%+11.4\%}_{-3.2\%-12.6\%} $&$ 0.03^{+1.4\%+11.4\%}_{-0.8\%-12.6\%}$\\
  & $\tilde\tau_1^+\tilde\tau_1^-$&$ 0.03^{+9.6\%}_{-20.3\%} $&$ 0.03^{+4.9\%+11.6\%}_{-3.3\%-12.7\%} $&$ 0.03^{+1.3\%+11.6\%}_{-0.8\%-12.7\%} $\\ 
\hline
\end{tabular}
\caption{\label{tab:7tev} Total production cross sections at the LHC, running
  at a center-of-mass energy of 7~TeV, for first or second generation left-handed
  and right-handed sleptons, as well as for maximally mixing staus. Results are presented
  together with the associated scale and PDF uncertainties.}
\end{center}
\end{table}

\begin{table}[!t]
\renewcommand{\arraystretch}{1.21}
\begin{center}
\begin{tabular}{| c | c || l | l | l |}
\hline
  $M_{\tilde \ell}$ [GeV] & Final state&  LO [fb] & NLO [fb] & NLL+NLO [fb] \\
\hline
  \multirow{3}{*}{$50$}&
 $\tilde\ell_L^+\tilde\ell_L^-$&$ 1623.00^{+7.7\%}_{-27.9\%} $&$ 2058.00^{+16.1\%+3.3\%}_{-1.6\%-4.4\%} $&$ 2039.00^{-0.8\%+3.3\%}_{-0.3\%-4.4\%} $\\
& $\tilde\ell_R^+\tilde\ell_R^- $&$ 727.10^{+7.8\%}_{-27.8\%} $&$ 921.90^{+16.1\%+3.4\%}_{-1.6\%-4.5\%} $&$ 913.70^{-0.8\%+3.4\%}_{-0.3\%-4.5\%} $\\
&$\tilde\tau_1^+\tilde\tau_1^-$&$ 478.40^{+6.9\%}_{-28.6\%} $&$ 606.70^{+16.8\%+3.9\%}_{-1.5\%-4.9\%} $&$ 600.10^{-0.8\%+3.9\%}_{-0.3\%-4.9\%} $\\
\hline

\multirow{3}{*}{$100$}&
 $\tilde\ell_L^+\tilde\ell_L^-$&$ 105.50^{+0.7\%}_{-22.6\%} $&$ 126.50^{+16.6\%+3.7\%}_{-1.7\%-4.8\%} $&$ 124.80^{+0.1\%+3.7\%}_{-0.3\%-4.8\%} $\\
& $\tilde\ell_R^+\tilde\ell_R^- $&$ 47.61^{+0.7\%}_{-22.6\%} $&$ 57.10^{+16.6\%+3.8\%}_{-1.7\%-5.0\%} $&$ 56.36^{+0.2\%+3.8\%}_{-0.3\%-5.0\%} $\\
&$\tilde\tau_1^+\tilde\tau_1^-$&$ 48.75^{+0.7\%}_{-22.6\%} $&$ 58.45^{+16.6\%+3.9\%}_{-1.7\%-5.2\%} $&$ 57.69^{+0.2\%+3.9\%}_{-0.3\%-5.2\%} $\\
\hline

  \multirow{3}{*}{$150$}&
 $\tilde\ell_L^+\tilde\ell_L^-$&$ 23.02^{+2.6\%}_{-20.0\%} $&$ 27.00^{+12.6\%+4.2\%}_{-1.9\%-6.0\%} $&$ 26.61^{+0.5\%+4.2\%}_{-0.2\%-6.0\%} $\\
& $\tilde\ell_R^+\tilde\ell_R^- $&$ 10.64^{+2.6\%}_{-20.0\%} $&$ 12.48^{+12.7\%+4.2\%}_{-1.9\%-5.7\%} $&$ 12.30^{+0.4\%+4.2\%}_{-0.2\%-5.7\%} $\\
&$\tilde\tau_1^+\tilde\tau_1^-$&$ 11.20^{+2.6\%}_{-20.1\%} $&$ 13.14^{+12.7\%+4.4\%}_{-1.9\%-5.9\%} $&$ 12.96^{+0.5\%+4.4\%}_{-0.2\%-5.9\%} $\\
\hline

  \multirow{3}{*}{$200$}&
 $\tilde\ell_L^+\tilde\ell_L^-$&$ 7.30^{+4.3\%}_{-18.8\%} $&$ 8.47^{+10.0\%+4.9\%}_{-1.9\%-7.0\%} $&$ 8.35^{+0.6\%+4.9\%}_{-0.0\%-7.0\%} $\\
& $\tilde\ell_R^+\tilde\ell_R^- $&$ 3.41^{+4.2\%}_{-18.8\%} $&$ 3.96^{+10.1\%+4.8\%}_{-1.9\%-6.8\%} $&$ 3.90^{+0.6\%+4.8\%}_{-0.1\%-6.8\%} $\\
&$\tilde\tau_1^+\tilde\tau_1^-$&$ 3.62^{+4.2\%}_{-18.8\%} $&$ 4.21^{+10.1\%+4.9\%}_{-1.9\%-6.8\%} $&$ 4.15^{+0.6\%+4.9\%}_{-0.1\%-6.8\%} $\\
\hline

  \multirow{3}{*}{$250$}&
 $\tilde\ell_L^+\tilde\ell_L^-$&$ 2.82^{+5.4\%}_{-18.3\%} $&$ 3.26^{+8.3\%+5.7\%}_{-1.8\%-7.8\%} $&$ 3.21^{+0.7\%+5.7\%}_{-0.1\%-7.8\%} $\\
& $\tilde\ell_R^+\tilde\ell_R^- $&$ 1.33^{+5.4\%}_{-18.3\%} $&$ 1.54^{+8.4\%+5.4\%}_{-1.8\%-7.6\%} $&$ 1.51^{+0.7\%+5.4\%}_{-0.1\%-7.6\%} $\\
&$\tilde\tau_1^+\tilde\tau_1^-$&$ 1.42^{+5.4\%}_{-18.3\%} $&$ 1.64^{+8.4\%+5.6\%}_{-1.8\%-7.4\%} $&$ 1.61^{+0.7\%+5.6\%}_{-0.1\%-7.4\%} $\\
\hline

  \multirow{3}{*}{$300$}&
 $\tilde\ell_L^+\tilde\ell_L^-$&$ 1.24^{+6.4\%}_{-18.2\%} $&$ 1.43^{+7.1\%+6.3\%}_{-1.8\%-8.8\%} $&$ 1.40^{+0.7\%+6.3\%}_{-0.0\%-8.8\%} $\\
& $\tilde\ell_R^+\tilde\ell_R^- $&$ 0.59^{+6.3\%}_{-18.2\%} $&$ 0.68^{+7.2\%+6.1\%}_{-1.8\%-8.2\%} $&$ 0.67^{+0.8\%+6.1\%}_{-0.0\%-8.2\%} $\\
&$\tilde\tau_1^+\tilde\tau_1^-$&$ 0.63^{+6.3\%}_{-18.1\%} $&$ 0.72^{+7.2\%+6.2\%}_{-1.8\%-8.2\%} $&$ 0.71^{+0.7\%+6.2\%}_{-0.0\%-8.2\%} $\\
\hline

  \multirow{3}{*}{$350$}&
 $\tilde\ell_L^+\tilde\ell_L^-$&$ 0.59^{+7.1\%}_{-18.3\%} $&$ 0.68^{+6.3\%+7.3\%}_{-2.1\%-9.5\%} $&$ 0.67^{+0.9\%+7.3\%}_{-0.1\%-9.5\%} $\\
& $\tilde\ell_R^+\tilde\ell_R^- $&$ 0.28^{+7.1\%}_{-18.2\%} $&$ 0.32^{+6.3\%+6.8\%}_{-2.0\%-8.9\%} $&$ 0.32^{+0.8\%+6.8\%}_{-0.1\%-8.9\%} $\\
&$\tilde\tau_1^+\tilde\tau_1^-$&$ 0.30^{+7.1\%}_{-18.2\%} $&$ 0.35^{+6.4\%+6.8\%}_{-2.0\%-9.1\%} $&$ 0.34^{+0.8\%+6.8\%}_{-0.1\%-9.1\%} $\\
\hline

  \multirow{3}{*}{$400$}&
 $\tilde\ell_L^+\tilde\ell_L^-$&$ 0.30^{+7.8\%}_{-18.6\%} $&$ 0.35^{+5.7\%+8.3\%}_{-2.2\%-10.2\%} $&$ 0.34^{+1.0\%+8.3\%}_{-0.3\%-10.2\%} $\\
& $\tilde\ell_R^+\tilde\ell_R^- $&$ 0.14^{+7.7\%}_{-18.5\%} $&$ 0.17^{+5.8\%+7.6\%}_{-2.0\%-9.6\%} $&$ 0.16^{+1.0\%+7.6\%}_{-0.2\%-9.6\%} $\\
&$\tilde\tau_1^+\tilde\tau_1^-$&$ 0.15^{+7.7\%}_{-18.4\%} $&$ 0.18^{+5.8\%+7.6\%}_{-2.1\%-9.9\%} $&$ 0.17^{+1.0\%+7.6\%}_{-0.2\%-9.9\%} $\\
\hline

  \multirow{3}{*}{$450$}&
 $\tilde\ell_L^+\tilde\ell_L^-$&$ 0.16^{+8.4\%}_{-18.9\%} $&$ 0.19^{+5.3\%+9.4\%}_{-2.5\%-11.2\%} $&$ 0.18^{+1.2\%+9.4\%}_{-0.3\%-11.2\%} $\\
& $\tilde\ell_R^+\tilde\ell_R^- $&$ 0.08^{+8.3\%}_{-18.8\%} $&$ 0.09^{+5.4\%+8.5\%}_{-2.5\%-10.5\%} $&$ 0.09^{+1.2\%+8.5\%}_{-0.3\%-10.5\%} $\\
&$\tilde\tau_1^+\tilde\tau_1^-$&$ 0.08^{+8.3\%}_{-18.8\%} $&$ 0.10^{+5.4\%+8.6\%}_{-2.5\%-10.6\%} $&$ 0.09^{+1.2\%+8.6\%}_{-0.3\%-10.6\%} $\\
\hline

  \multirow{3}{*}{$500$}&
  $\tilde\ell_L^+\tilde\ell_L^-$&$ 0.09^{+8.9\%}_{-19.4\%} $&$ 0.10^{+5.1\%+10.8\%}_{-2.5\%-11.9\%} $&$ 0.10^{+1.3\%+10.8\%}_{-0.6\%-11.9\%}$\\ 
& $\tilde\ell_R^+\tilde\ell_R^- $&$ 0.04^{+8.8\%}_{-19.3\%} $&$ 0.05^{+5.1\%+9.6\%}_{-2.7\%-11.3\%} $&$ 0.05^{+1.3\%+9.6\%}_{-0.5\%-11.3\%}$\\ 
&$\tilde\tau_1^+\tilde\tau_1^-$&$ 0.05^{+8.8\%}_{-19.3\%} $&$ 0.05^{+5.1\%+9.7\%}_{-2.8\%-11.4\%} $&$ 0.05^{+1.3\%+9.7\%}_{-0.5\%-11.4\%} $\\ 
\hline
\end{tabular}
\caption{\label{tab:8tev}Same as Table~\ref{tab:7tev}, but for a center-of-mass
  energy of 8~TeV.}
\end{center}
\end{table}

\section{Sensitivity to slepton pair production at the LHC}
\label{sec:MC}
Until recently, both LHC collaborations were mainly
focusing on searches for the strongly interacting superpartners.
As a consequence of the associated negative search results,
the interest in the production of the electroweak superpartners, and in
particular in the production of a pair of charged sleptons, has increased over the
last few years. Dedicated direct slepton ATLAS and CMS analyses
have hence allow one to improve, for the first time from the
end of LEP experiments,
the bounds on the slepton masses.

We dedicate this section to the reinterpretation of the
recent ATLAS results of Ref.~\cite{Aad:2012pxa} and CMS
results of Ref.~\cite{CMS:aro}. These two analyses both contain
a signal region focusing on final states
featuring an electron-positron or muon-antimuon pair
produced in association with missing
transverse energy. They are therefore suitable for the extraction of bounds
on the first and second generation sleptons from their direct pair production
at the LHC, followed by their subsequent decays into the corresponding Standard
Model partner and missing energy. The limits
on the slepton masses obtained from those analyses
have been initially deduced after considering simplified
models such as those introduced in Section~\ref{sec:bench12} but under a very
specific setup
in terms of the nature of the sleptons and neutralinos.
In Section~\ref{sec:mc12a} and
Section~\ref{sec:mc12b}, we generalize these results
in the context of arbitrary neutralino (bino, wino or mixed bino-wino\footnote{Due to
the negligible first and second generation Yukawa couplings, selectrons
and smuons do not couple to the Higgsino components of the neutralinos that
are therefore not considered.})
and slepton (left-handed or right-handed) compositions, and also study their dependence
on the slepton flavor.

The ATLAS
analysis of Ref.~\cite{ATLAS:2013yla} is based on the production of two
hadronically-decaying taus with missing energy, so that it could
be in principle recast in the context of the production
of a pair of staus in the framework of
any of the benchmark scenario of the class of simplified models designed in
Section~\ref{sec:bench3}. However, a complicated tau reconstruction together with
small signal cross sections, in particular once we include the tau decay branching ratios,
render the task of constraining the simplified model parameter space impossible.
This would indeed requires cross sections larger by at least factors of 10-50 in the best cases
(\ie, in the low mass region of
the parameter space) in order
to imply at least a few visible events assuming the currently available luminosity.

In order to simulate signal events at the LHC, 
we employ the Monte Carlo event generator \madgraph~5~\cite{Alwall:2011uj}
and use its interface to \pythia~6 \cite{Sjostrand:2006za} to generate
hadron-level events from the merging of
parton-level events associated with matrix elements containing up to two
additional hard jets~\cite{Mangano:2002ea,Mangano:2006rw,Alwall:2008qv}.
While the {\sc Tauola}~\cite{Jadach:1993hs} package is employed
for the handling of tau decays, detector simulation
is performed by means of the {\sc Delphes} program~\cite{Ovyn:2009tx},
using the recent CMS detector description of Ref.~\cite{Agram:2013koa}
and the built-in ATLAS detector setup. The simplified scenarios that have been
designed in Section~\ref{sec:bench} have been implemented in \madgraph~5
via the UFO interface~\cite{Degrande:2011ua} of \feynrules~\cite{Christensen:2008py,
Christensen:2009jx,Duhr:2011se,Fuks:2012im,Alloul:2013bka} and the generated events
are reweighted according to the NLL+NLO predictions as computed
by {\sc Resummino}~\cite{Fuks:2013vua}. Event analysis
is finally performed by means of
the {\sc MadAnalysis~5} program~\cite{Conte:2012fm,Conte:2013mea}, after having
reconstructed the jets with the
anti-$k_T$ algorithm (using a radius parameter set to
$R=0.5$)~\cite{Cacciari:2008gp} as implemented in the {\sc FastJet}
package~\cite{Cacciari:2011ma}.

By the time this work was being completed,
additional simplified models with different slepton/neutralino
definitions have been investigated
by both the ATLAS and CMS collaborations~\cite{ATLAS-CONF-2013-049, CMS-PAS-SUS-13-006},
extending the earlier
analyses of Refs.~\cite{Aad:2012pxa,CMS:aro}. When a comparison is possible,
the obtained results have been found to fairly
agree with our predictions.

\subsection{Revisiting ATLAS searches for first and second generation sleptons}
\label{sec:mc12a}
In this section, we start by recasting the ATLAS analysis of Ref.~\cite{Aad:2012pxa}
dedicated to first and second generation slepton searches in LHC collisions
at a center-of-mass energy of 7~TeV. After generating
signal events by means of the above-mentioned simulation setup, we apply the
selection strategy designed by the ATLAS collaboration and demand the following
criteria to be satisfied.
\begin{itemize}
  \item We require the event final state to contain exactly two isolated
    leptons of the same flavor, their transverse momentum being imposed to be
    greater than 10~GeV and their pseudorapidity
    to fulfill $|\eta| \leq 2.47$ and  $|\eta| \leq 2.4$ for electrons and muons,
    respectively. The isolation is enforced by constraining the transverse activity
    in a cone of radius $R = \sqrt{\Delta\varphi^2 +
    \Delta\eta^2} = 0.2$ centered on the lepton, $\varphi$ being the azimuthal
    angle with respect to the beam direction, to be less than 10\% of the lepton $p_T$
    for electrons and less than 1.8~GeV for muons.
  \item Events featuring
    at least one jet with a transverse momentum $p_T \geq 30$~GeV and a pseudorapidity
    $|\eta| \leq 2.5$ are vetoed.
  \item The lepton pair  is asked not to be compatible with a $Z$-boson,
    the dilepton invariant mass $m_{\ell\ell}$ being constrained to be off
    the $Z$-boson peak, $m_{\ell\ell}~\not\in~[80,100]$~GeV.
  \item We ask the final state to contain a significant quantity of relative missing
    transverse energy $\met^{\rm ~rel} \geq 40$~GeV, where the relative
    missing transverse energy is defined as the missing transverse energy $\met$
    when the azimuthal angle $\tilde\varphi$
    between the direction of the missing momentum and that of the nearest reconstructed
    object is larger than $\pi/2$, and by $\met^{\rm ~rel} = \met \sin\tilde\varphi$
    otherwise.
  \item The properties of the $m_{T2}$ variable~\cite{Barr:2003rg}
    are finally exploited and we select events for which
    $m_{T2} \geq 90$~GeV.
\end{itemize}

\begin{figure}
  \centering
  \includegraphics[width=.32\columnwidth]{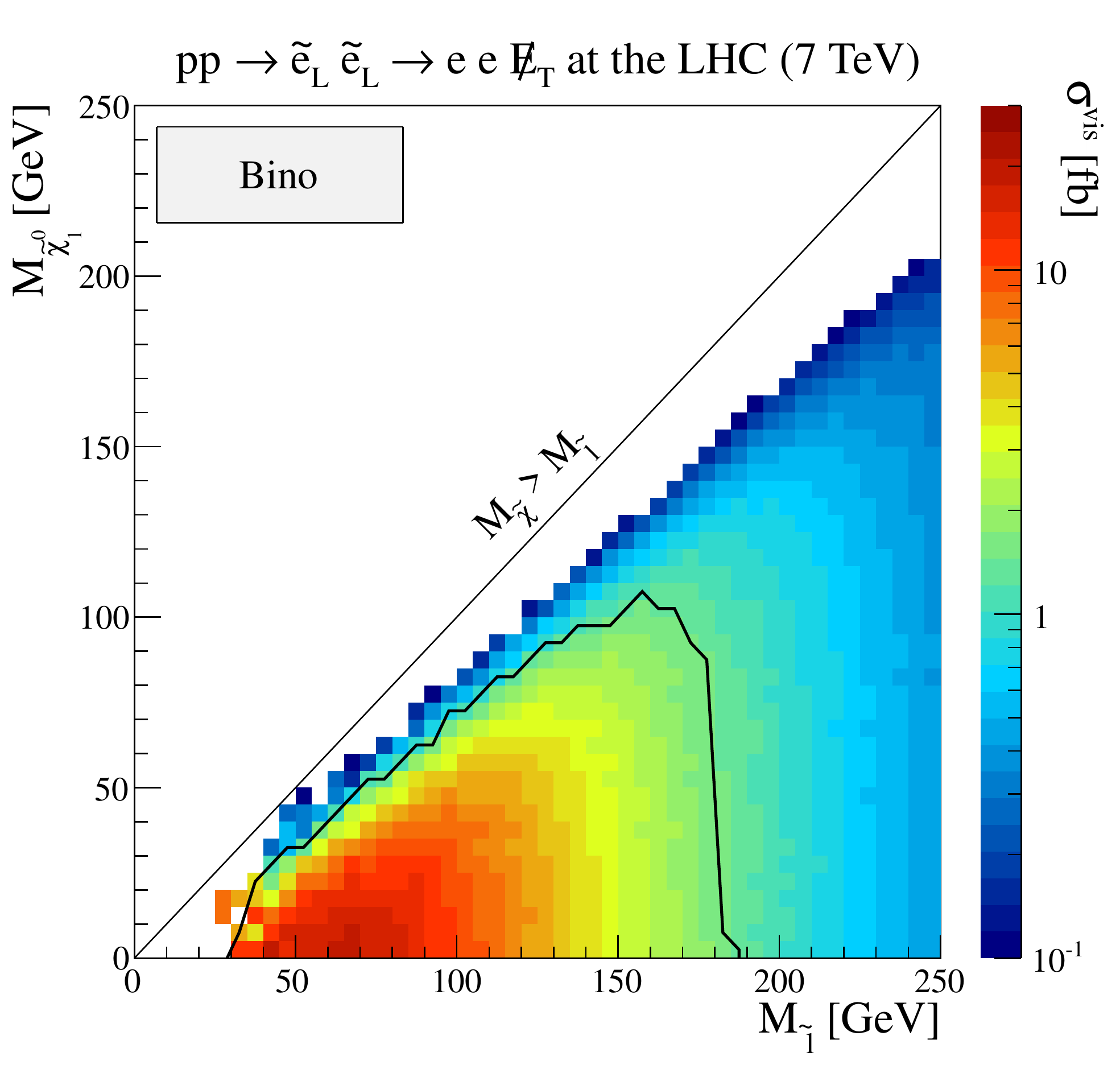}
  \includegraphics[width=.32\columnwidth]{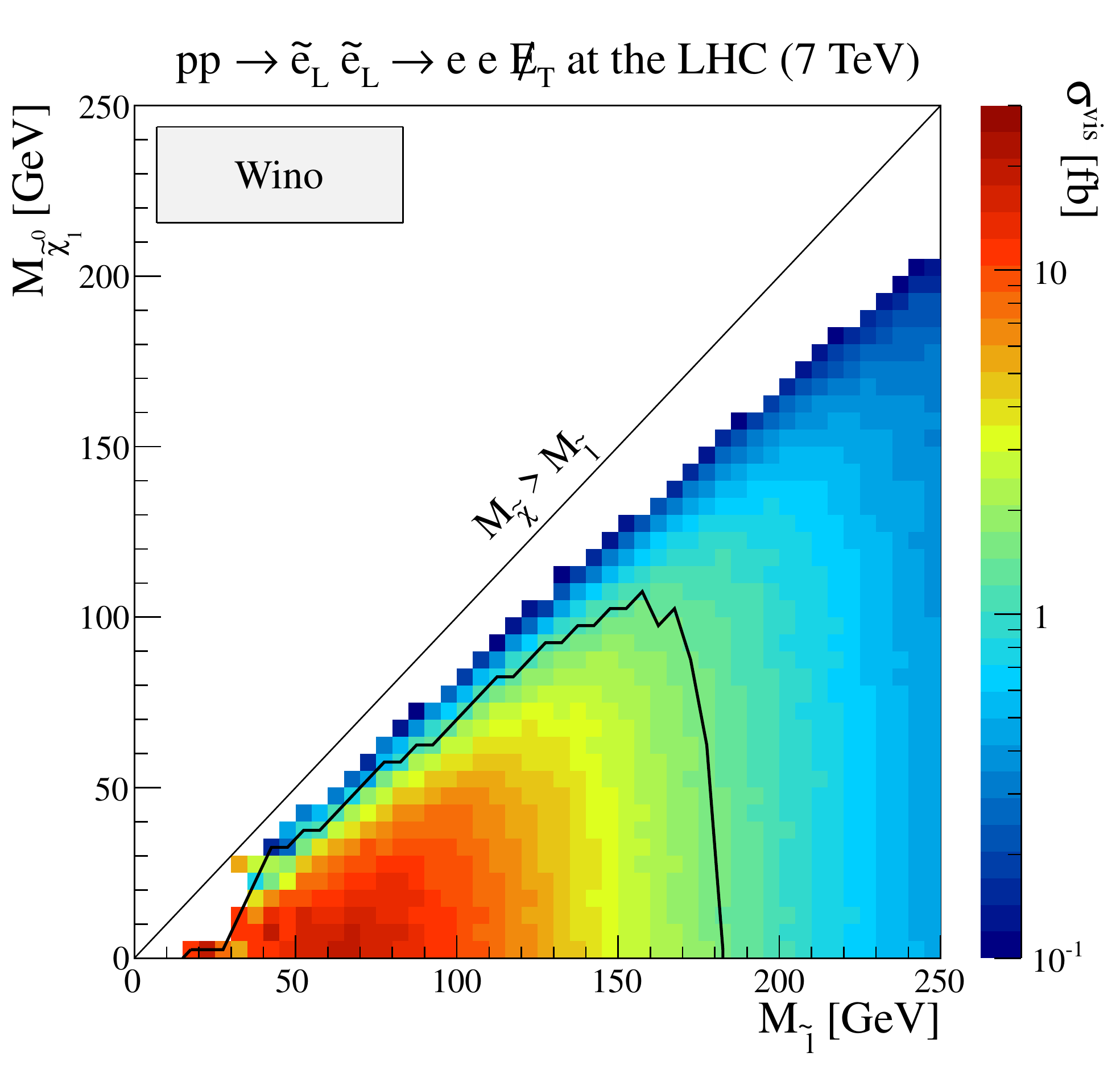}
  \includegraphics[width=.32\columnwidth]{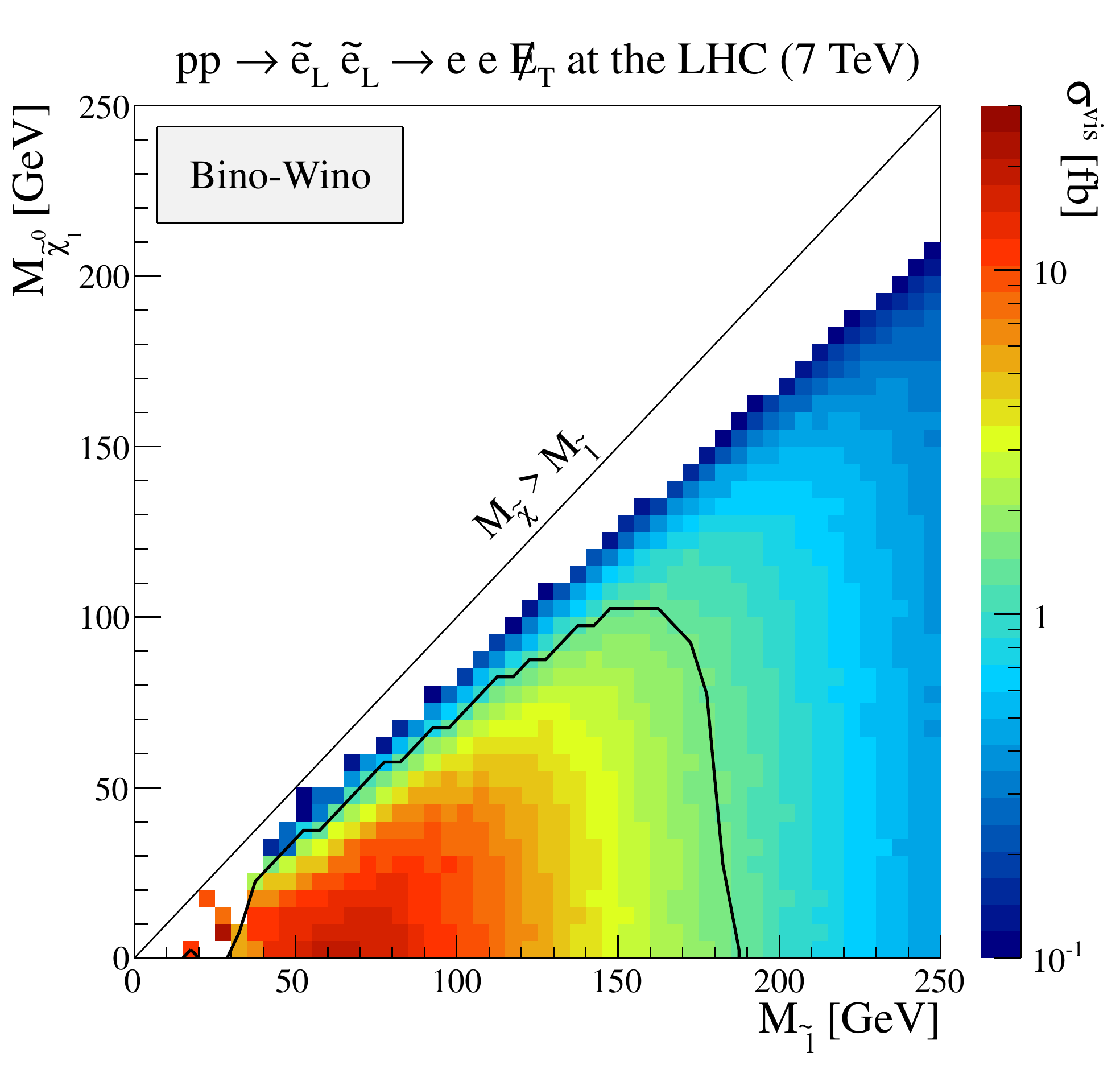}\\
\caption{\label{fig:atlas7}95\% confidence exclusion limit for left-handed
selectron pair production,
given in the $(M_\sl,M_{\tilde\chi^0_1})$ mass plane of the simplified
model of Section~\ref{sec:bench12} for different choices of the neutralino nature
taken as bino (left), wino (center) and mixed (right). We present the visible cross
section after applying the ATLAS selection strategy depicted in the text.
The limits are extracted for 4.7~fb$^{-1}$ of LHC collisions at a center-of-mass
energy of 7~TeV.}
\end{figure}

\begin{figure}
  \centering
  \includegraphics[width=.47\columnwidth]{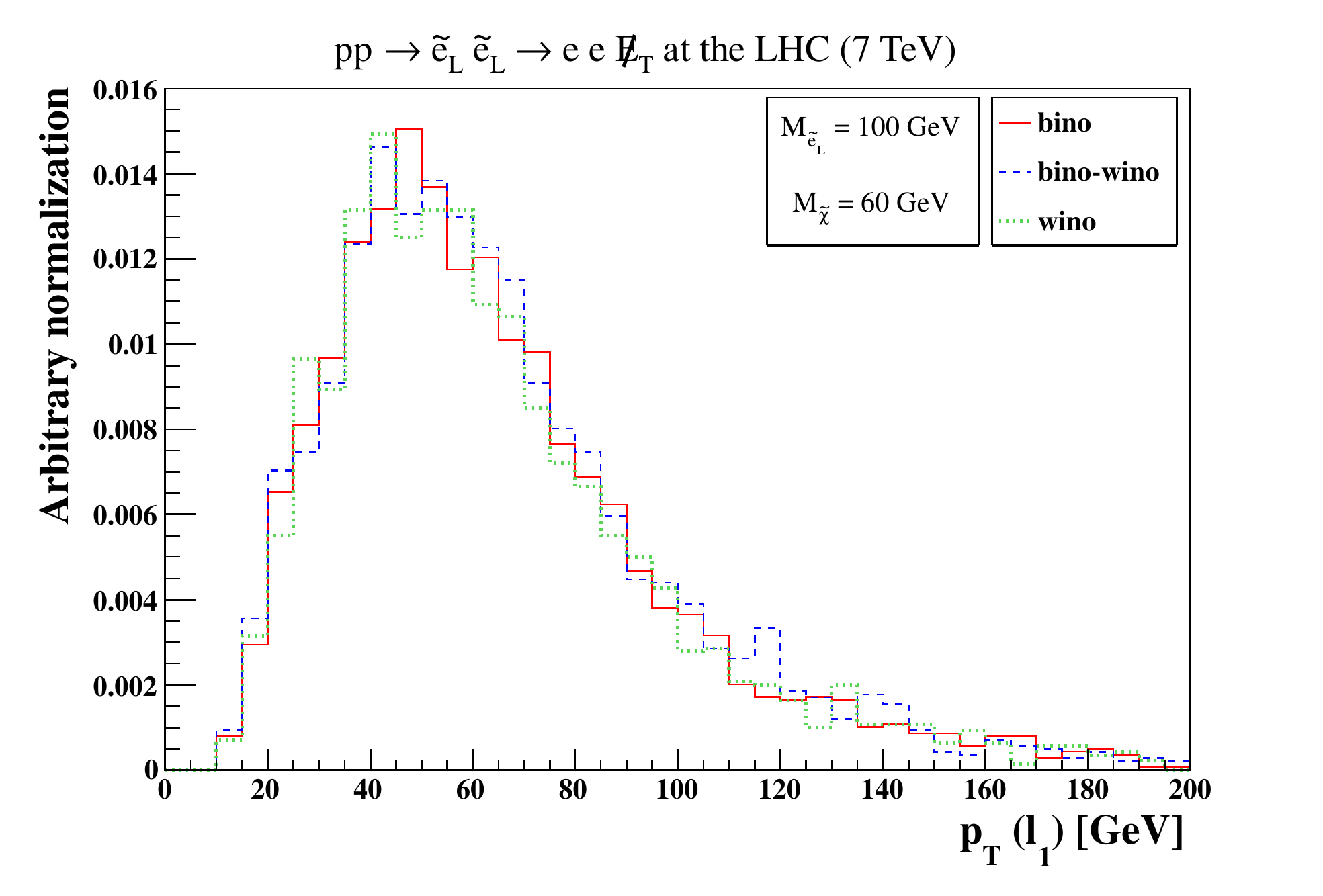}
  \includegraphics[width=.47\columnwidth]{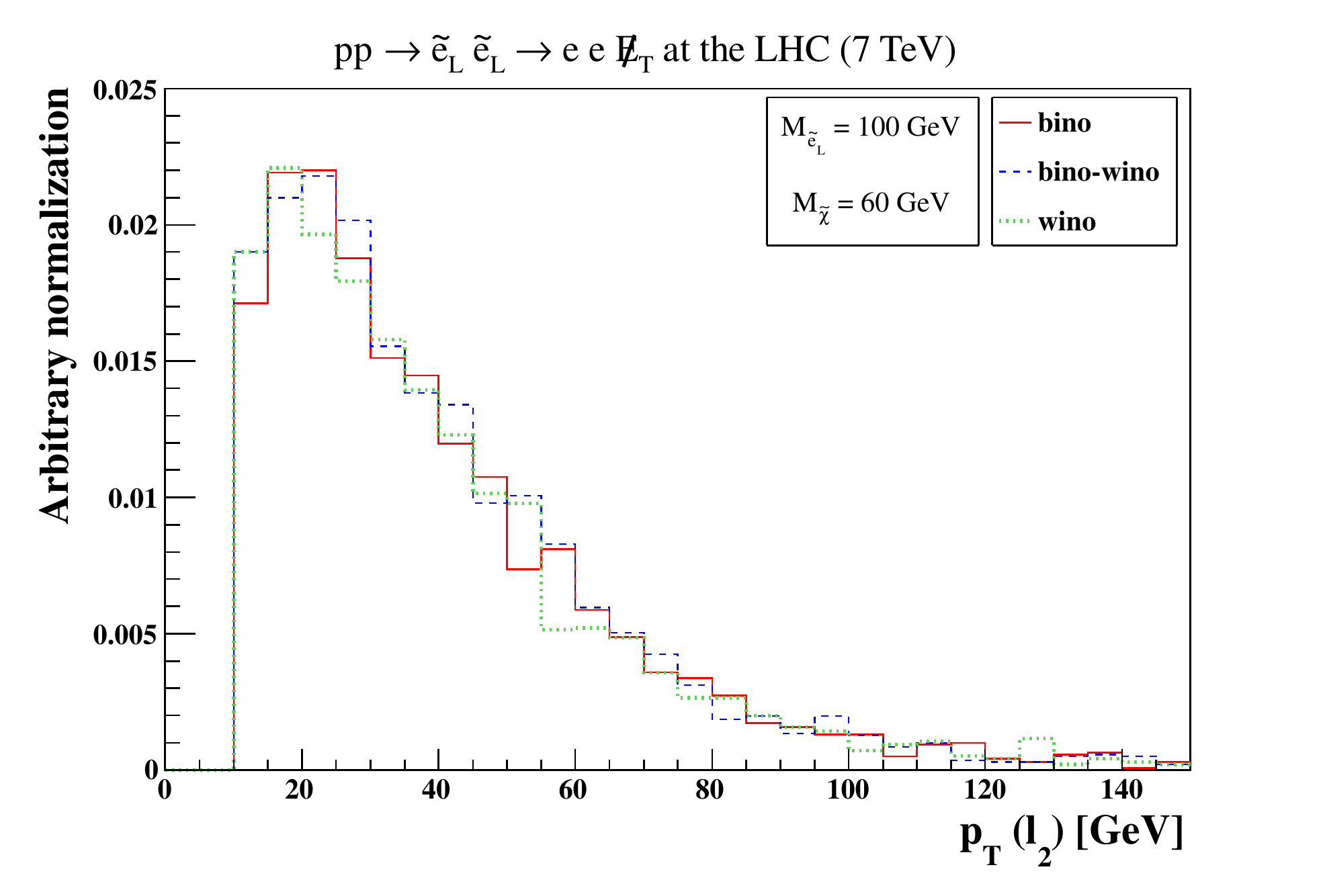}\\
  \includegraphics[width=.47\columnwidth]{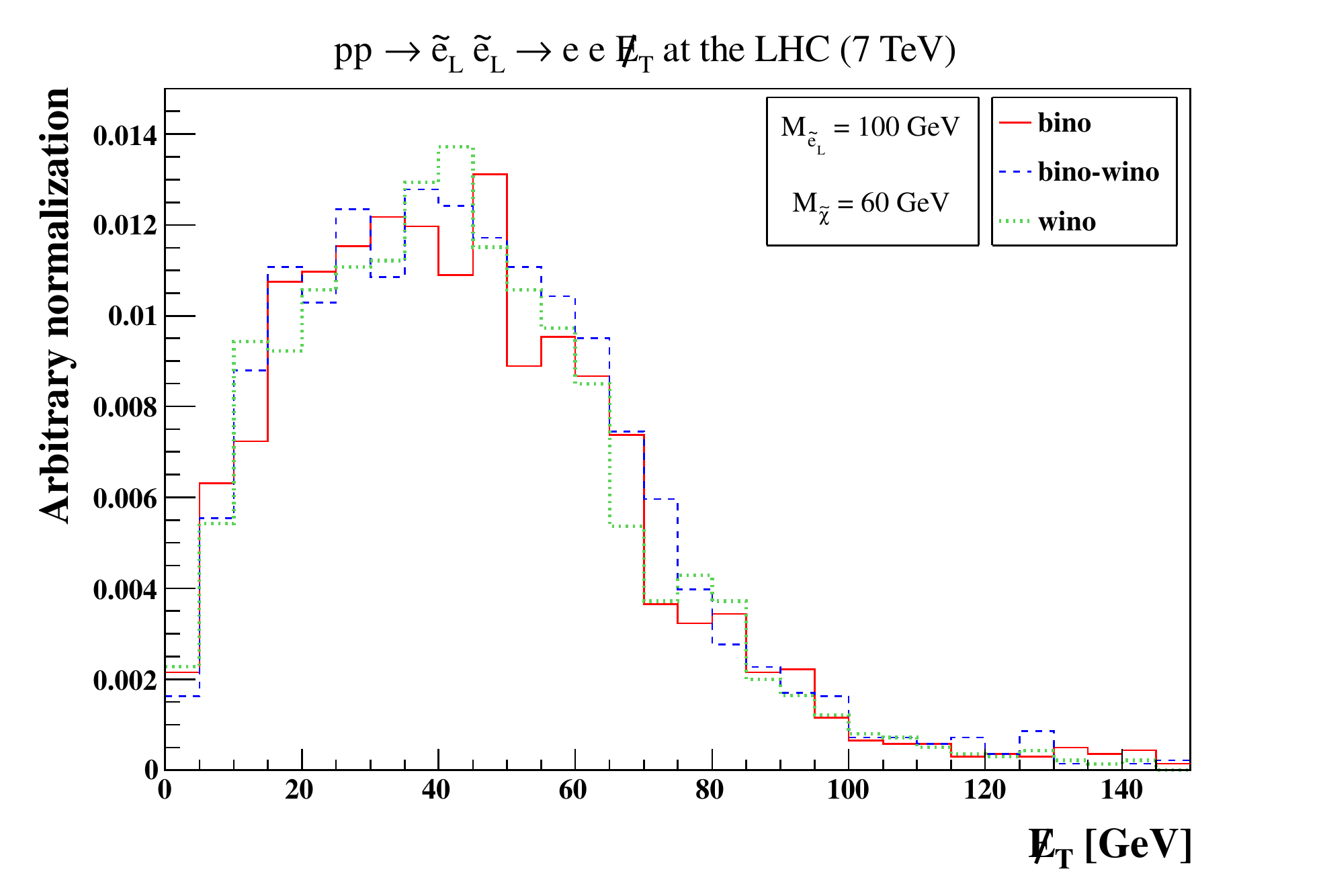}
  \includegraphics[width=.47\columnwidth]{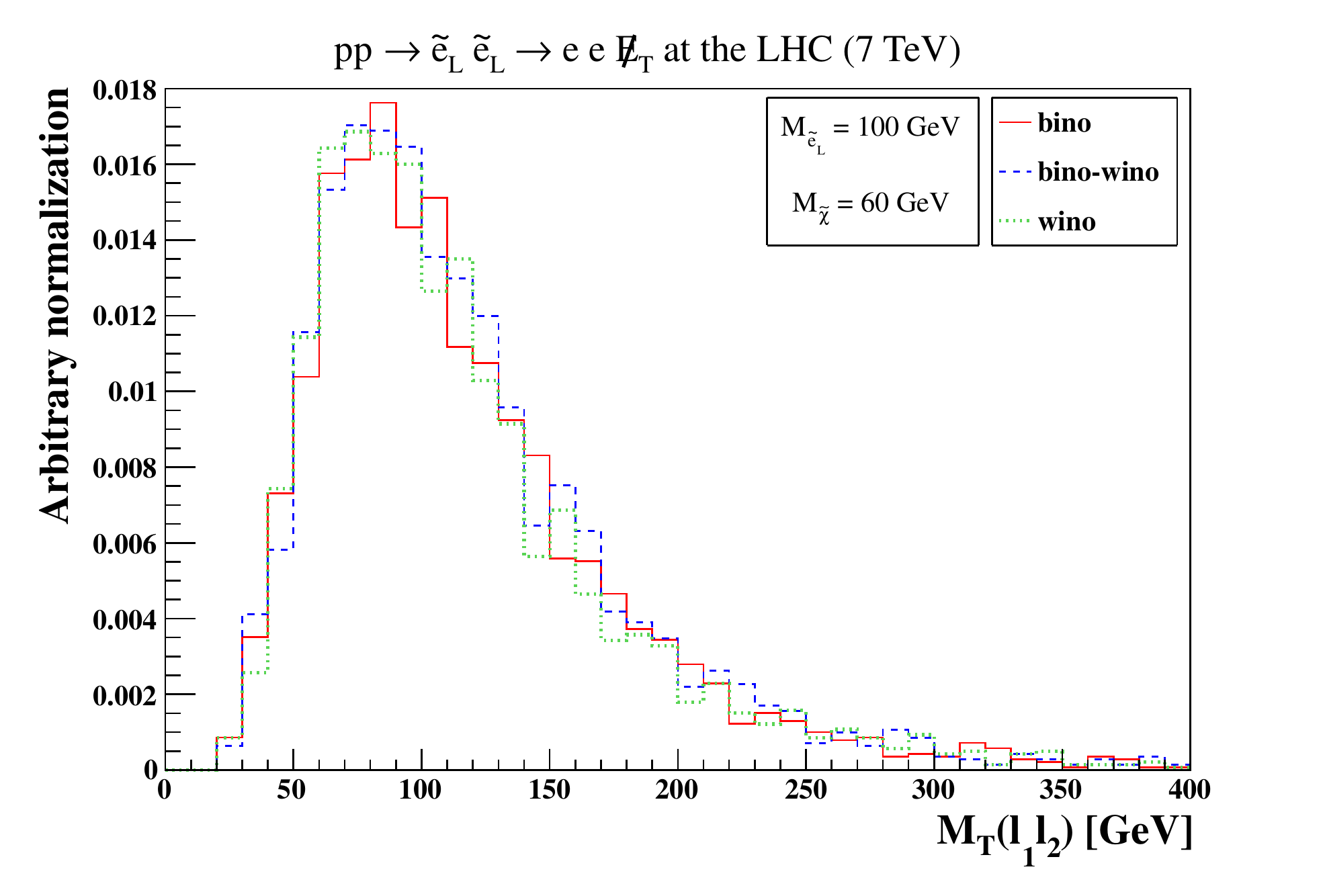}\\
  \includegraphics[width=.47\columnwidth]{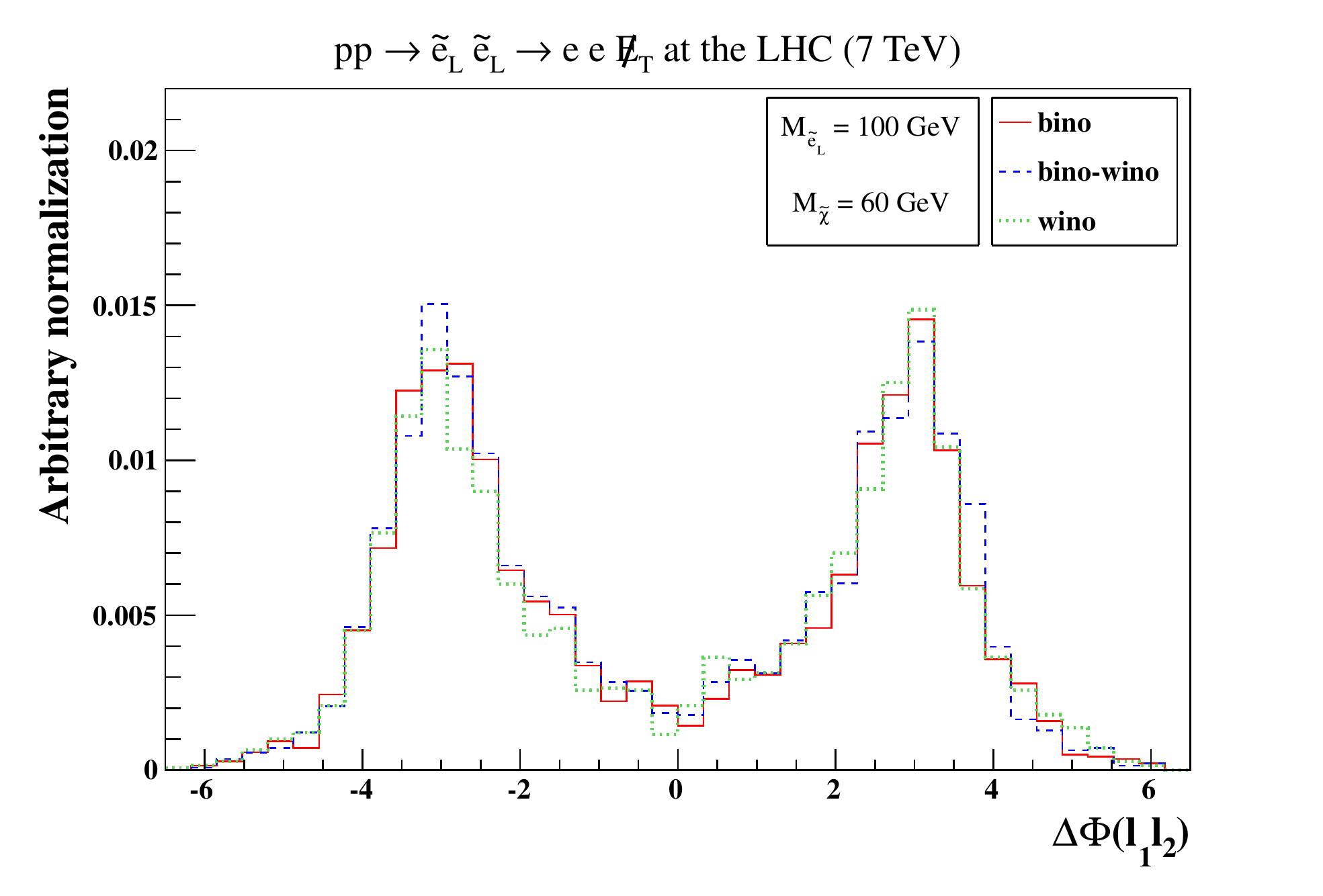}
  \includegraphics[width=.47\columnwidth]{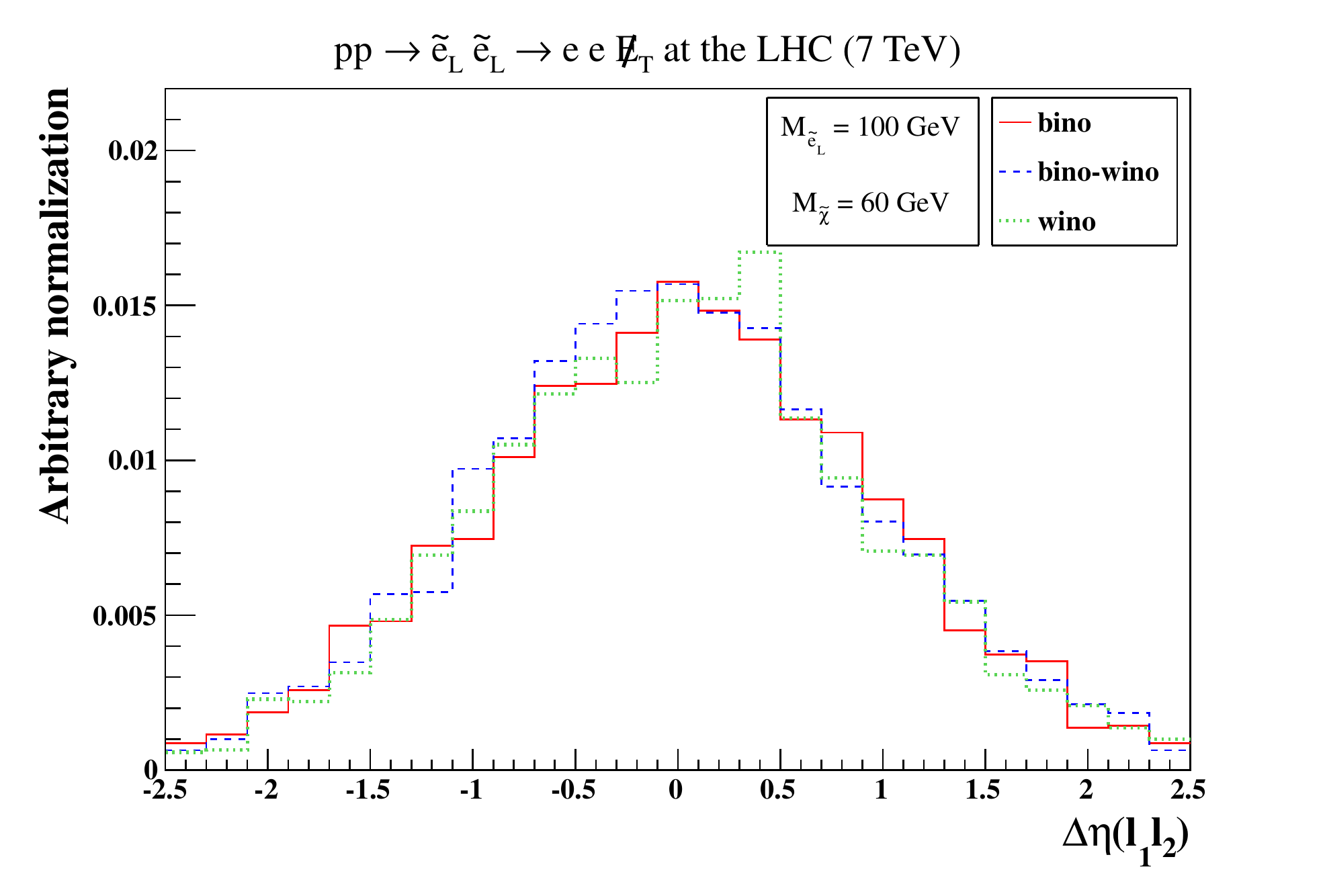}\\
  \includegraphics[width=.47\columnwidth]{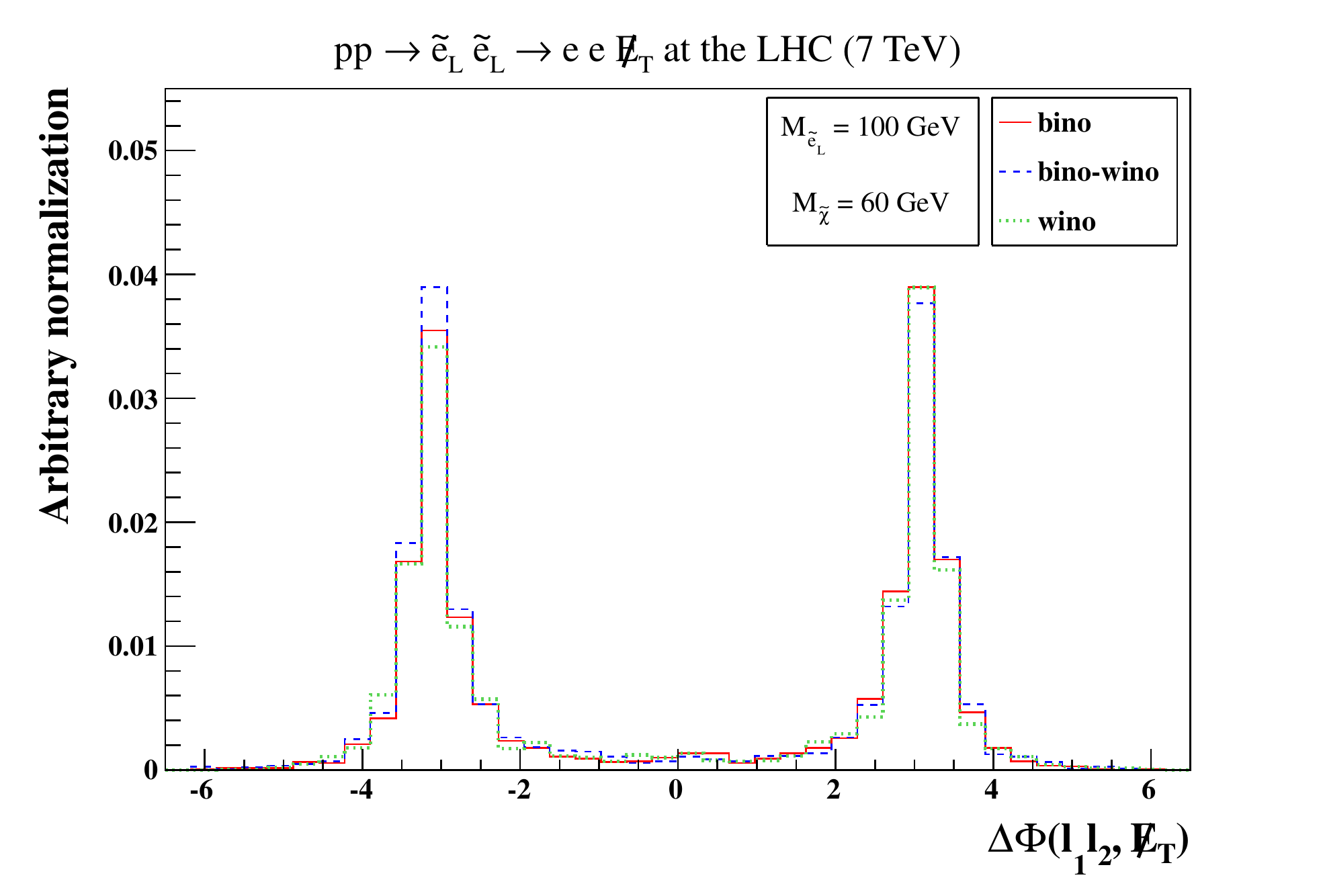}
  \includegraphics[width=.47\columnwidth]{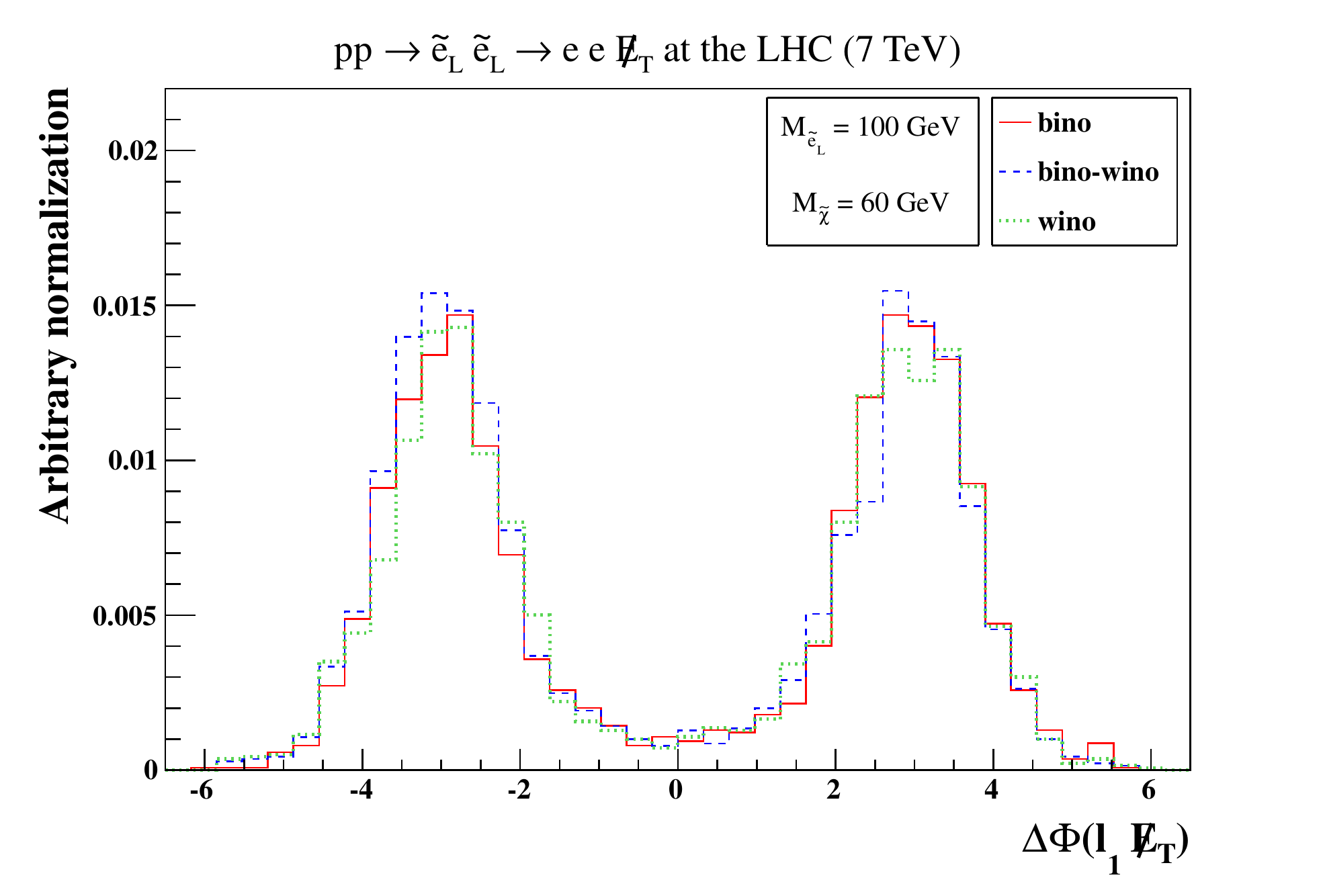}
\caption{\label{fig:atlas7c}Properties of the dilepton plus missing energy
  final state arising
  from the production and decay of a pair of left-handed selectrons, for a
  benchmark scenario where the neutralino mass is set to 60~GeV and the left-handed
  selectron mass to 100~GeV. The distributions are shown for different choices of the
  neutralino nature.}
\end{figure}

Focusing on different scenarios from the class of simplified models
introduced in Section~\ref{sec:bench12}, the results are
presented in Figure~\ref{fig:atlas7}. We first restrict ourselves to the
production of a pair of left-handed selectrons decaying into a pair of leptons
and missing energy. We consider three choices for the nature of the lightest
neutralino, namely a pure bino state ($N_1=1$; $N_2=N_3=N_4=0$;
left panel of the figure),
a pure wino state ($N_2=1$; $N_1=N_3=N_4=0$; central panel of the figure)
and a mixed state ($N_1= N_2 = 1/\sqrt{2}$; $N_3 = N_4=0$; right panel
of the figure). We show the visible cross section $\sigma^{\rm vis}$
for a center-of-mass energy of 7~TeV, defined as the
fraction of the cross section giving rise to events that can be observed
by the ATLAS experiment when using the analysis
described above, in  $(M_\sl,M_{\tilde\chi^0_1})$ mass planes.
In addition, we superimpose to the cross section
results
the 95\% confidence level contours that can be extracted from the ATLAS
bounds on the visible cross section
$\sigma^{\rm vis} \leq 1.5$~fb.

\begin{figure}
  \centering
  \includegraphics[width=.32\columnwidth]{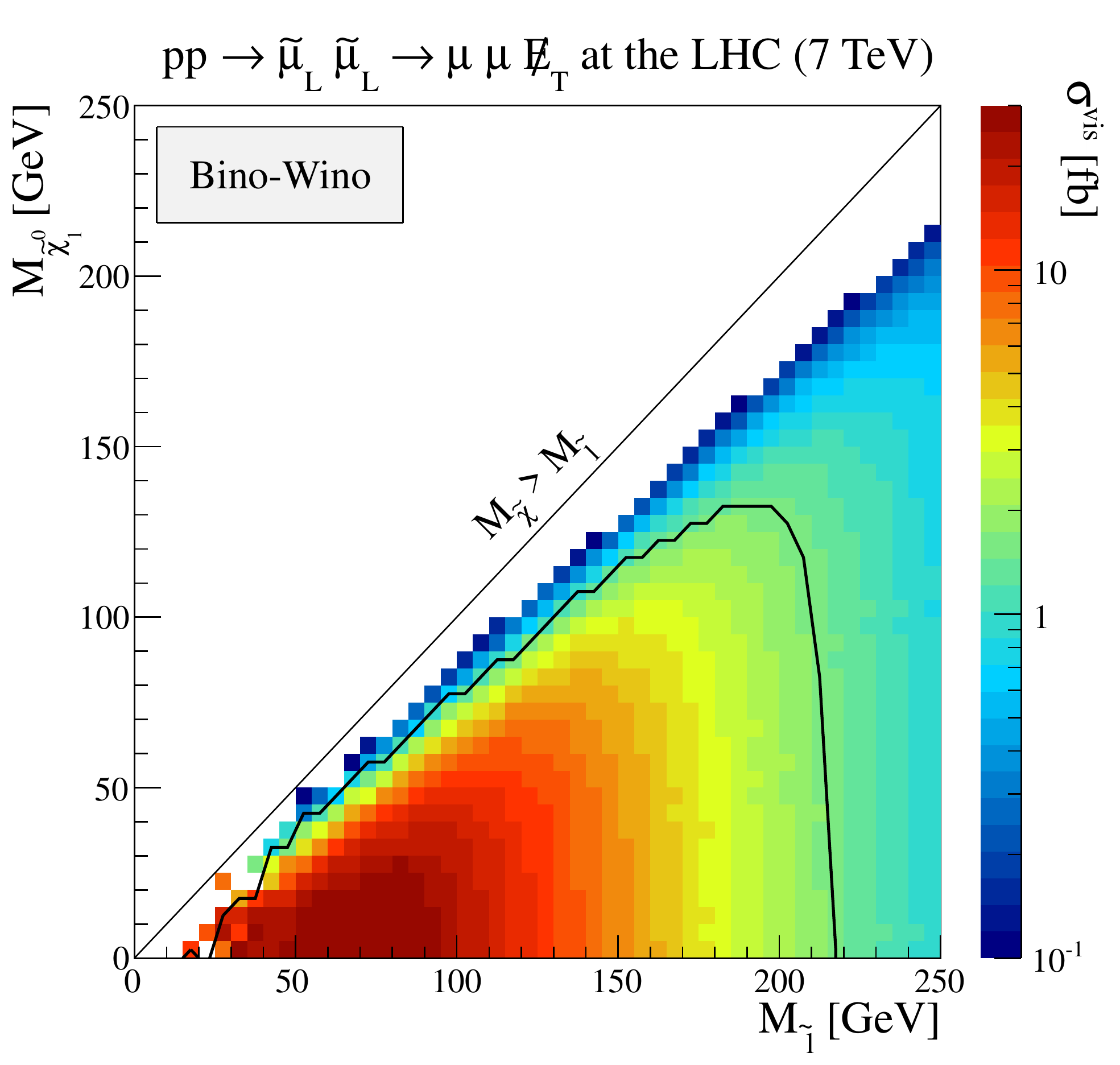}
  \includegraphics[width=.32\columnwidth]{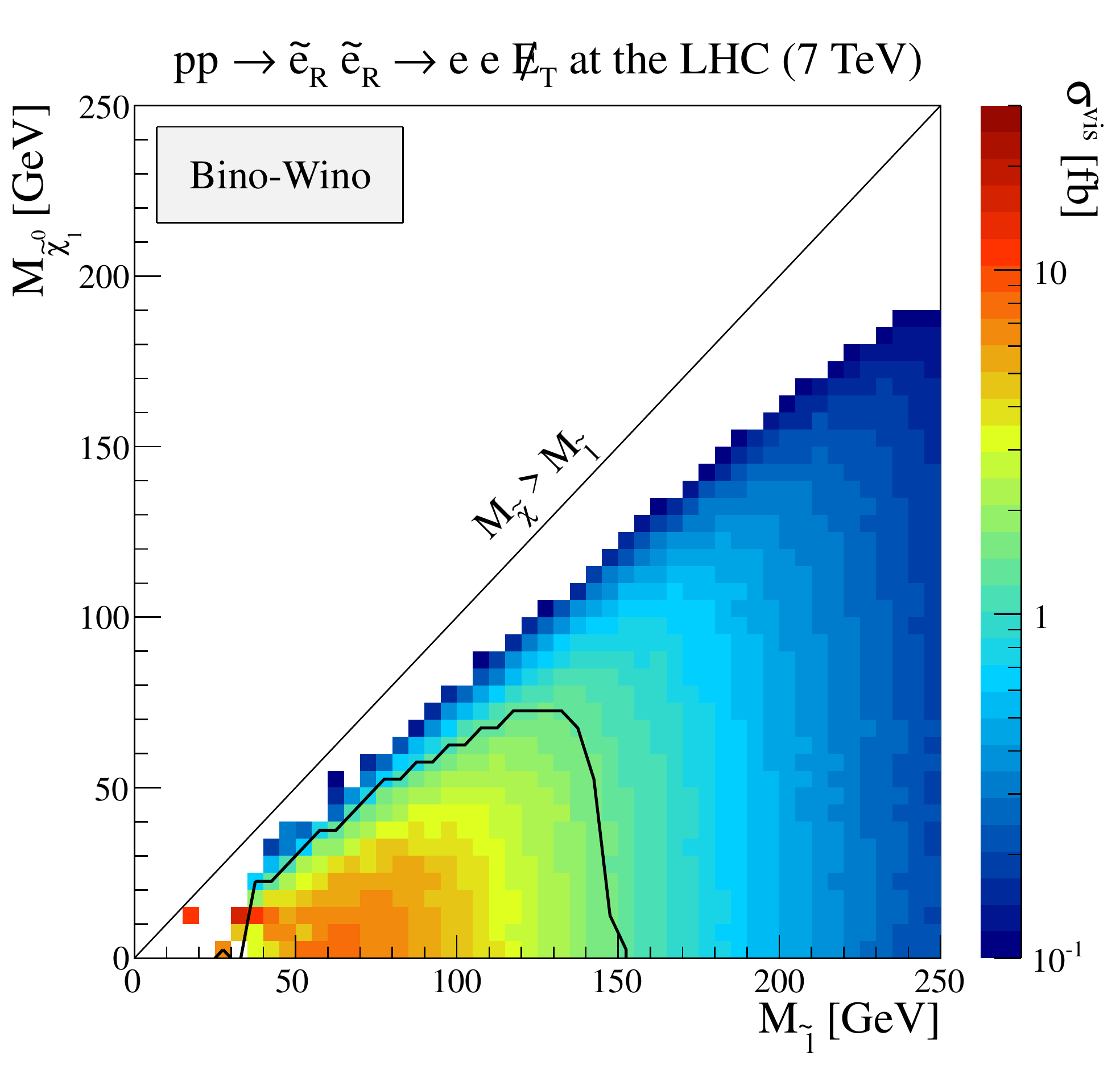}
  \includegraphics[width=.32\columnwidth]{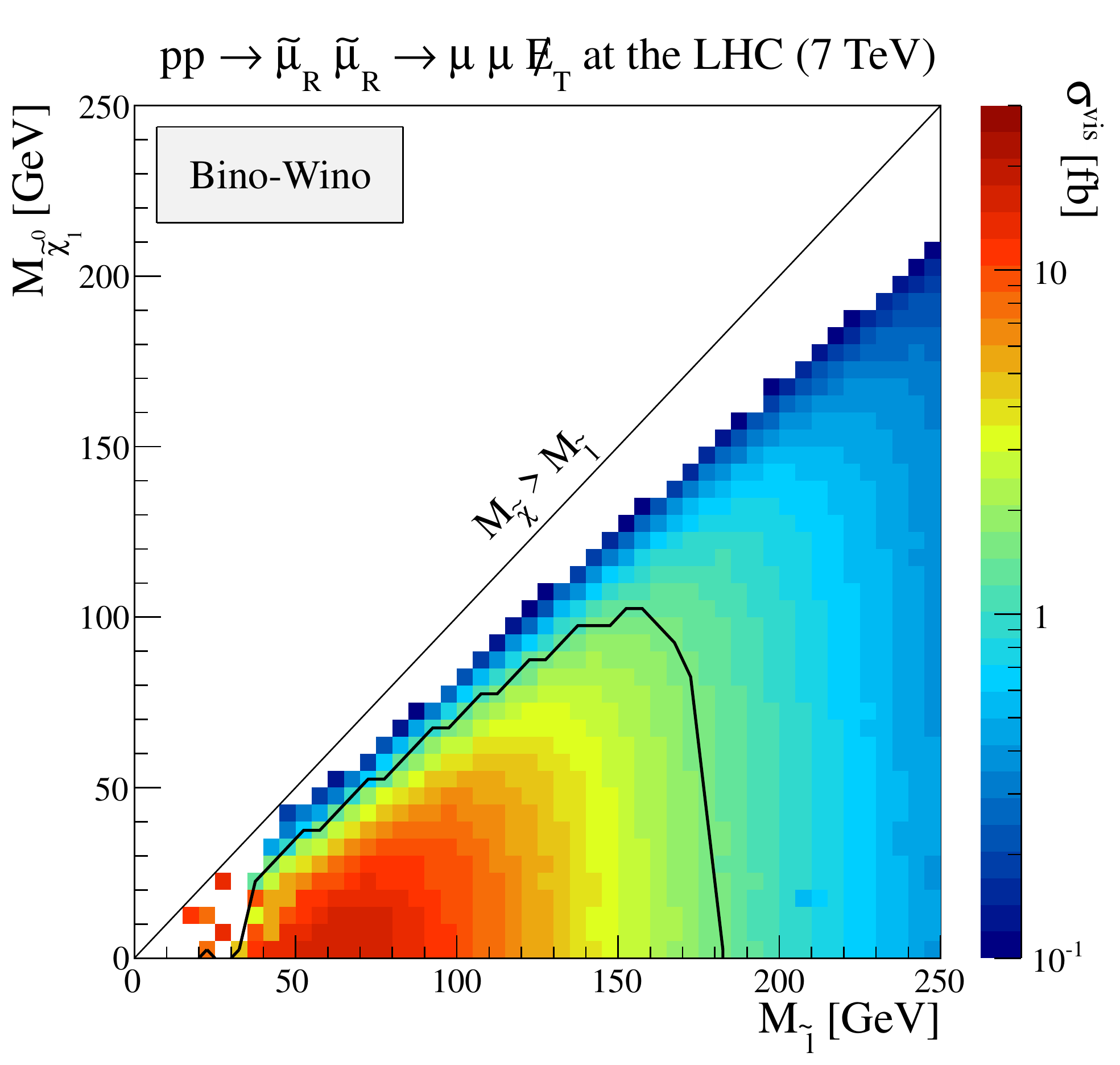}
\caption{\label{fig:atlas7b}Same as in Figure~\ref{fig:atlas7}, but for
  the production of a pair of left-handed smuons (left), right-handed selectrons
  (center) and right-handed smuons (right).}
\end{figure}

The selectron
being a scalar field, we expect the direct production of a pair of selectrons
followed by their decay into electrons and lightest neutralinos to
be non-sensitive to the neutralino nature. This is illustrated
by the three panels of Figure~\ref{fig:atlas7},
that show that the LHC reach is independent
of the neutralino nature.
For almost massless neutralinos,
selectrons with masses ranging up to about 175~GeV
are found excluded by 4.7~fb$^{-1}$ of 7~TeV LHC data.
This upper bound holds
for any neutralino mass smaller than 100~GeV,
a value from which the sensitivity drops so that no constraints can be derived
from data. In addition, benchmark models featuring a compressed spectrum
still offer a way to evade LHC constraints since final state leptons originating
from slepton decays are in this case too soft for being detected.

This independence of the results
on the neutralino nature is also depicted on Figure~\ref{fig:atlas7c},
where we present different kinematical observables.
We focus on the transverse-momentum
spectra of the two detected leptons $\ell_1$ and $\ell_2$
(first line of the figure), the missing transverse energy distribution
(left panel of the second line of the figure), the transverse mass of the lepton
pair $M_T(\ell_1\ell_2)$ (right panel of the second line of the figure), the angular distance
in the azimuthal plane between the two leptons $\Delta\Phi(\ell_1\ell_2)$
(left panel of the third line of the figure), their pseudorapidity separation
$\Delta\eta(\ell_1\ell_2)$ (right panel of the third line of the figure),
the angular distance in the azimuthal plane
of the missing momentum with the lepton pair $\Delta\Phi(\ell_1\ell_2,
\met)$ as well as the one with the hardest lepton $\Delta\Phi(\ell_1,
\met)$ (last line of the figure). On the different
subfigures, we superimpose results obtained for simplified models
exhibiting different neutralino natures and fix, for the sake of the example,
the slepton and neutralino masses to 100~GeV and 60~GeV, respectively.
From now on, we consider
the neutralino to be a mixed wino-bino state.

\begin{figure}
  \centering
  \includegraphics[width=.48\columnwidth]{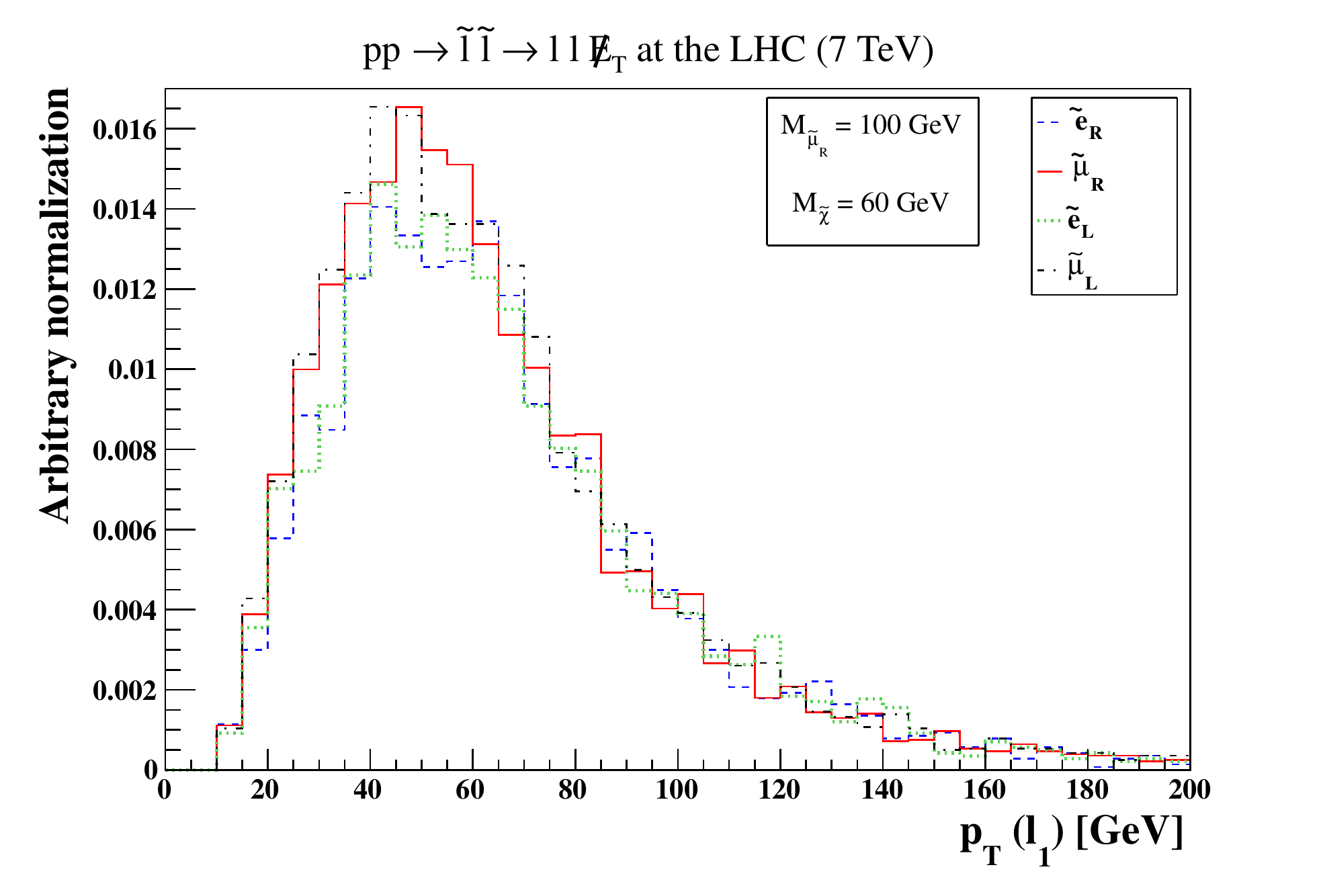}
  \includegraphics[width=.48\columnwidth]{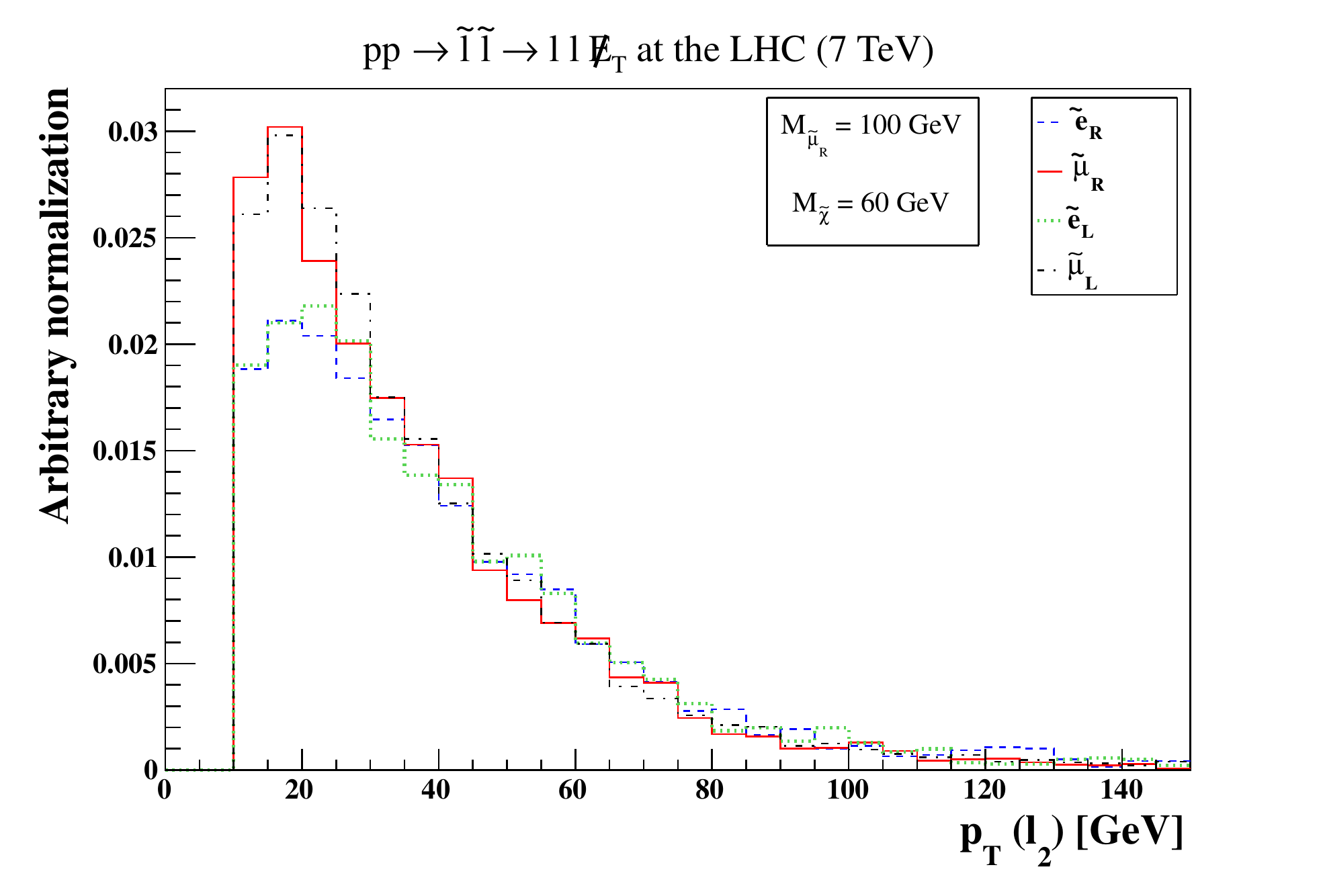}\\
  \includegraphics[width=.48\columnwidth]{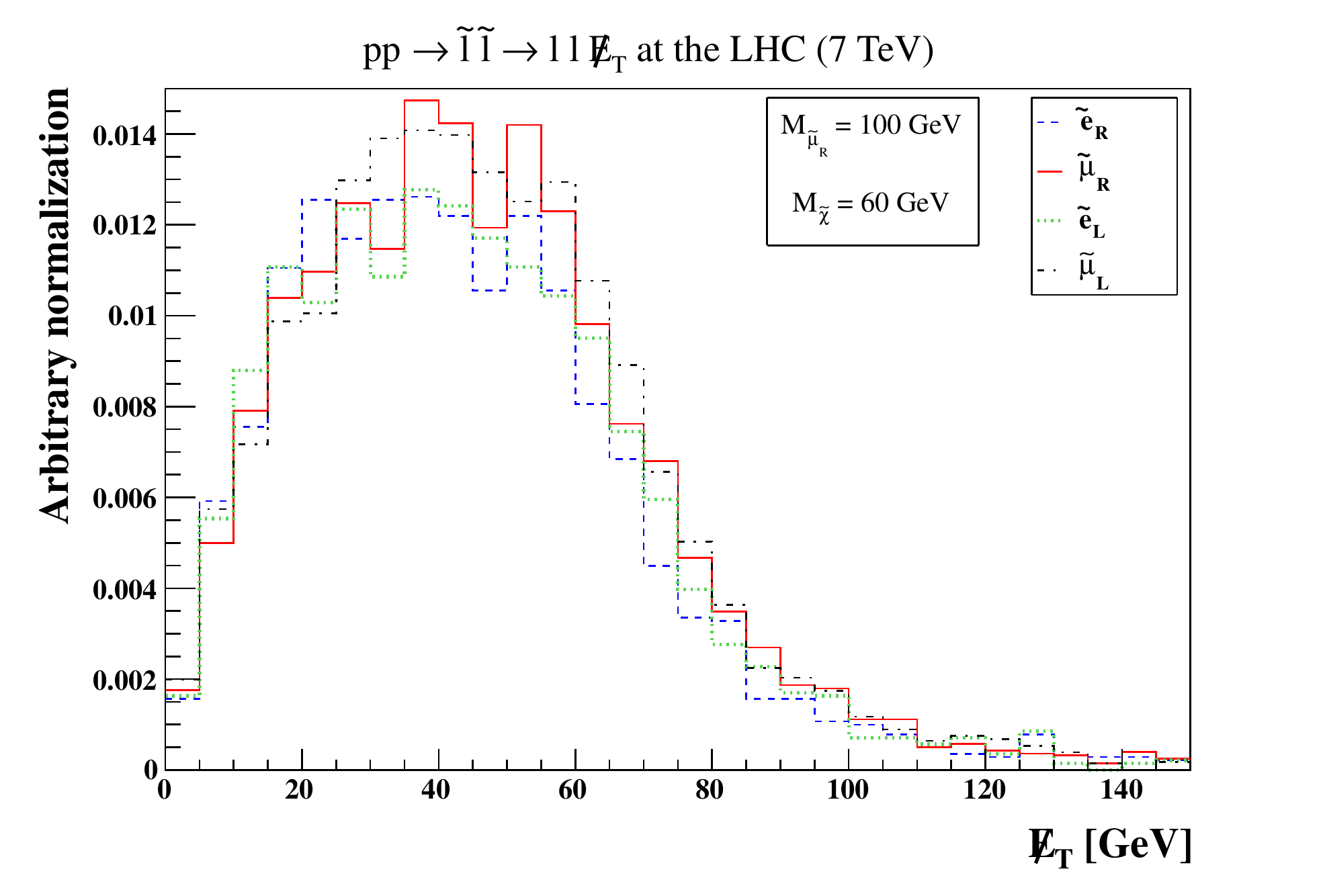}
  \includegraphics[width=.48\columnwidth]{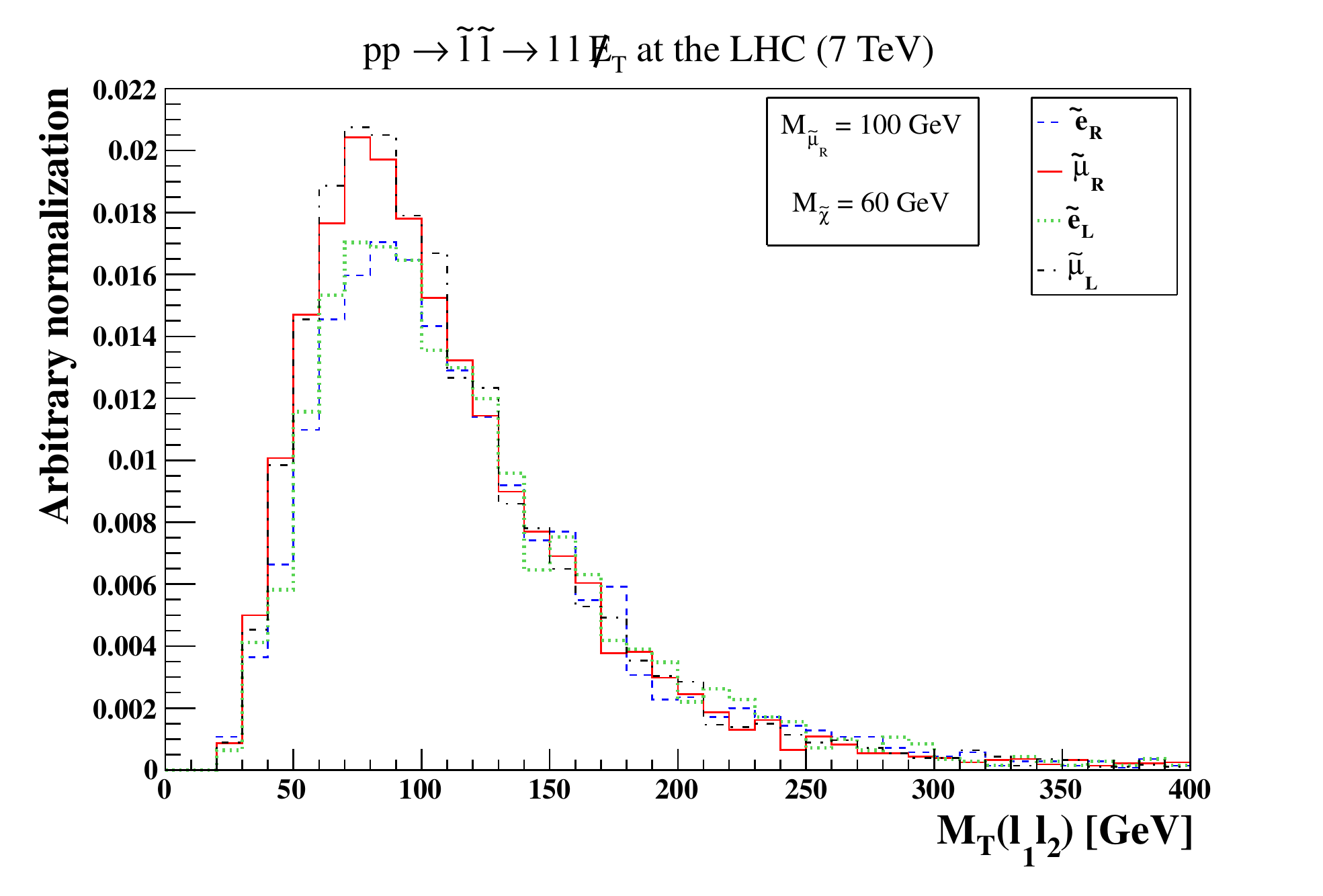}\\
  \includegraphics[width=.48\columnwidth]{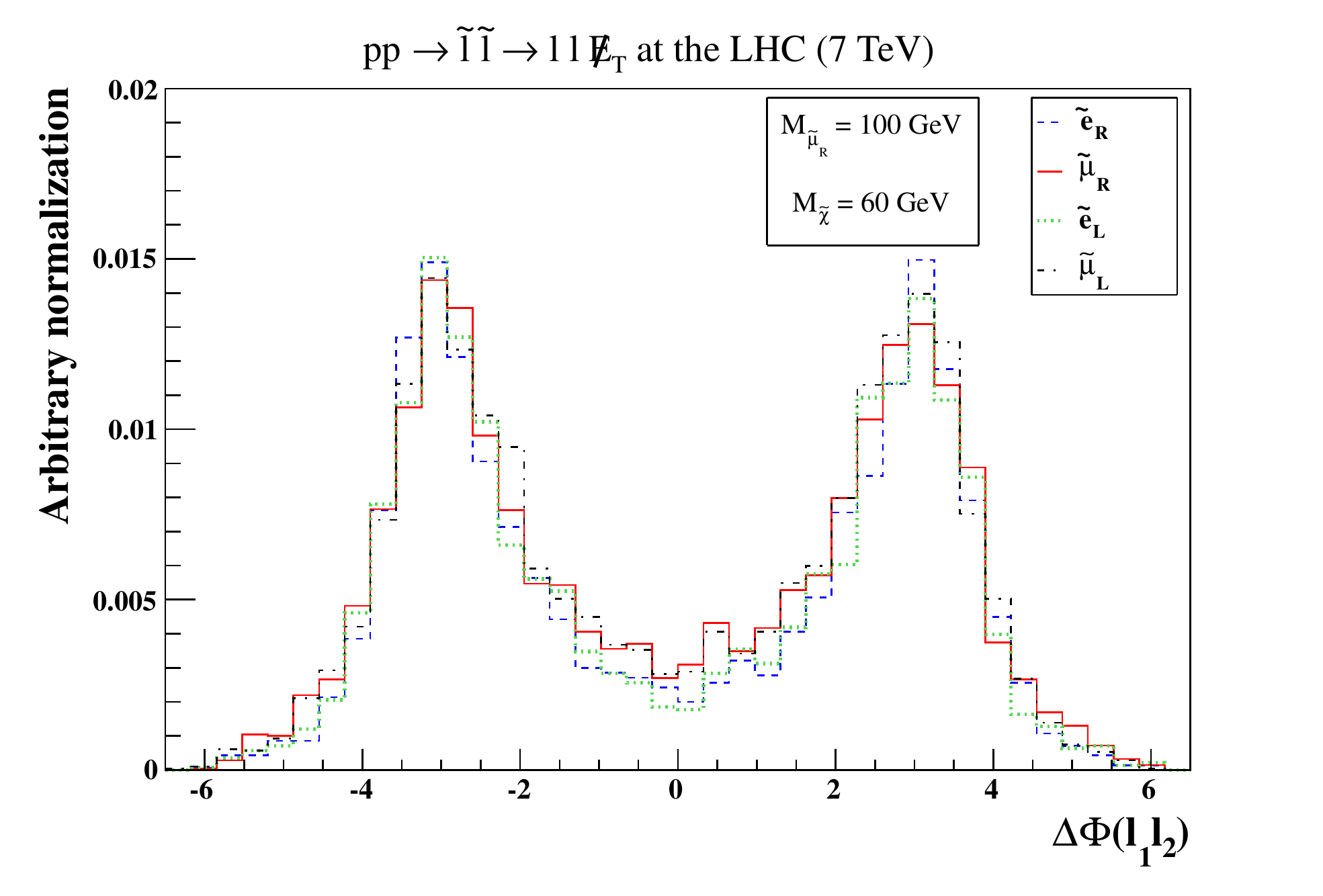}
  \includegraphics[width=.48\columnwidth]{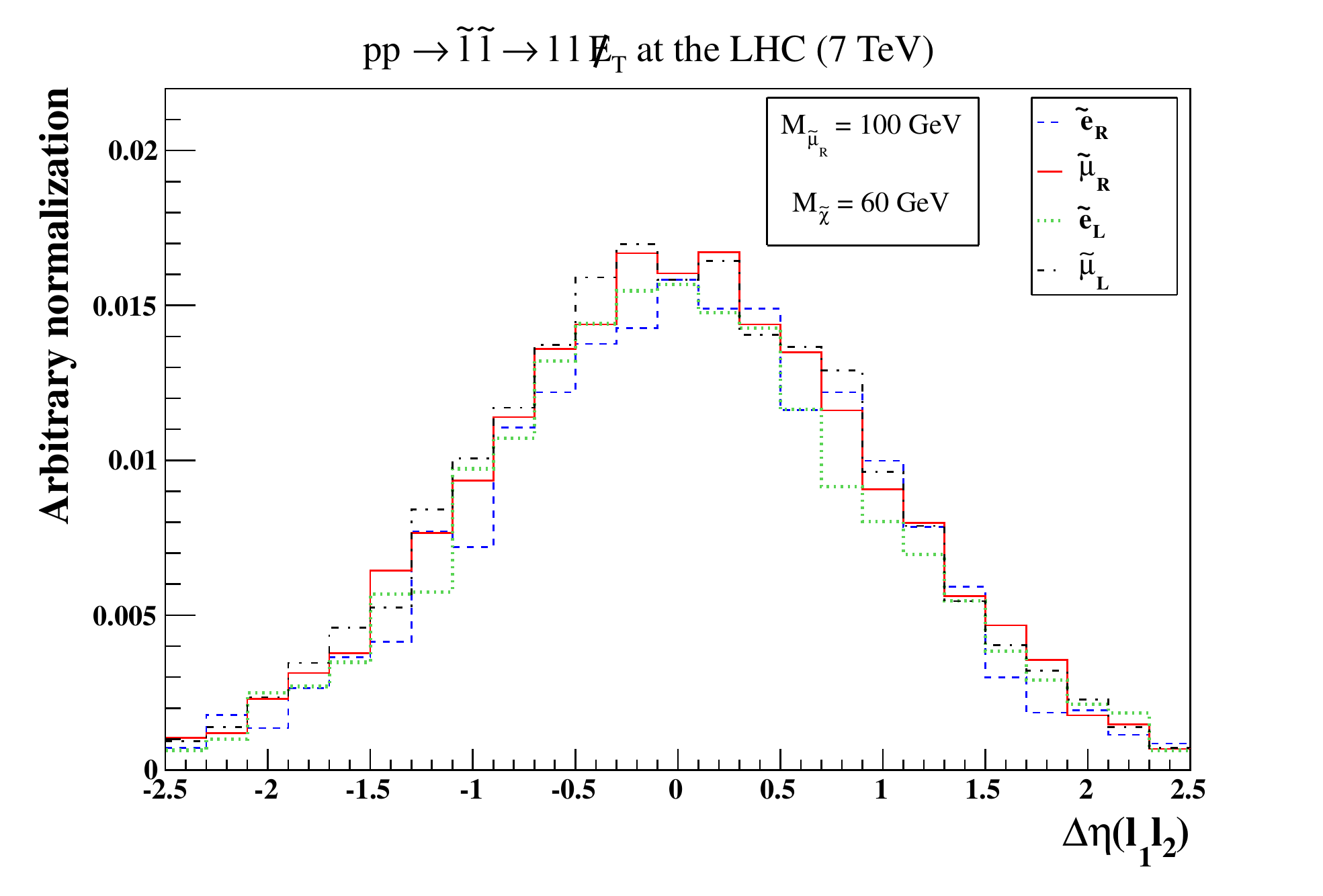}\\
  \includegraphics[width=.48\columnwidth]{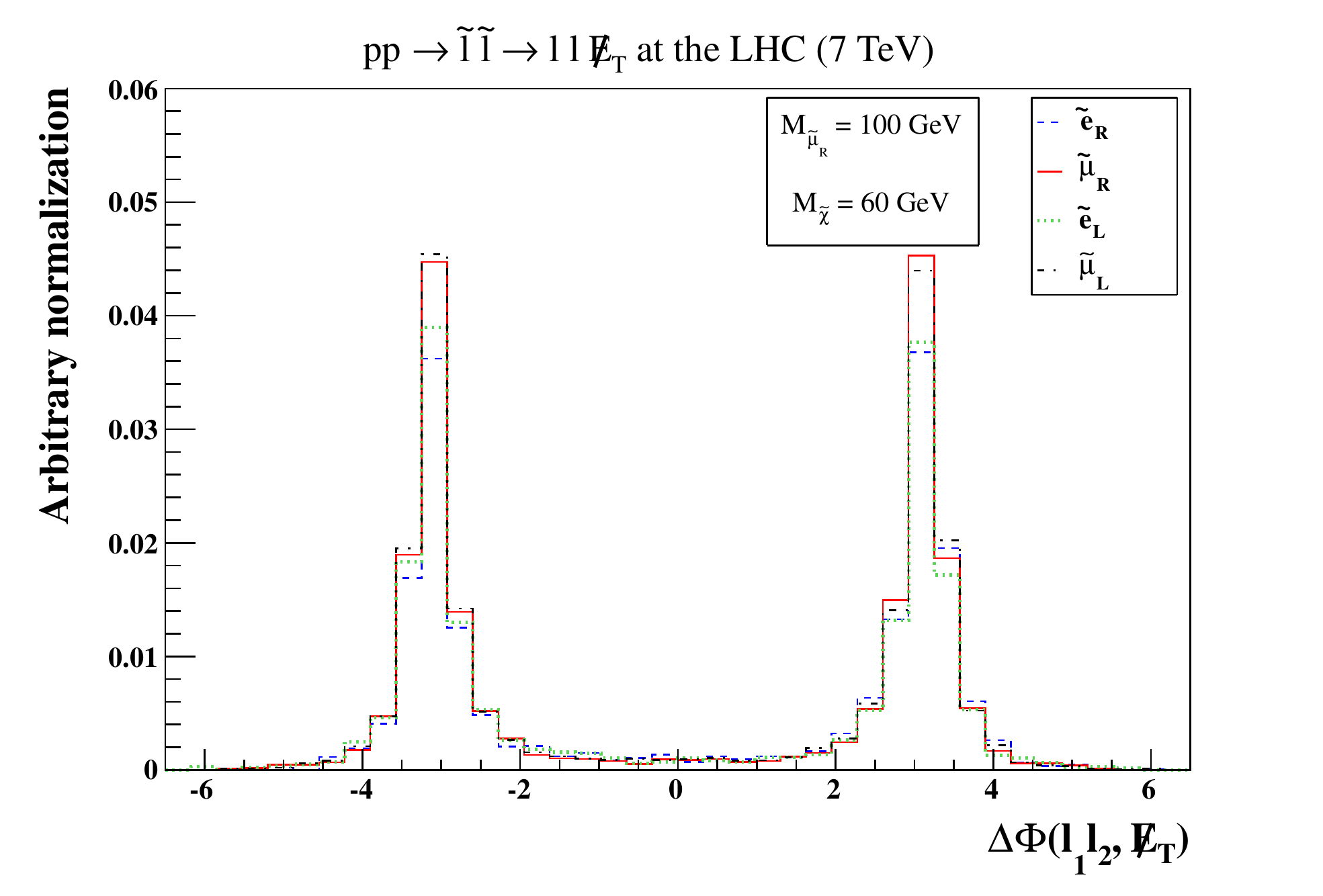}
  \includegraphics[width=.48\columnwidth]{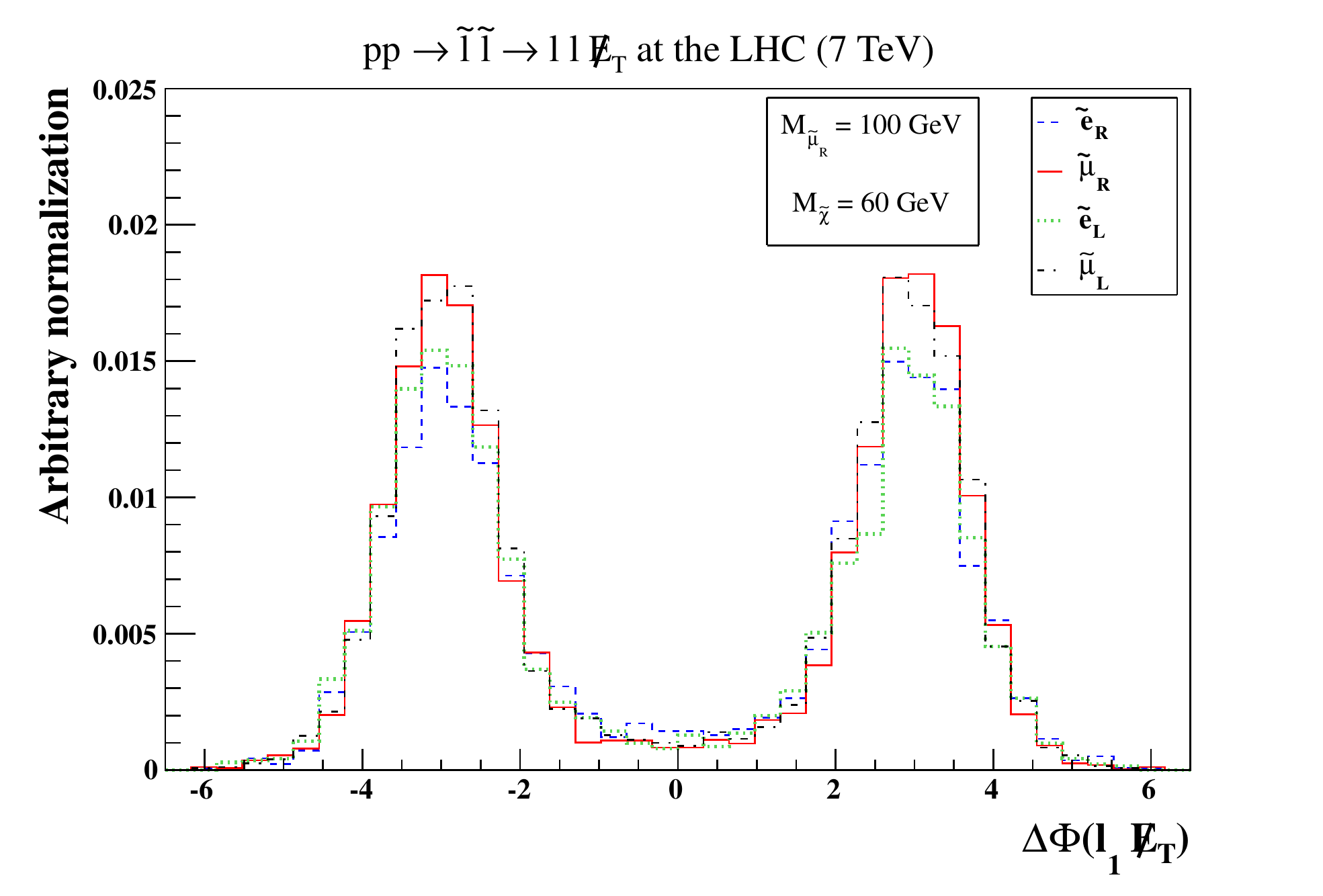}
\caption{\label{fig:atlas7d}Same as in Figure~\ref{fig:atlas7c} but when the neutralino
  nature is fixed to a bino-wino mixed state and for the production of different
  types of sleptons.}
\end{figure}

In Figure~\ref{fig:atlas7b}, we study the dependence of the results on the
left-handed and right-handed nature of the produced sleptons, as well as on
their flavor. Concerning muonic final states,
visible cross sections larger than 1.6~fb have been
excluded by the ATLAS collaboration. The resulting 95\% exclusion limits are presented
on the left and right panels of the figure for left-handed and right-handed muons,
respectively.
The central panel of the figure is dedicated to the only species
of sleptons not considered yet, \ie, right-handed selectrons. The different
selections of muon and electron candidates, together with the slightly different
detector acceptance for both types of particles, imply that the ATLAS results of
Ref.~\cite{Aad:2012pxa} more strongly constrain the simplified model parameter space
related to smuon models than the one related to selectron models.
In addition, the larger left-handed slepton
production rate allows to reach a larger mass range as when considering
right-handed sleptons.
Consequently, masses ranging up to 175~GeV (220~GeV) and 150~GeV (180~GeV) are
found excluded for models containing a
left-handed and right-handed selectrons (smuons), respectively, when the
neutralino is assumed massless. Contrary, when the neutralino is
heavier than 100~GeV and 60~GeV in the left-handed and right-handed selectron cases,
and heavier than 130~GeV and 100~GeV in the left-handed and right-handed
smuon cases, respectively, all sensitivity is lost.
These properties are also illustrated on Figure~\ref{fig:atlas7d} where we
present the variations of the kinematical distributions already
considered in Figure~\ref{fig:atlas7c}, but this time when we modify the nature of the
produced sleptons. We compare the shapes of the various distributions
and observe differences for electron and muon spectra, the figures
having been obtained prior the $Z$-boson veto, missing energy selection and the $m_{T2}$
selection steps of the analysis.

\subsection{Revisiting CMS searches for first and second generation sleptons}
\label{sec:mc12b}
First and second generation sleptons have also
been searched for in LHC collisions at a center-of-mass energy of 8~TeV,
for instance in the CMS analysis of Ref.~\cite{CMS:aro} which
we focus on in this subsection. This analysis
contains several signal regions, one of them being
expected to be sensitive to direct slepton pair production followed
by their decays into the associated Standard Model partners and missing energy.
The corresponding selection strategy is defined as follows.
\begin{itemize}
  \item We select jets whose transverse momentum is larger than 30~GeV
    and pseudorapidity $|\eta| \leq 2.5$. A veto on events containing at least
    one $b$-tagged jet is then applied, our $b$-tagging efficiency and mistagging
   rates being described in Ref.~\cite{Agram:2013koa}.
  \item Signal electrons (muons) are defined as isolated electron (muon) candidates with
    a transverse momentum larger than 20~GeV and a pseudorapidity such that
    $|\eta| \leq 2.4$. The isolation criterion is imposed by requiring
    that the activity in a cone of radius $R=0.3$ around the lepton is smaller
    than 15\% of the lepton $p_T$.
  \item We select events containing two leptons of the same flavor and
    with an opposite electric charge and impose the dilepton invariant
    mass $m_{\ell\ell}$
    to be first larger than 12~GeV (to veto events with a leptonically-decaying
    hadronic resonance)
    and next not compatible with the mass of the $Z$-boson,
    $m_{\ell\ell}\not\in[75,105]$~GeV.
  \item Each jet lying within a cone of radius $R=0.4$ around an electron is
    removed from the analysis.
  \item We require the event final state to feature at least 60~GeV of missing
    transverse energy and a transverse contransverse mass
    $M_{CT_\perp}$~\cite{Matchev:2009ad} larger than 100~GeV.
\end{itemize}

\begin{figure}
  \centering
  \includegraphics[width=.48\columnwidth]{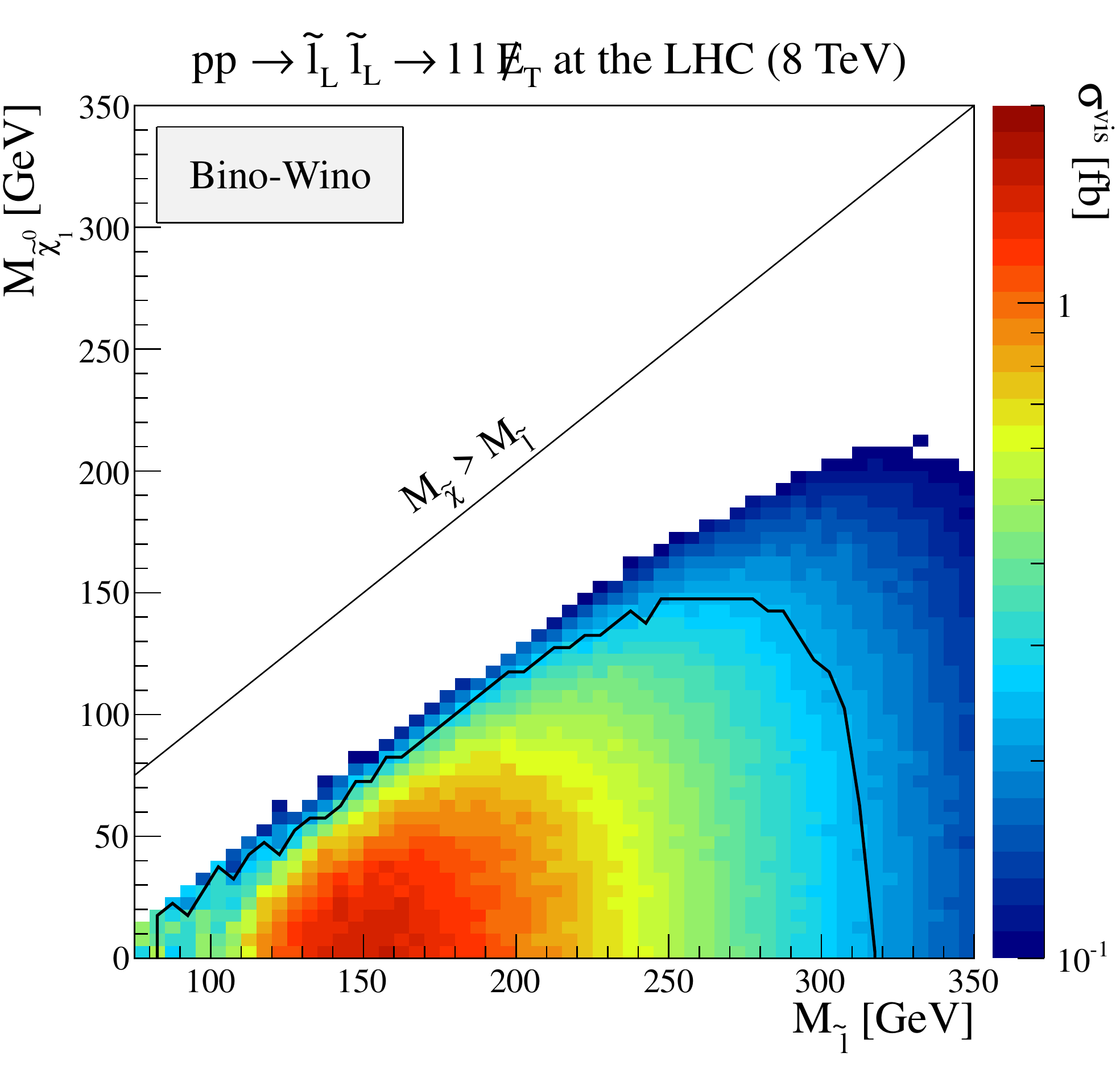}
  \includegraphics[width=.48\columnwidth]{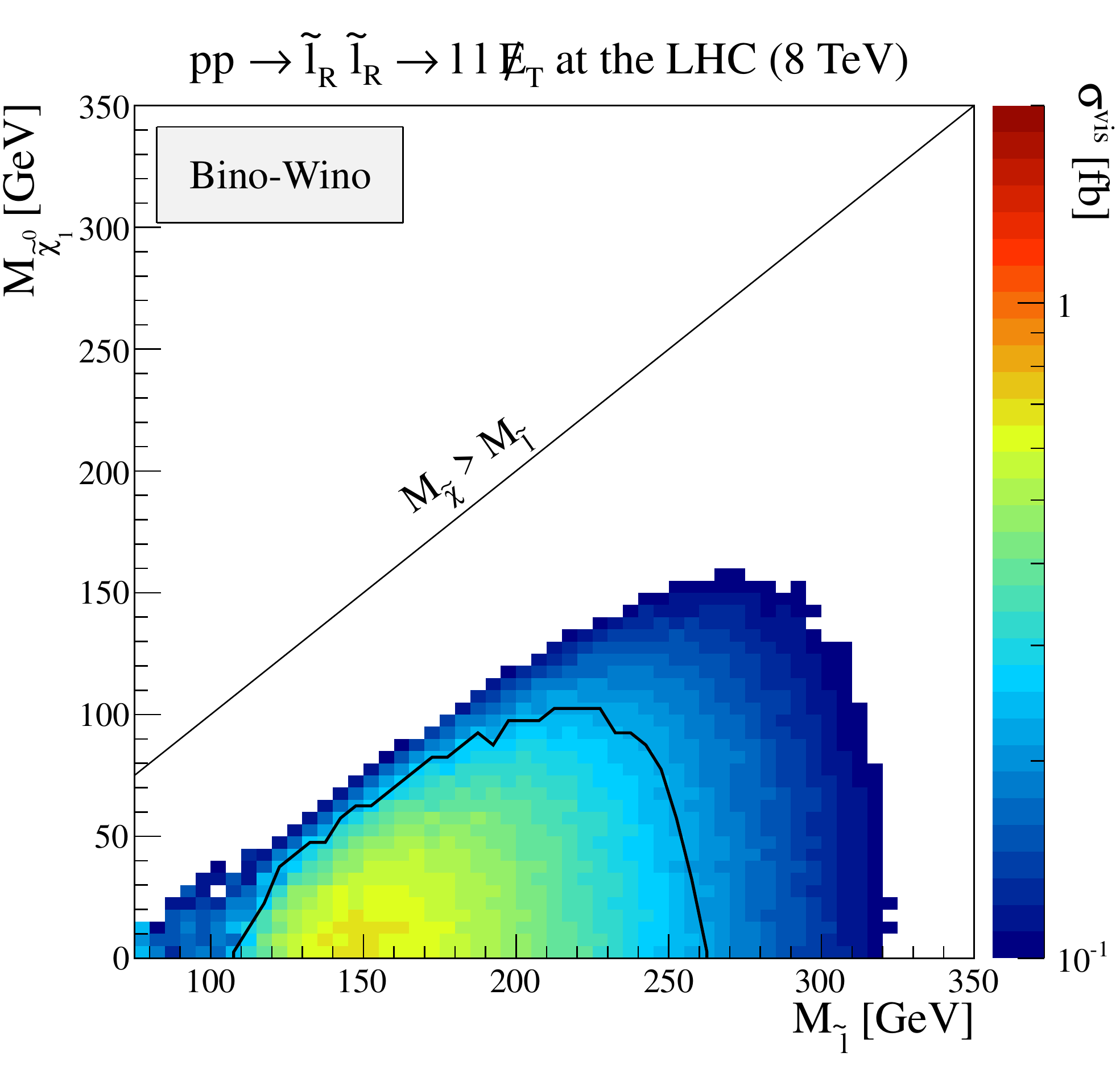}
\caption{\label{fig:cms8} 95\% confidence exclusion limit for left-handed (left)
and right-handed (right) slepton
pair production, given in the $(M_\sl,M_{\tilde\chi^0})$ mass plane of the simplified
model of Section~\ref{sec:bench12}, for a mixed bino-wino neutralino and after
summing over both first and second slepton generations.
We present signal cross sections $\sigma^{\rm vis}$ that are
visible after having applied the CMS selection strategy depicted in the text.
The limits are extracted for 9.2 fb$^{-1}$ of LHC collisions at a center-of-mass
energy of 8~TeV.}
\end{figure}

We reinterpret this analysis in the framework of the
pair production of left-handed and right-handed sleptons in the left
and right panels of Figure~\ref{fig:cms8}, respectively. Those results
are extracted after summing over both selectron
and smuon channels, considering the two types of superpartners mass-degenerate, and
our predictions are given
for an integrated luminosity of 9.2 fb$^{-1}$ of LHC collisions at a center-of-mass
energy of 8~TeV. We present, in both panels of the figure,
$(M_\sl, M_{\tilde\chi^0_1})$ planes with the visible cross section
defined as the fraction of the cross section remaining
after applying the CMS selection described above. We then superimpose
95\% confidence level exclusion contours extracted from the observation
by the CMS experiment of six data events. This number is compared
to our computations of the signal contribution, summed to the Standard Model
expectation as calculated by the CMS experiment,
which is given by $14.2 \pm 4.5$ events.

We observe that
left-handed and right-handed sleptons of masses ranging up to 310~GeV and 250~GeV
are excluded, respectively, in the case the lightest neutralino is massless.
These upper bounds hold, like for the results shown in Section~\ref{sec:mc12a},
for heavier neutralinos with masses smaller than 150~GeV (100~GeV) in the
case of left-handed (right-handed) sleptons. Moreover,
the CMS analysis is found to be insensitive to sleptons via their direct production
mechanism both when the neutralino mass is larger than those values, and
when the mass
difference $M_\sl-M_{\tilde\chi_1^0}$ becomes smaller
than 70~GeV, the bulk of the Standard Model leptons originating from the slepton decays
being soft enough not to populate the signal region.

\section{Conclusion}
\label{sec:conclusions}

In conclusion, we have updated in this paper our precision predictions at next-to-leading
order of perturbative QCD matched to resummation at the next-to-leading logarithmic accuracy
for direct slepton pair production in proton-proton collisions. We used simplified models,
which are now commonly adopted by the experimental collaborations for selectrons and
smuons as well as mixing staus, as benchmarks for total cross sections at past
($\sqrt{S}=7$~TeV and 8~TeV) and future ($\sqrt{S}=13$~TeV and 14~TeV) center-of-mass energies.
Our calculations stabilized the cross sections considerably with respect to renormalization
and factorization scale uncertainties, but also with respect to parton density variations.
The most precise slepton-pair production cross section values have in this
way been made available so that they can now be used in all future LHC
experimental analyses dedicated to the direct search for sleptons.

They were then
employed in combination with modern Monte Carlo techniques to reanalyze recent ATLAS and CMS
slepton searches for various assumptions on the decomposition of the sleptons and their
neutralino decay products, giving rise to more precise and more specific current exclusion
limits for first- and second-generation sleptons. We conclude from our results that exhibit different limits for selectron
and smuon scenarios and for both chiralities, that it might be valuable to
present, in the future, slepton results without combining left-handed and
right-handed channels, as well as selectron and smuon ones.
However, a reanalysis of an ATLAS
stau search within our simplified model parameter space proved to be impossible due
to a complicated tau reconstruction together with small signal cross sections, in particular
once tau decay branching ratios were included.

\appendix
\section{Total cross sections at center-of-mass energies of 13~TeV and 14~TeV}
\label{app:highcms}
This appendix complements the summary tables of Section~\ref{sec:summarytables}
and contains total cross sections for slepton pair production at the LHC. We focus
on the future runs at center-of-mass energies of 13~TeV and 14~TeV and respectively
present the results, in Table~\ref{tab:13tev} and Table~\ref{tab:14tev},
for specific choices of the slepton mass. Our predictions
are given together with the uncertainties
arising from the variation of the unphysical scales by a factor
of two around their central value chosen to be the produced slepton mass,
as well as those issued from the choice of the parton densities. The latter
have been computed by using the 52 parton density fits obtained from
variations along the 26 eigenvectors of the
covariance matrix associated with the CT10 NLO best fit of the CTEQ collaboration,
adopted as our central choice.

\begin{table}[!t]
\renewcommand{\arraystretch}{1.195} 
 \begin{center}
 \begin{tabular}{| c | c || l | l | l |} 
 \hline 
  $M_{\tilde \ell}$ [GeV] & Final state &  LO [fb] & NLO [fb] & NLO+NLL [fb] \\ 
 \hline
\multirow{3}{*}{$50$}
 & $\tilde\ell_L^+\tilde\ell_L^-$  & $3194.60_{-11.2\%}^{+9.8\%}$ & $4139.50_{-0.7\%}^{+1.2\%}{}_{-3.7\%}^{+3.1\%}$ &  $4104.90_{-1.6\%}^{+1.4\%}{}_{-3.6\%}^{+3.1\%}$  \\
 & $\tilde\ell_R^+\tilde\ell_R^-$  & $1071.70_{-11.1\%}^{+9.8\%}$ & $1389.10_{-0.8\%}^{+1.1\%}{}_{-3.8\%}^{+3.2\%}$ &  $1377.60_{-1.6\%}^{+1.4\%}{}_{-3.7\%}^{+3.3\%}$  \\
 & $\tilde\tau_1^+\tilde\tau_1^-$  & $731.01_{-10.2\%}^{+8.8\%}$ & $952.84_{-0.2\%}^{+0.7\%}{}_{-4.5\%}^{+3.8\%}$ &  $943.29_{-1.3\%}^{+1.2\%}{}_{-4.4\%}^{+3.8\%}$  \\
\hline
\multirow{3}{*}{$100$}
 & $\tilde\ell_L^+\tilde\ell_L^-$  & $223.80_{-3.9\%}^{+2.9\%}$ & $273.89_{-0.8\%}^{+1.5\%}{}_{-3.5\%}^{+3.2\%}$ &  $270.79_{-0.4\%}^{+0.0\%}{}_{-3.4\%}^{+3.3\%}$  \\
 & $\tilde\ell_R^+\tilde\ell_R^-$  & $79.68_{-3.9\%}^{+2.9\%}$ & $97.59_{-0.8\%}^{+1.5\%}{}_{-4.1\%}^{+3.5\%}$ &  $96.51_{-0.3\%}^{+0.0\%}{}_{-4.0\%}^{+3.6\%}$  \\
 & $\tilde\tau_1^+\tilde\tau_1^-$  & $84.85_{-3.9\%}^{+2.9\%}$ & $103.94_{-0.9\%}^{+1.5\%}{}_{-4.3\%}^{+3.7\%}$ &  $102.77_{-0.3\%}^{+0.0\%}{}_{-4.3\%}^{+3.7\%}$  \\
\hline
\multirow{3}{*}{$150$}
 & $\tilde\ell_L^+\tilde\ell_L^-$  & $53.88_{-0.7\%}^{+0.1\%}$ & $64.12_{-1.3\%}^{+1.9\%}{}_{-3.9\%}^{+3.4\%}$ &  $63.34_{-0.3\%}^{+0.1\%}{}_{-3.9\%}^{+3.4\%}$  \\
 & $\tilde\ell_R^+\tilde\ell_R^-$  & $19.81_{-0.7\%}^{+0.1\%}$ & $23.60_{-1.3\%}^{+1.9\%}{}_{-4.4\%}^{+3.7\%}$ &  $23.32_{-0.3\%}^{+0.1\%}{}_{-4.5\%}^{+3.7\%}$  \\
 & $\tilde\tau_1^+\tilde\tau_1^-$  & $21.69_{-0.7\%}^{+0.1\%}$ & $25.83_{-1.3\%}^{+1.9\%}{}_{-4.6\%}^{+3.7\%}$ &  $25.53_{-0.3\%}^{+0.1\%}{}_{-4.6\%}^{+3.8\%}$  \\
\hline
\multirow{3}{*}{$200$}
 & $\tilde\ell_L^+\tilde\ell_L^-$  & $18.84_{-1.6\%}^{+1.3\%}$ & $22.07_{-1.5\%}^{+2.0\%}{}_{-4.4\%}^{+3.6\%}$ &  $21.81_{-0.4\%}^{+0.1\%}{}_{-4.4\%}^{+3.6\%}$  \\
 & $\tilde\ell_R^+\tilde\ell_R^-$  & $7.03_{-1.6\%}^{+1.3\%}$ & $8.25_{-1.5\%}^{+2.0\%}{}_{-4.9\%}^{+3.9\%}$ &  $8.15_{-0.4\%}^{+0.1\%}{}_{-4.9\%}^{+3.9\%}$  \\
 & $\tilde\tau_1^+\tilde\tau_1^-$  & $7.77_{-1.6\%}^{+1.3\%}$ & $9.11_{-1.5\%}^{+2.0\%}{}_{-5.0\%}^{+3.9\%}$ &  $9.00_{-0.4\%}^{+0.1\%}{}_{-5.0\%}^{+3.9\%}$  \\
\hline
\multirow{3}{*}{$250$}
 & $\tilde\ell_L^+\tilde\ell_L^-$  & $8.04_{-2.9\%}^{+2.8\%}$ & $9.33_{-1.7\%}^{+2.0\%}{}_{-4.9\%}^{+3.9\%}$ &  $9.21_{-0.4\%}^{+0.3\%}{}_{-5.0\%}^{+3.9\%}$  \\
 & $\tilde\ell_R^+\tilde\ell_R^-$  & $3.03_{-2.8\%}^{+2.8\%}$ & $3.52_{-1.7\%}^{+2.0\%}{}_{-5.3\%}^{+4.2\%}$ &  $3.47_{-0.4\%}^{+0.2\%}{}_{-5.3\%}^{+4.2\%}$  \\
 & $\tilde\tau_1^+\tilde\tau_1^-$  & $3.36_{-2.8\%}^{+2.8\%}$ & $3.90_{-1.7\%}^{+2.0\%}{}_{-5.4\%}^{+4.2\%}$ &  $3.85_{-0.4\%}^{+0.1\%}{}_{-5.4\%}^{+4.3\%}$  \\
\hline
\multirow{3}{*}{$300$}
 & $\tilde\ell_L^+\tilde\ell_L^-$  & $3.89_{-3.8\%}^{+4.0\%}$ & $4.49_{-1.8\%}^{+2.0\%}{}_{-5.4\%}^{+4.2\%}$ &  $4.43_{-0.3\%}^{+0.1\%}{}_{-5.4\%}^{+4.2\%}$  \\
 & $\tilde\ell_R^+\tilde\ell_R^-$  & $1.48_{-3.8\%}^{+3.9\%}$ & $1.70_{-1.8\%}^{+2.0\%}{}_{-5.7\%}^{+4.5\%}$ &  $1.68_{-0.3\%}^{+0.2\%}{}_{-5.7\%}^{+4.5\%}$  \\
 & $\tilde\tau_1^+\tilde\tau_1^-$  & $1.64_{-3.8\%}^{+3.9\%}$ & $1.89_{-1.8\%}^{+2.0\%}{}_{-5.8\%}^{+4.6\%}$ &  $1.87_{-0.3\%}^{+0.2\%}{}_{-5.8\%}^{+4.6\%}$  \\
\hline
\multirow{3}{*}{$350$}
 & $\tilde\ell_L^+\tilde\ell_L^-$  & $2.05_{-4.6\%}^{+4.9\%}$ & $2.36_{-2.0\%}^{+2.1\%}{}_{-5.9\%}^{+4.6\%}$ &  $2.33_{-0.3\%}^{+0.0\%}{}_{-5.9\%}^{+4.6\%}$  \\
 & $\tilde\ell_R^+\tilde\ell_R^-$  & $0.78_{-4.5\%}^{+4.8\%}$ & $0.90_{-2.0\%}^{+2.1\%}{}_{-6.1\%}^{+4.9\%}$ &  $0.89_{-0.3\%}^{+0.0\%}{}_{-6.1\%}^{+4.9\%}$  \\
 & $\tilde\tau_1^+\tilde\tau_1^-$  & $0.87_{-4.5\%}^{+4.8\%}$ & $1.00_{-2.0\%}^{+2.0\%}{}_{-6.2\%}^{+5.0\%}$ &  $0.99_{-0.3\%}^{+0.0\%}{}_{-6.3\%}^{+4.9\%}$  \\
\hline
\multirow{3}{*}{$400$}
 & $\tilde\ell_L^+\tilde\ell_L^-$  & $1.15_{-5.2\%}^{+5.7\%}$ & $1.32_{-2.1\%}^{+2.1\%}{}_{-6.3\%}^{+5.0\%}$ &  $1.31_{-0.3\%}^{+0.0\%}{}_{-6.5\%}^{+5.0\%}$  \\
 & $\tilde\ell_R^+\tilde\ell_R^-$  & $0.44_{-5.2\%}^{+5.7\%}$ & $0.51_{-2.1\%}^{+2.1\%}{}_{-6.6\%}^{+5.3\%}$ &  $0.50_{-0.4\%}^{+0.0\%}{}_{-6.7\%}^{+5.2\%}$  \\
 & $\tilde\tau_1^+\tilde\tau_1^-$  & $0.49_{-5.2\%}^{+5.6\%}$ & $0.56_{-2.1\%}^{+2.1\%}{}_{-6.6\%}^{+5.4\%}$ &  $0.56_{-0.3\%}^{+0.0\%}{}_{-6.8\%}^{+5.3\%}$  \\
\hline
\multirow{3}{*}{$450$}
 & $\tilde\ell_L^+\tilde\ell_L^-$  & $0.68_{-5.8\%}^{+6.4\%}$ & $0.78_{-2.2\%}^{+2.1\%}{}_{-6.8\%}^{+5.4\%}$ &  $0.77_{-0.3\%}^{+0.0\%}{}_{-6.8\%}^{+5.4\%}$  \\
 & $\tilde\ell_R^+\tilde\ell_R^-$  & $0.26_{-5.7\%}^{+6.4\%}$ & $0.30_{-2.2\%}^{+2.1\%}{}_{-7.0\%}^{+5.7\%}$ &  $0.30_{-0.3\%}^{+0.0\%}{}_{-7.0\%}^{+5.7\%}$  \\
 & $\tilde\tau_1^+\tilde\tau_1^-$  & $0.29_{-5.7\%}^{+6.3\%}$ & $0.33_{-2.2\%}^{+2.1\%}{}_{-7.0\%}^{+5.8\%}$ &  $0.33_{-0.3\%}^{+0.0\%}{}_{-7.1\%}^{+5.7\%}$  \\
\hline
\multirow{3}{*}{$500$}
 & $\tilde\ell_L^+\tilde\ell_L^-$  & $0.42_{-6.3\%}^{+7.1\%}$ & $0.48_{-2.3\%}^{+2.1\%}{}_{-7.2\%}^{+5.8\%}$ &  $0.47_{-0.4\%}^{+0.0\%}{}_{-7.1\%}^{+5.9\%}$  \\
 & $\tilde\ell_R^+\tilde\ell_R^-$  & $0.16_{-6.2\%}^{+7.0\%}$ & $0.18_{-2.3\%}^{+2.1\%}{}_{-7.4\%}^{+6.1\%}$ &  $0.18_{-0.3\%}^{+0.0\%}{}_{-7.4\%}^{+6.2\%}$  \\
 & $\tilde\tau_1^+\tilde\tau_1^-$  & $0.18_{-6.2\%}^{+7.0\%}$ & $0.21_{-2.3\%}^{+2.1\%}{}_{-7.5\%}^{+6.2\%}$ &  $0.20_{-0.3\%}^{+0.0\%}{}_{-7.4\%}^{+6.3\%}$  \\
\hline
\end{tabular} 
\caption{\label{tab:13tev} Total production cross sections at the LHC, running
  at a center-of-mass energy of 13~TeV, for first or second generation left-handed
  and right-handed sleptons, as well as for maximally mixing staus. Results are presented
  together with the associated scale and PDF uncertainties.}
\end{center}
\end{table}

\begin{table}[!t] \renewcommand{\arraystretch}{1.21} 
 \begin{center} 
 \begin{tabular}{| c | c || l | l | l |} 
 \hline 
  $M_{\tilde \ell}$ [GeV] & Final state &  LO [fb] & NLO [fb] & NLO+NLL [fb] \\ 
 \hline
\multirow{3}{*}{$50$}
 & $\tilde\ell_L^+\tilde\ell_L^-$  & $3467.30_{-11.7\%}^{+10.3\%}$ & $4507.50_{-0.9\%}^{+1.3\%}{}_{-3.8\%}^{+3.1\%}$ &  $4470.80_{-1.6\%}^{+1.6\%}{}_{-3.7\%}^{+3.2\%}$  \\
 & $\tilde\ell_R^+\tilde\ell_R^-$  & $1162.40_{-11.6\%}^{+10.3\%}$ & $1511.40_{-1.0\%}^{+1.3\%}{}_{-3.8\%}^{+3.2\%}$ &  $1499.50_{-1.7\%}^{+1.6\%}{}_{-3.8\%}^{+3.2\%}$  \\
 & $\tilde\tau_1^+\tilde\tau_1^-$  & $792.71_{-10.7\%}^{+9.3\%}$ & $1037.10_{-0.4\%}^{+0.9\%}{}_{-4.5\%}^{+3.8\%}$ &  $1027.10_{-1.4\%}^{+1.3\%}{}_{-4.4\%}^{+3.8\%}$  \\
\hline
\multirow{3}{*}{$100$}
 & $\tilde\ell_L^+\tilde\ell_L^-$  & $246.85_{-4.4\%}^{+3.4\%}$ & $303.04_{-0.7\%}^{+1.3\%}{}_{-3.4\%}^{+3.2\%}$ &  $299.78_{-0.5\%}^{+0.0\%}{}_{-3.4\%}^{+3.2\%}$  \\
 & $\tilde\ell_R^+\tilde\ell_R^-$  & $87.75_{-4.4\%}^{+3.3\%}$ & $107.81_{-0.7\%}^{+1.4\%}{}_{-4.0\%}^{+3.5\%}$ &  $106.68_{-0.4\%}^{+0.0\%}{}_{-4.1\%}^{+3.5\%}$  \\
 & $\tilde\tau_1^+\tilde\tau_1^-$  & $93.41_{-4.3\%}^{+3.3\%}$ & $114.79_{-0.7\%}^{+1.4\%}{}_{-4.3\%}^{+3.6\%}$ &  $113.58_{-0.4\%}^{+0.0\%}{}_{-4.3\%}^{+3.6\%}$  \\
\hline
\multirow{3}{*}{$150$}
 & $\tilde\ell_L^+\tilde\ell_L^-$  & $60.19_{-1.2\%}^{+0.5\%}$ & $71.82_{-1.2\%}^{+1.8\%}{}_{-3.8\%}^{+3.3\%}$ &  $70.98_{-0.3\%}^{+0.0\%}{}_{-3.8\%}^{+3.3\%}$  \\
 & $\tilde\ell_R^+\tilde\ell_R^-$  & $22.09_{-1.2\%}^{+0.5\%}$ & $26.38_{-1.2\%}^{+1.8\%}{}_{-4.3\%}^{+3.6\%}$ &  $26.09_{-0.4\%}^{+0.0\%}{}_{-4.3\%}^{+3.6\%}$  \\
 & $\tilde\tau_1^+\tilde\tau_1^-$  & $24.18_{-1.2\%}^{+0.5\%}$ & $28.88_{-1.2\%}^{+1.8\%}{}_{-4.5\%}^{+3.7\%}$ &  $28.55_{-0.4\%}^{+0.0\%}{}_{-4.5\%}^{+3.7\%}$  \\
\hline
\multirow{3}{*}{$200$}
 & $\tilde\ell_L^+\tilde\ell_L^-$  & $21.29_{-1.2\%}^{+0.9\%}$ & $25.00_{-1.4\%}^{+1.9\%}{}_{-4.2\%}^{+3.5\%}$ &  $24.70_{-0.4\%}^{+0.1\%}{}_{-4.3\%}^{+3.5\%}$  \\
 & $\tilde\ell_R^+\tilde\ell_R^-$  & $7.94_{-1.2\%}^{+0.8\%}$ & $9.32_{-1.4\%}^{+1.9\%}{}_{-4.7\%}^{+3.8\%}$ &  $9.22_{-0.4\%}^{+0.1\%}{}_{-4.8\%}^{+3.8\%}$  \\
 & $\tilde\tau_1^+\tilde\tau_1^-$  & $8.76_{-1.2\%}^{+0.8\%}$ & $10.29_{-1.4\%}^{+1.9\%}{}_{-4.9\%}^{+3.9\%}$ &  $10.17_{-0.4\%}^{+0.1\%}{}_{-4.9\%}^{+3.9\%}$  \\
\hline
\multirow{3}{*}{$250$}
 & $\tilde\ell_L^+\tilde\ell_L^-$  & $9.19_{-2.5\%}^{+2.4\%}$ & $10.69_{-1.6\%}^{+2.0\%}{}_{-4.7\%}^{+3.8\%}$ &  $10.55_{-0.4\%}^{+0.0\%}{}_{-4.8\%}^{+3.8\%}$  \\
 & $\tilde\ell_R^+\tilde\ell_R^-$  & $3.46_{-2.4\%}^{+2.3\%}$ & $4.02_{-1.6\%}^{+2.0\%}{}_{-5.2\%}^{+4.0\%}$ &  $3.97_{-0.4\%}^{+0.0\%}{}_{-5.2\%}^{+4.1\%}$  \\
 & $\tilde\tau_1^+\tilde\tau_1^-$  & $3.83_{-2.4\%}^{+2.3\%}$ & $4.46_{-1.6\%}^{+2.0\%}{}_{-5.2\%}^{+4.1\%}$ &  $4.41_{-0.4\%}^{+0.0\%}{}_{-5.3\%}^{+4.1\%}$  \\
\hline
\multirow{3}{*}{$300$}
 & $\tilde\ell_L^+\tilde\ell_L^-$  & $4.50_{-3.4\%}^{+3.5\%}$ & $5.20_{-1.7\%}^{+2.0\%}{}_{-5.2\%}^{+4.1\%}$ &  $5.13_{-0.4\%}^{+0.2\%}{}_{-5.3\%}^{+4.1\%}$  \\
 & $\tilde\ell_R^+\tilde\ell_R^-$  & $1.70_{-3.4\%}^{+3.5\%}$ & $1.97_{-1.7\%}^{+2.0\%}{}_{-5.5\%}^{+4.3\%}$ &  $1.94_{-0.4\%}^{+0.2\%}{}_{-5.6\%}^{+4.4\%}$  \\
 & $\tilde\tau_1^+\tilde\tau_1^-$  & $1.89_{-3.4\%}^{+3.4\%}$ & $2.19_{-1.7\%}^{+2.0\%}{}_{-5.6\%}^{+4.4\%}$ &  $2.16_{-0.4\%}^{+0.2\%}{}_{-5.7\%}^{+4.4\%}$  \\
\hline
\multirow{3}{*}{$350$}
 & $\tilde\ell_L^+\tilde\ell_L^-$  & $2.40_{-4.2\%}^{+4.5\%}$ & $2.76_{-1.9\%}^{+2.0\%}{}_{-5.6\%}^{+4.4\%}$ &  $2.73_{-0.3\%}^{+0.0\%}{}_{-5.8\%}^{+4.3\%}$  \\
 & $\tilde\ell_R^+\tilde\ell_R^-$  & $0.91_{-4.1\%}^{+4.4\%}$ & $1.05_{-1.9\%}^{+2.0\%}{}_{-5.9\%}^{+4.7\%}$ &  $1.04_{-0.3\%}^{+0.0\%}{}_{-6.1\%}^{+4.7\%}$  \\
 & $\tilde\tau_1^+\tilde\tau_1^-$  & $1.02_{-4.1\%}^{+4.4\%}$ & $1.17_{-1.9\%}^{+2.0\%}{}_{-6.0\%}^{+4.8\%}$ &  $1.16_{-0.3\%}^{+0.1\%}{}_{-6.0\%}^{+4.8\%}$  \\
\hline
\multirow{3}{*}{$400$}
 & $\tilde\ell_L^+\tilde\ell_L^-$  & $1.36_{-4.9\%}^{+5.3\%}$ & $1.56_{-2.0\%}^{+2.0\%}{}_{-6.1\%}^{+4.8\%}$ &  $1.54_{-0.3\%}^{+0.0\%}{}_{-6.1\%}^{+4.9\%}$  \\
 & $\tilde\ell_R^+\tilde\ell_R^-$  & $0.52_{-4.8\%}^{+5.2\%}$ & $0.60_{-2.0\%}^{+2.0\%}{}_{-6.3\%}^{+5.1\%}$ &  $0.59_{-0.3\%}^{+0.0\%}{}_{-6.3\%}^{+5.2\%}$  \\
 & $\tilde\tau_1^+\tilde\tau_1^-$  & $0.58_{-4.8\%}^{+5.2\%}$ & $0.67_{-2.0\%}^{+2.0\%}{}_{-6.4\%}^{+5.1\%}$ &  $0.66_{-0.3\%}^{+0.0\%}{}_{-6.4\%}^{+5.2\%}$  \\
\hline
\multirow{3}{*}{$450$}
 & $\tilde\ell_L^+\tilde\ell_L^-$  & $0.81_{-5.4\%}^{+6.0\%}$ & $0.93_{-2.1\%}^{+2.1\%}{}_{-6.5\%}^{+5.1\%}$ &  $0.92_{-0.2\%}^{+0.0\%}{}_{-6.5\%}^{+5.3\%}$  \\
 & $\tilde\ell_R^+\tilde\ell_R^-$  & $0.31_{-5.4\%}^{+5.9\%}$ & $0.36_{-2.1\%}^{+2.0\%}{}_{-6.7\%}^{+5.4\%}$ &  $0.35_{-0.2\%}^{+0.0\%}{}_{-6.7\%}^{+5.6\%}$  \\
 & $\tilde\tau_1^+\tilde\tau_1^-$  & $0.35_{-5.3\%}^{+5.9\%}$ & $0.40_{-2.1\%}^{+2.0\%}{}_{-6.7\%}^{+5.5\%}$ &  $0.39_{-0.3\%}^{+0.0\%}{}_{-6.7\%}^{+5.6\%}$  \\
\hline
\multirow{3}{*}{$500$}
 & $\tilde\ell_L^+\tilde\ell_L^-$  & $0.50_{-5.9\%}^{+6.6\%}$ & $0.58_{-2.2\%}^{+2.1\%}{}_{-6.9\%}^{+5.5\%}$ &  $0.57_{-0.3\%}^{+0.0\%}{}_{-7.0\%}^{+5.4\%}$  \\
 & $\tilde\ell_R^+\tilde\ell_R^-$  & $0.19_{-5.8\%}^{+6.5\%}$ & $0.22_{-2.2\%}^{+2.1\%}{}_{-7.1\%}^{+5.8\%}$ &  $0.22_{-0.3\%}^{+0.0\%}{}_{-7.2\%}^{+5.6\%}$  \\
 & $\tilde\tau_1^+\tilde\tau_1^-$  & $0.22_{-5.8\%}^{+6.5\%}$ & $0.25_{-2.2\%}^{+2.1\%}{}_{-7.1\%}^{+5.9\%}$ &  $0.24_{-0.3\%}^{+0.0\%}{}_{-7.2\%}^{+5.8\%}$  \\
\hline
\end{tabular} 
 \caption{\label{tab:14tev} Same as Table~\ref{tab:13tev}, but for a center-of-mass energy
  of 14~Tev.}
 \end{center} 
 \end{table}

\acknowledgments

This work has been supported by the BMBF Theorie-Verbund and by the Theory-LHC-France initiative of the
CNRS/IN2P3.

\bibliographystyle{JHEP}
\bibliography{biblio}

\end{document}